\documentclass[a4paper,11pt]{article}
\pdfoutput=1
\usepackage[no-natbib-sort]{jheppub}

\usepackage[utf8]{inputenc}
\usepackage{etoolbox}
\usepackage{hhline}
\usepackage{array}
\usepackage{amssymb,amsmath,amsthm,amsfonts,graphicx}
\usepackage{latexsym}
\usepackage{bm}
\usepackage[]{hyperref}
\usepackage{cleveref}
\usepackage{nccmath}	
\usepackage{color}

\makeatletter
    \patchcmd{\maketitle}{\@fpheader}{}{}{}
    \makeatother

\def\be{\begin{equation}}
\def\ee{\end{equation}}
\def\ba{\begin{eqnarray}}
\def\ea{\end{eqnarray}}

\title{The BTZ black hole violates strong cosmic censorship}

\author[a]{Oscar~J.~C.~Dias,}
\emailAdd{ojcd1r13@soton.ac.uk}
\affiliation[a]{STAG research centre and Mathematical Sciences, Highfield Campus, University of Southampton, Southampton SO17 1BJ, UK}

\author[b]{Harvey~S.~Reall}
\emailAdd{hsr1000@cam.ac.uk}
\affiliation[b]{Department of Applied Mathematics and Theoretical Physics, University of Cambridge, Wilberforce Road, Cambridge CB3 0WA, UK} 

\author[b]{and Jorge~E.~Santos}
\emailAdd{jss55@cam.ac.uk}

\abstract{
We investigate the stability of the inner horizon of a rotating BTZ black hole. We show that linear perturbations arising from smooth initial data are arbitrarily differentiable at the inner horizon if the black hole is sufficiently close to extremality. This is demonstrated for scalar fields, for massive Chern-Simons fields, for Proca fields, and for massive spin-2 fields. Thus the strong cosmic censorship conjecture is violated by a near-extremal BTZ black hole in a large class of theories. However, we show that a weaker ``rough" version of the conjecture is respected. We calculate the renormalized energy-momentum tensor of a scalar field in the Hartle-Hawking state in the BTZ geometry. We show that the result is finite at the inner horizon of a near-extremal black hole. Hence the backreaction of vacuum polarization does not enforce strong cosmic censorship.}

\begin{document}

\maketitle

\section{Introduction}

The Reissner-Nordstr\"om and Kerr black holes have the property that they can be analytically extended beyond an inner horizon. This inner horizon is a Cauchy horizon bounding the region of spacetime in which physics can be predicted uniquely from initial data. Even the extension of the spacetime beyond the Cauchy horizon is not predictable: there are infinitely many smooth (but non-analytic) extensions that satisfy Einstein's equation. Thus the presence of a Cauchy horizon represents a failure of determinism in classical physics. 

Fortunately it is well-known that the Cauchy horizons of Reissner-Nordstr\"om or Kerr black holes are {\it unstable} \cite{Simpson:1973ua,McNamara499,chandra,Poisson:1990eh,Dafermos:2003wr}. If the initial data is perturbed then the resulting spacetime can still be continuously extended beyond a Cauchy horizon \cite{Ori:1991zz,Dafermos:2003wr,Dafermos:2017dbw}. But, generically, this extension is not $C^2$ \cite{Poisson:1990eh,Luk:2017jxq} and it is believed that, generically, no extension can satisfy Einstein's equation, even in a weak sense, at the Cauchy horizon (see \cite{Dafermos:2012np} for a discussion). Thus, in the perturbed spacetime, there is strong evidence that the Cauchy horizon is replaced by a curvature singularity and determinism is restored. The claim that this must happen is called the strong cosmic censorship conjecture \cite{penrose}.  

There has been recent interest in strong cosmic censorship with a positive cosmological constant. It has been argued that strong cosmic censorship is violated by near-extremal Reissner-Nordstr\"om-de Sitter black holes \cite{Cardoso:2017soq,Dias:2018etb,Luna:2018jfk} (see \cite{Mellor:1989ac} for earlier work). In particular, it has been shown that linearized gravitational and electromagnetic perturbations of such a hole can be arbitrarily differentiable at the Cauchy horizon if the black hole is near-extremal and large enough \cite{Dias:2018etb}. On the other hand, it has been shown that linear gravitational perturbations of a Kerr-de Sitter black hole exhibit behaviour that is in agreement with strong cosmic censorship \cite{Dias:2018ynt}. 

What about the case of negative cosmological constant? In four spacetime dimensions, linear perturbations of asymptotically anti-de Sitter (AdS) black holes exhibit very slow decay. This was first noticed by studying quasinormal modes of such black holes \cite{Festuccia:2008zx}. This slow decay arises from the {\it stable trapping} of null geodesics in such spacetimes \cite{Holzegel:2011uu}: there exist null geodesics which orbit the black hole, and such orbits are stable against perturbations.  The very slow decay of perturbations outside the black hole is expected to strengthen the instability of the Cauchy horizon.\footnote{But not enough to prevent the existence of a continuous extension across the Cauchy horizon \cite{Kehle:2018zws}. See also \cite{Bhattacharjee:2016zof} for a discussion of tidal forces at the Cauchy horizon.} Thus, in 4d, AdS black holes are expected to respect strong cosmic censorship.

In this paper, we will consider the rotating BTZ black hole, which is a 3-dimensional vacuum solution with negative cosmological constant \cite{Banados:1992wn,Banados:1992gq}. In contrast with the 4-dimensional case, the BTZ black hole does not exhibit stable trapping, and its quasinormal modes do not exhibit slow decay. Hence it is not obvious whether or not the BTZ black hole will respect strong cosmic censorship. 

We will start by considering the behaviour of a free massive scalar field in the BTZ geometry. The initial perturbation is taken to be a smooth outgoing wavepacket emanating from the white hole region of the geometry. We also allow for outgoing waves in the black hole interior and a ``non-normalizable" wavepacket incident from infinity. The resulting solution is not smooth at the Cauchy horizon. The degree of differentiability at the Cauchy horizon depends on how close the black hole is to extremality. 
If the black hole is far from extremality then the scalar field is not $C^1$ at the Cauchy horizon, which implies that its energy-momentum tensor diverges there, in agreement with strong cosmic censorship. However, if the black hole is close to extremality then the scalar field is smoother. We will show that the scalar field is $C^k$ at the Cauchy horizon if $\beta > k$ with
\be
\label{betadef1}
 \beta \equiv \frac{\Delta}{\displaystyle \frac{r_+}{r_-} -1}\,,
\ee
where $\Delta$ is the conformal dimension of the operator dual to the scalar field in the AdS/CFT correspondence \cite{Aharony:1999ti},\footnote{
Our results do not depend in any way on AdS/CFT but it is convenient to use the parameter $\Delta$  as it encodes both the mass of the scalar field and the boundary conditions that it satisfies.} and $r_\pm$ are the radii of the BTZ horizons. This implies that the energy-momentum tensor of the scalar field is finite at the Cauchy horizon if $\beta>1$. Note that $\beta$ diverges in the extremal limit. 
Therefore, for any given $\Delta$ and $k$, {\it scalar field perturbations are $C^k$ at the Cauchy horizon if the black hole is close enough to extremality}. If there are multiple scalar fields then the one with the smallest value of $\Delta$ exhibits the least smooth behaviour at the Cauchy horizon. 

This result relies on a surprising coincidence. The analysis reduces to considering quasinormal modes of the BTZ black holes. These can be divided into two classes: prograde ({\it i.e.}, co-rotating) and retrograde ({\it i.e.}, counter-rotating) \cite{Birmingham:2001hc}. The coincidence is that the prograde quasinormal frequencies coincide with the frequencies of a class of quasinormal modes of the black hole {\it interior}. (We will define below precisely what we mean by interior quasinormal modes.) It turns out that this implies that the prograde modes do not contribute to the non-smooth part of the field at the Cauchy horizon, which is determined by the much faster-decaying retrograde modes. This rapid decay leads to the enhanced differentiability near extremality.

We will also discuss linear fields with higher spin. These include massive Chern-Simons fields ({\it i.e.} Maxwell fields with a gauge-invariant mass term), Proca fields and massive spin-2 fields. In all cases we find that the equations of motion can be reduced to the massive scalar equation and so we obtain the same result, {\it i.e.}, a violation of strong cosmic censorship near extremality.

We emphasize that the discussion above concerns perturbations arising from {\it smooth} initial data. The similar failure of the strong cosmic censorship conjecture for Reissner-Nordstr\"om de Sitter black holes \cite{Cardoso:2017soq,Dias:2018etb} led to the formulation of a weaker version of the conjecture in which one allows perturbations arising from non-smooth initial data \cite{Dafermos:2018tha} (see also \cite{Dias:2018etb}). We will show that this weaker form of strong cosmic censorship is respected by the BTZ black hole. However, it is not clear whether one should allow non-smooth initial data in, say, the AdS/CFT correspondence.

Since (for smooth initial data) strong cosmic censorship appears to be violated classically, we examine whether quantum effects might help. If we consider a BTZ black hole formed by gravitational collapse then, at late time, it is natural to expect the quantum state of fields in this spacetime to approach the Hartle-Hawking state. In this state, quantum fields will exhibit vacuum polarization, {\it i.e.}, the expectation value of the energy-momentum tensor $\langle T_{ab} \rangle$ will be non-vanishing. Therefore we can ask how  $\langle T_{ab} \rangle$ behaves near the Cauchy horizon of the black hole. If it always diverges at the Cauchy horizon (as in a 2-dimensional toy model \cite{Birrell:1978th}) then maybe the gravitational backreaction of vacuum polarization could enforce strong cosmic censorship. 

We have computed $\langle T_{ab} \rangle$ for a massive scalar field in the Hartle-Hawking state in the BTZ black hole spacetime. Our result is that, for a near-extremal black hole,  $\langle T_{ab} \rangle$ is {\it finite} on the Cauchy horizon. More precisely, it {\it extends continuously} to the Cauchy horizon of a black hole with $\beta>1$. Thus vacuum polarization does not save strong cosmic censorship for the BTZ black hole. Surprisingly, the condition $\beta>1$ is the same as the condition for {\it classical} scalar field perturbations to have finite energy-momentum tensor at the Cauchy horizon. 

Our results contradict claims in the literature. A previous study of classical scalar field perturbations of the BTZ black hole reported results in agreement with strong cosmic censorship \cite{Balasubramanian:2004zu}. We will explain below why we think this study contains a subtle error in its treatment of certain poles on the real axis in the complex frequency plane. In the quantum case, there seems to be a belief that a calculation of Steif \cite{Steif:1993zv} demonstrates that $\langle T_{ab} \rangle$  always diverges at the Cauchy horizon. We will explain below why this conclusion is invalid because it involves considering the behaviour of quantum fields {\it behind} the Cauchy horizon where we can predict neither the classical geometry nor the behaviour of quantum fields. The correct approach is to consider the limiting behaviour of $\langle T_{ab} \rangle$ as the Cauchy horizon is approached from {\it outside}. 

This paper is organized as follows. In Section \ref{sec:background} we review some background material that will be used in the following sections. 
In Section~\ref{sec:sccBTZ} we discuss classical scalar field perturbations of the BTZ black hole and demonstrate that they violate strong cosmic censorship. In Section~\ref{sec:fields} we find that strong cosmic censorship cannot be restored by other classical fields (Chern-Simons, Proca, Kaluza-Klein gravitons) that naturally appear when we embed the BTZ solution in supergravity theories. In Section \ref{sec:quantum} we will show that $\langle T_{ab} \rangle$ is finite at the Cauchy horizon of a near-extremal BTZ black hole. Finally, we discuss our results further in Section~\ref{sec:discussion}.

{\bf Notation.} Latin indices on tensor equations are abstract indices, indicating that the equation holds in any basis. Greek indices refer to a specific basis.

\section{Background material}
\label{sec:background}

\subsection{The BTZ solution}\label{sec:BTZ}

The BTZ black hole is a solution of Einstein's gravity in three dimensions with a negative cosmological constant $\Lambda=-1/L^2$, where $L$ is the AdS radius. This solution (which is locally AdS$_3$) has metric \cite{Banados:1992wn,Banados:1992gq}
\begin{eqnarray}
&& \mathrm{d}s^2 = -f \mathrm{d}t^2 + \frac{\mathrm{d}r^2}{f}+r^2 \bigl( \mathrm{d}\phi - \Omega\, \mathrm{d}t \bigr)^2\,, \nonumber\\
&& f=\frac{(r^2-r_+^2)(r^2-r_-^2)}{L^2\, r^2}\,,\qquad  \Omega=\frac{r_+
r_-}{L \,r^2}\,.
\label{BTZmetric}
\end{eqnarray}
The function $f$ has two positive roots $0< r_-\le r_+$ corresponding to the Cauchy horizon and event horizon, respectively. The asymptotic timelike AdS$_3$ boundary is at $r\to \infty$. The (positive) surface gravities $\kappa_{\pm}$ and the angular velocities associated to each of the two horizons are, respectively,
\be\label{kappaOmega}
\kappa_{\pm} = \frac{r_+^2 - r_-^2}{L^2 \, r_\pm} \,, \qquad\Omega_{\pm}=\frac{r_\mp}{L \,r_{\pm}}\,. 
\ee
It will be important to note that any non-extremal BTZ black hole satisfies
\be
\label{kappaineq}
\kappa_- > \kappa_+\,, 
\ee 
and  $\Omega_{-} >\Omega_{+}$. The extremal configuration occurs when $\kappa_+=\kappa_-=0$, {\it i.e.} when $r_+=r_-$.
For completeness, the mass and angular momentum are $M= (r_+^2 + r_-^2)/L^2$ and $J= 2r_+ r_-/L$.

\begin{figure}[ht]
	\centering
	\includegraphics[width=0.6\textwidth]{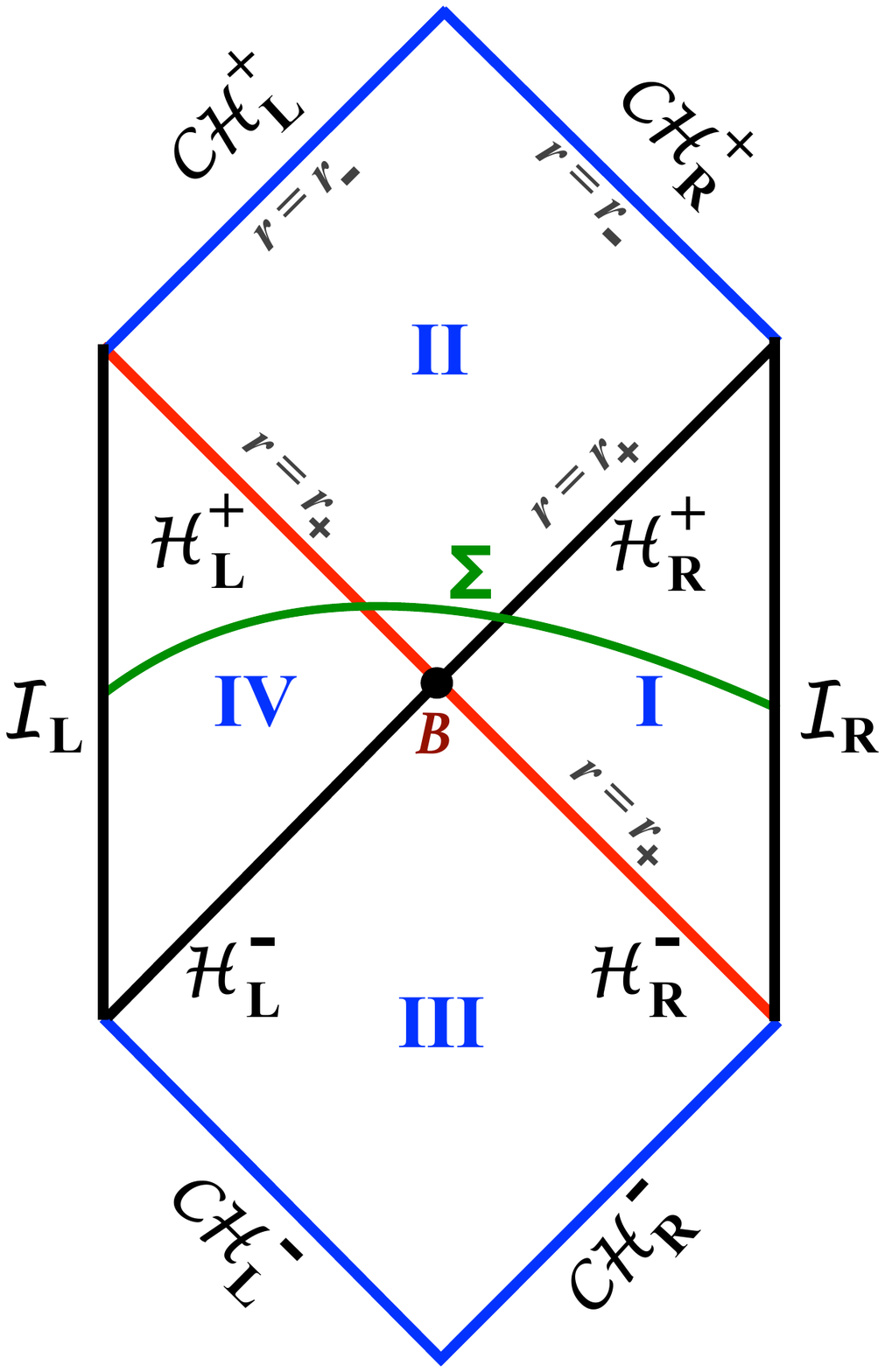}
	\caption{Penrose diagram for the non-extremal BTZ black hole. Region I is the black hole exterior and region II the interior. The asymptotic timelike boundary is $\mathcal{I}_R$, the future event horizon is $\mathcal{H}^+_{R}$ and the past event horizon is at $\mathcal{H}^-_R$. Region III is the white hole region. Region IV is another asymptotically AdS region with timelike boundary $\mathcal{I}_L$ and future event horizon $\mathcal{H}^+_{L}$. $\mathcal{CH}_{L,R}^+$ are the ``left" and ``right" future Cauchy horizons, respectively.}
	\label{fig:penrose}
\end{figure}

The causal structure of a non-extremal BTZ black hole is displayed in Fig. \ref{fig:penrose}. Region I is the region with $r_+ < r < \infty$ between the event horizon and the timelike boundary, {\it i.e.} the black hole exterior. Region II is the black hole interior, where $r_-<r<r_+$. Region III is the white hole region and region IV is another asymptotically AdS region.

As we will explain below, the geometry can be analytically continued beyond the surfaces $r=r_-$ into new regions with $r<r_-$. In the extended spacetime, the surfaces $r=r_-$ are Cauchy horizons.\footnote{Here we use the term Cauchy horizon in the context of the initial-boundary value problem appropriate to asymptotically AdS spacetimes.} Consider any spacelike surface $\Sigma$ extending from infinity in region I to infinity in region IV, {\it e.g.} the one shown in Fig. \ref{fig:penrose}. Then, given initial conditions on $\Sigma$ and appropriate boundary conditions at timelike infinity, physics is uniquely determined throughout regions I to IV. However what happens beyond the Cauchy horizon cannot be predicted from initial data on $\Sigma$. 

We should emphasize that even the metric is not uniquely determined beyond the Cauchy horizon. Analytic continuation can be used to extend the metric beyond the Cauchy horizon. But there is no reason why the metric should be analytic. 
If we merely demand smoothness then there are infinitely many ways of extending the metric. To see this, consider 3d gravity coupled to matter obeying suitable boundary conditions at timelike infinity (see below). Start from exactly BTZ initial data on a surface such as $\Sigma$. Exactly BTZ initial data implies that no matter fields are present initially. The matter fields will remain zero, and the solution will be exactly BTZ, all the way up to the Cauchy horizon. However, there is no reason why matter fields should remain zero beyond the Cauchy horizon. Non-zero matter fields will affect the geometry beyond the Cauchy horizon so that it differs from the metric obtained via analytic continuation of BTZ. It would be interesting to know whether one also has non-uniqueness of smooth extensions in the absence of matter fields. This is less obvious in 3d than in higher dimensions owing to the absence of propagating gravitational degrees of freedom. 

So far we have been discussing the eternal BTZ black hole. If we consider a black hole formed in gravitational collapse that, at late times, ``settles down" to BTZ then the geometry of this black hole will approach the BTZ geometry in a neighbourhood of the intersection of ${\cal I}_R$, ${\cal H}_R^+$ and ${\cal CH}_R^+$ on Fig. \ref{fig:penrose}. This implies that it is the early time portion of ${\cal CH}_R^+$ that is relevant for a black hole formed in gravitational collapse. For this reason, most of our discussion will focus on ${\cal CH}_R^+$ rather than ${\cal CH}_L^+$. 

We need to recall the definitions of various coordinate systems in the BTZ geometry \cite{Banados:1992gq}. Define the ``tortoise" radial coordinate $r_*$ by
\be
 r_* = \frac{1}{2 \kappa_+} \log | F_+(r) |\,,
\ee
where
\be
\label{Fpdef}
 F_+(r) \equiv \frac{r-r_+}{r+r_+} \left( \frac{r+r_-}{r-r_-} \right)^{r_-/r_+}
\ee
and this definition implies $dr_* = dr/f$. Note that $r_* \rightarrow -\infty$ as $ r\rightarrow r_+$ and $r_* \rightarrow + \infty$ as $r \rightarrow \infty$ or $r\rightarrow r_-$. We then define advanced and retarded time coordinates in region I as
\be
\label{uvdef}
 u = t-r_*\,,\qquad v=t+r_*\,.
\ee
We now define Kruskal coordinates $(U_+,V_+, \phi_{+})$ in region I (where $U_+<0$, $V_+>0$) by \be \label{KruskalEvent}
 U_+ = -e^{-\kappa_+ u}\,,  \qquad V_+ = e^{\kappa_+ v}\,, \qquad \phi_+ = \phi - \Omega_+ t\,.
\ee
In these coordinates we have
\be
\label{tdef}
 t = \frac{1}{2\kappa_+} \log \left| \frac{V_+}{U_+} \right|
\ee
and $r$ is given by solving
\be
\label{rUVplus}
F_+(r) =  - U_+ V_+ \,.
\ee
The metric becomes
\be
\mathrm{d}s^2=-\frac{f(r)}{\kappa_+^2 F_+(r)} \mathrm{d}U_+ \mathrm{d}V_+ + r^2 \left[ \mathrm{d}\phi_+ + \frac{(\Omega_+ - \Omega(r))}{2 \kappa_+ F_+(r)} \left( V_+ \mathrm{d}U_+ - U_+ \mathrm{d}V_+ \right) \right]^2.
\ee
These coordinates also allow the metric to be analytically extended into region II (where $U_+>0$, $V_+>0$) as well as two further regions not displayed in Fig. \ref{fig:penrose}. The future event horizon ${\cal H}_R^+$ is the surface $U_+ = 0$ (and $V_+>0$). An observer with constant $ \phi_{+}$ co-rotates with the event horizon, {\it i.e.},  $\phi_{+}$ is constant along the generators of the event horizon.

In region II we define $t$ by \eqref{tdef} and $(u,v)$ by \eqref{uvdef}. We then have
\be
 U_+ = +e^{- \kappa_+ u}\,,\qquad V_+ = e^{\kappa_+ v}\,, \qquad {\rm region \;\;\; II.}
\ee
Note that $u \rightarrow + \infty$ as we approach ${\cal H}_R^+$ from either region I or II. We define $\phi$ in region II by $\phi = \phi_+ + \Omega_+ t$. 

To extend across the Cauchy horizon we need to define another set of Kruskal coordinates in region II by
\be\label{KruskalCauchy}
 U_- = - e^{ \kappa_- u}\,,  \qquad V_- = - e^{- \kappa_- v}\,, \qquad \phi_- = \phi - \Omega_- t\,.
\ee 
In these coordinates,  region II has $U_-<0,V_-<0$ and $r$ is given by
\be
\label{rUVminus}
 F_-(r) = U_- V_-\,,
\ee
where
\be
\label{Fmdef}
 F_-(r) \equiv \frac{r-r_-}{r+r_-} \left( \frac{r_+ + r}{r_+ - r} \right)^{r_+/r_-}.
\ee
The metric becomes
\be\label{metricKruskalCauchy}
\mathrm{d}s^2=\frac{f(r)}{\kappa_{-}^2 F_-(r)} {\rm d}U_-{\rm d}V_-+r^2\left[ {\rm d}\phi_- +\frac{\Omega_--\Omega(r)}{2\kappa_- F_-(r)}\left( V_-{\rm d}U_- - U_- {\rm d}V_-\right)\right]^2.
\ee
This metric can now be analytically continued to $U_->0$ and/or $V_->0$. The ``right" Cauchy horizon ${\cal CH}_R^+$ is the surface $V_-=0$ (and $U_-<0$) while the ``left" Cauchy horizon ${\cal CH}_R^+$ is the surface  $U_-=0$ (and $V_-<0$). 

Finally, we will also make use of Eddington-Finkelstein coordinates. Ingoing Eddington-Finkelstein coordinates are $(v,r,\phi')$ where
\be
\label{vphi'}
 \mathrm{d}\phi =  \mathrm{d}\phi' - \frac{\Omega}{f}  \mathrm{d}r \,.
\ee
In these coordinates the metric is 
\be
 \mathrm{d}s^2 = -f  \mathrm{d}v^2 + 2  \mathrm{d}v  \mathrm{d}r + r^2 \bigl(  \mathrm{d}\phi'-\Omega  \mathrm{d}v \bigr)^2\,. \label{inBTZ}
\ee
This metric can be analytically extended across the future event horizon $\mathcal{H}^+_{R}$  (at $r=r_+$) into region II so these coordinates cover regions I and II of Fig. \ref{fig:penrose}. These coordinates are also smooth at ${\cal CH}_L^+$ ({\it i.e.} at $r=r_-$).


Outgoing Eddington-Finkelstein coordinates are $(u,r,\phi'')$ where 
\be
 \label{uphi''}
  \mathrm{d}\phi =  \mathrm{d}\phi''+ \frac{\Omega}{f} \mathrm{d} r \,.
\ee
The metric in these coordinates is
\be 
 \mathrm{d}s^2 = -f \mathrm{d}u^2 - 2 \mathrm{d}u \mathrm{d}r + r^2 \bigl( \mathrm{d}\phi''-\Omega \mathrm{d}u \bigr)^2\,. \label{outBTZ}
\ee
These coordinates cover region I and III. However they are not smooth at ${\cal H}_R^+$. We can also define these coordinates in region II using the above expressions. In region II these coordinates are smooth at ${\cal H}_L^+$ (where $r=r_+$) and ${\cal CH}_R^+$ (where $r=r_-$).

\subsection{Scalar field boundary conditions}\label{sec:Delta} 
Consider a test scalar field $\Phi(t,r,\phi)$ of mass $\mu$ satisfying the Klein-Gordon equation:
\be
\Box \Phi -\mu^2 \Phi=0\,. \label{KG}
\ee 
A Frobenius analysis of \eqref{KG} about the asymptotic boundary $r\to \infty$ finds two possible linearly independent decays\footnote{When $\mu^2>\mu_{\rm BF}^2$. For $\mu=\mu_{\rm BF}$ one of the independent solutions is a power of $r$ and the other is logarithmic.}
\be
\Phi \big|_{r\sim \infty}=r^{-\Delta_-}\left( A(t,\phi) +\cdots \right) + r^{-\Delta_+}\left( B(t,\phi) +\cdots \right)\,, \qquad \Delta_\pm=1\pm \sqrt{1+\mu^2L^2} \label{KGdecays}
\ee
where the ellipses denote terms that decay as $r \rightarrow \infty$. Note that $\Delta_+ \geq \Delta_-$. Stability of an asymptotically AdS$_3$ solution requires that $\Delta_\pm \in \mathbb{R}$, which occurs if the mass of the scalar field is above the Breitenl\"ohner-Freedman (BF) bound $\mu_{\rm BF}^2L^2=-1$ \cite{Breitenlohner:1982jf,Mezincescu:1984ev}.

For $\mu^2\geq 0$, only the mode $r^{-\Delta_+}$ with the faster fall-off  is normalizable. In the AdS/CFT correspondence, the non-normalizable mode $A(t,\phi)$ is said to be the source of a boundary operator $\mathcal{O}_\Phi$ since it determines the (deformation) of the boundary theory action. On the other hand, the normalizable modes $B(t,\phi)$ are identified with states of the theory and $B(t,\phi)$ is proportional to the expectation value $\langle \mathcal{O}_\Phi \rangle$ of the boundary operator (in the presence of the source $A$). $\Delta_+\equiv \Delta$ is then the (mass) conformal dimension of the boundary operator $\mathcal{O}_\Phi$ dual to $\Phi$. The undeformed boundary theory corresponds thus to the Dirichlet boundary condition choice whereby the source vanishes, $A(t,\phi)=0$, and we have a pure normalizable solution.

 \begin{figure}[hb]
	\centering
	\includegraphics[width=0.5\textwidth]{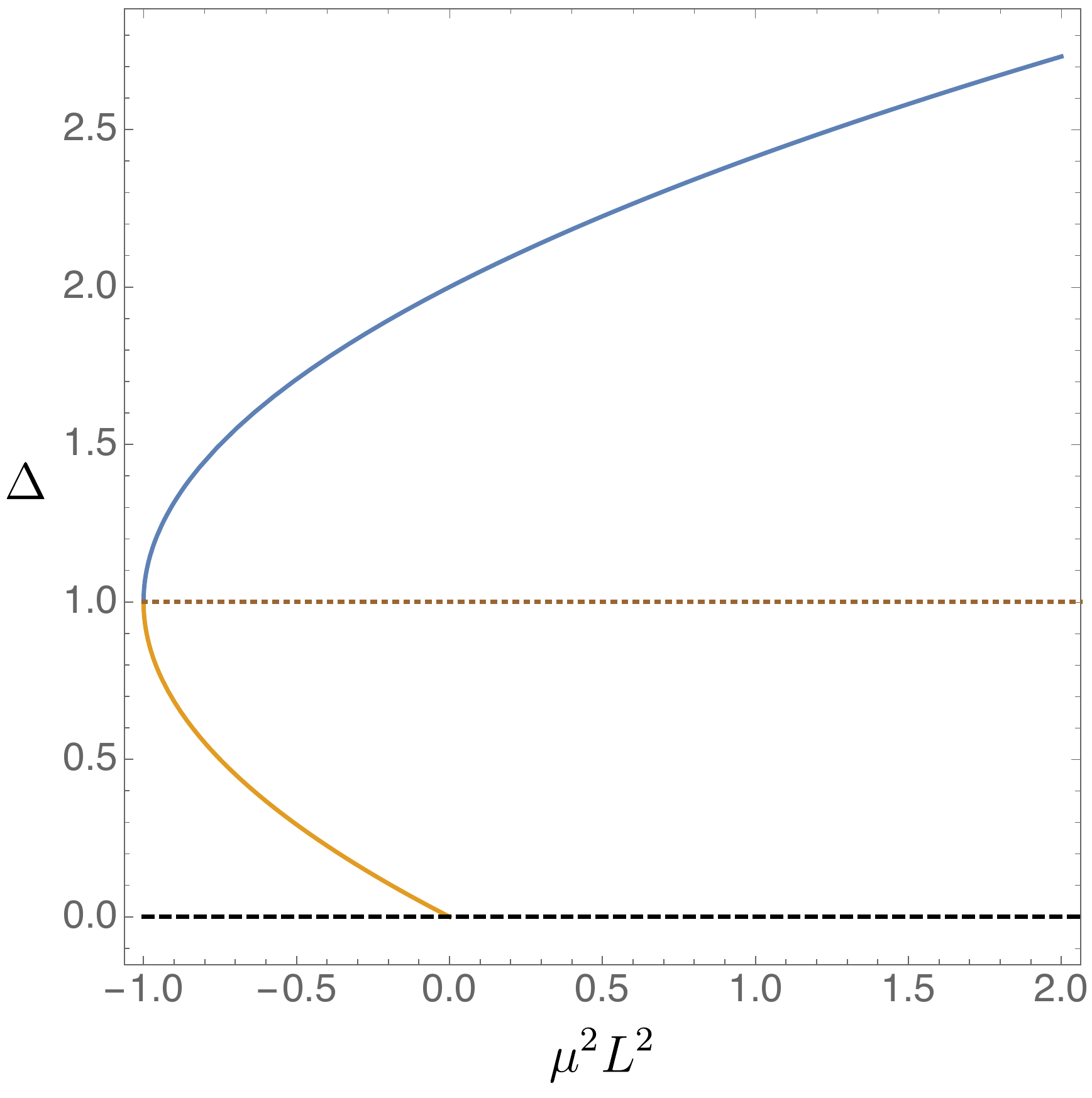}
	\caption{The conformal dimension $\Delta$ as a function of the scalar field mass $\mu^2 L^2$. The upper branch (blue) corresponds to $\Delta_+$, the lower branch (orange) to $\Delta_-$. Note $\Delta_\pm(\mu_{\rm BF})=1$ and $\Delta_-(0)=0$.
}
	\label{fig:Delta}
\end{figure} 

For masses in the range $\mu_{\rm BF}^2 < \mu^2<0$, the two modes in \eqref{KGdecays} are normalizable \cite{Klebanov:1999tb}. We then have two choices of boundary conditions: 1) the so called standard quantization where $A(t,\phi)$ is identified with the source and we are often interested on the Dirichlet boundary condition $A(t,\phi)=0$, or 2) the alternative quantization where it is instead $B(t,\phi)$ that is identified with the source (absence sources then  requires the Neumann boundary condition, $B(t,\phi)=0$). These two choices correspond, respectively, to dual operators with conformal dimension $\Delta_+$ and $\Delta_-$ (since the mass scaling dimensions of $B(t,\phi)$ and $A(t,\phi)$ are $\Delta_+$ and $\Delta_-$, respectively).


We can write the scalar mass $\mu$ as a function of the conformal dimension $\Delta$ of ${\cal O}_\Phi$:
\be\label{mDelta}
\mu^2 L^2=\Delta(\Delta-2)\,.
\ee
 This relationship is shown in Fig. \ref{fig:Delta}. 
In the next section we will use \eqref{mDelta} to present our results in terms $\Delta$ (instead of $\mu$) because $\Delta$ uniquely determines both the mass and the boundary conditions. 

\section{Scalar field perturbations \label{sec:sccBTZ}}

\subsection{Introduction}

In this section we will consider the behaviour of linear scalar field perturbations of the BTZ spacetime. Ideally we would like to specify smooth initial data for the scalar field on a surface such as $\Sigma$ in Fig. \ref{fig:penrose}. We then want to determine the behaviour of the resulting solution at the Cauchy horizon. In particular we want to know: for generic initial data, how smooth is the scalar field at the Cauchy horizon? 

\begin{figure}[ht]
	\centering
	\includegraphics[width=0.6\textwidth]{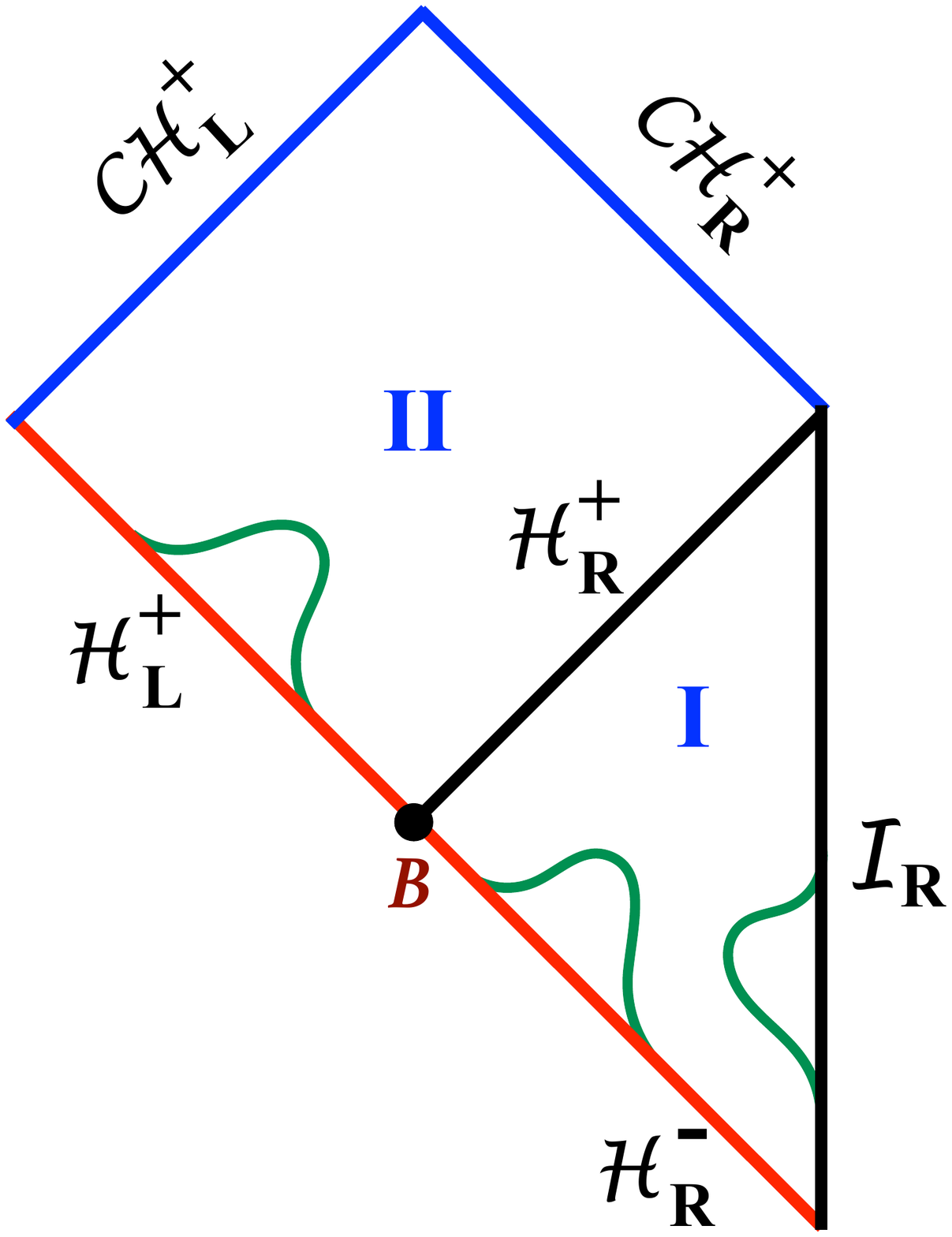}
	\caption{Our (characteristic) initial-boundary data for solving the Klein-Gordon equation consists of smooth (outgoing) wavepackets on ${\cal H}_L^+$ and ${\cal H}_R^-$ and a smooth wavepacket source at ${\cal I}_R$. We will determine the solution in regions I and II and investigate its smoothness at ${\cal CH}_R^+$.}
	\label{fig:scattering}
\end{figure}

Instead of specifying initial data on a spacelike surface, it is more convenient to specify (characteristic) initial data on the null surface ${\cal H}_L^+\cup {\cal H}_R^-$ along with suitable boundary conditions at timelike infinity ${\cal I}_R$. Given such data one expects a unique solution throughout regions I and II of Fig. \ref{fig:penrose}. As shown in Fig. \ref{fig:scattering}, we will take the data on ${\cal H}_L^+\cup {\cal H}_R^-$ to consist of a smooth wavepacket on ${\cal H}_L^+$ and a smooth wavepacket on ${\cal H}_R^-$. The boundary conditions at ${\cal I}_R$ will allow for a ``non-normalizable" perturbation corresponding to a wavepacket on ${\cal I}_R$. 

The advantage of formulating the problem as above is that we can write the solution as a superposition of mode solutions of the Klein-Gordon equation. Thus, in subsection~\ref{sec:bases} we review mode solutions in regions I and II and natural bases to construct wavepackets. The discussion of strong cosmic censorship will turn out to depend crucially on properties of quasinormal frequencies (just as for the de Sitter case \cite{Hintz:2015jkj,Cardoso:2017soq,Dias:2018etb}). Surprisingly, it depends not only on the quasinormal frequencies of the black hole exterior (region I) but also on quasinormal frequencies of the black hole interior (region II).  Therefore, in subsection \ref{sec:extQNM} we review the values of the {\it ``exterior"} quasinormal mode frequencies of scalar field perturbations of BTZ \cite{Birmingham:2001hc} and, in subsection \ref{sec:intQNM}, we will introduce the notion of  {\it ``interior"} quasinormal modes and find the associated frequencies. A remarkable coincidence between one of these interior frequencies and one of the exterior quasinormal mode frequencies will be observed. This coincidence will have far-reaching consequences for the fate of strong cosmic censorship in BTZ which we will explain in subsection \ref{sec:sccBTZviolation}.

We should note that there is significant overlap between our analysis and the previous analysis of Ref. \cite{Balasubramanian:2004zu}. However, a certain amount of repetition is necessary in order for us to explain carefully why we disagree with the conclusion of Ref. \cite{Balasubramanian:2004zu}. 

\subsection{Bases for mode solutions \label{sec:bases}}

  
To study solutions of the massive Klein-Gordon equation it is convenient to replace the original coordinate $r$ by the new radial coordinate $z$,
\be
z=\frac{r^2-r_-^2}{r_+^2-r_-^2}\,, \label{zcoord}
\ee
which has the nice property of locating the Cauchy horizon at $z=0$, the event horizon at $z=1$ and the asymptotic boundary at $z=+\infty$. A mode solution of the Klein-Gordon equation \eqref{KG} has the separable form
\begin{eqnarray}\label{ansatzScalarField}
 \Phi&=& e^{-i \omega t} e^{i m \phi} R(z)\nonumber\\
 &\equiv& e^{-i \omega t} e^{i m \phi}z^{-i\,\frac{\omega-m\,\Omega_{-}}{2\kappa_-}}(1-z)^{-i\,\frac{\omega-m\,\Omega_{+}}{2\kappa_+}}F(z) \,,
\end{eqnarray}
where $\omega$ and $m$ are the frequency and azimuthal  number of the mode and $\kappa_\pm$ and $\Omega_\pm$ are the surface gravities and angular velocities of BTZ already introduced in \eqref{kappaOmega}. Regularity of the solution requires that $m$ is an integer. 
In \eqref{ansatzScalarField} we have redefined the radial function $R(z)$ in terms of $F(z)$.  This satisfies the hypergeometric equation
\be \label{hypergeometric}
z (1-z) F''(z)+[c-z (a+b+1)]F'(z)-a b \,F(z)=0\,, 
\ee
with (after using \eqref{mDelta} to replace the scalar field mass $\mu$ by the conformal dimension $\Delta$)\footnote{Note that the hypergeometric function has the property $\, _2F_1(a,b;c;z)=\, _2F_1(b,a;c;z)$ which, together with the property $|\Delta-1|+1=\Delta$ if $\Delta>1$ but $|\Delta-1|+1=2-\Delta$ if $0<\Delta<1$,  allows us to treat the $\Delta<1$ and $\Delta>1$ cases in a unified way.}
\begin{eqnarray}\label{abc}
&& a=\frac{1}{2}\left(\Delta -i\,\frac{\omega-m\,\Omega_{-}}{\kappa_-}   -i\,\frac{\omega-m\,\Omega_{+}}{\kappa_+} \right),\nonumber\\
&& b=\frac{1}{2}\left(2-\Delta -i\,\frac{\omega-m\,\Omega_{-}}{\kappa_-}   -i\,\frac{\omega-m\,\Omega_{+}}{\kappa_+} \right), \nonumber \\
&& c=1-i\,\frac{\omega-m\,\Omega_{-}}{\kappa_-} \,.
\end{eqnarray}
The general solution of equation \eqref{hypergeometric} can be written in terms of a basis consisting of two linearly independent solutions. Different choices of basis are convenient for describing the properties of the solution near each singular point $z=0,1,\infty$. This gives a natural trio of bases for $R(z)$. We will describe each of these three bases, starting from the inner boundary and moving towards the asymptotic one.

We  start by considering solutions in region II ($r_-<r<r_+$ \emph{i.e.} $0<z<1$). In the vicinity of the Cauchy horizon ($r=r_-$ or $z=0$) two linearly independent solutions for $R(z)$ are \cite{Abramowitz:1964}\footnote{The two solutions in all our bases are strictly independent only when $a-b=\Delta-1$ is {\it not} an integer, \emph{i.e.} when $\Delta\geq 0$ is not an integer \cite{Abramowitz:1964}. However, our final physical results should be valid for any $\Delta > 0$. 
 } 
\begin{subequations}\label{Cauchy:basis}
\begin{align}
& R_{{\rm out},-}=z^{-\frac{1}{2} (1-c)} (1-z)^{\frac{1}{2} (a+b-c)} \, _2F_1(a,b;c;z)\,, \\
& R_{{\rm in},-}=z^{+\frac{1}{2} (1-c)} (1-z)^{\frac{1}{2} (a+b-c)} \, _2F_1(a-c+1,b-c+1;2-c;z)\,.
\end{align}
\end{subequations}
Using the hypergeometric property $F(\alpha,\beta,\gamma;0)=1$ one finds that 
the two solutions behave at the Cauchy horizon $z=0$ as 
\begin{subequations}\label{Cauchy:decay}
\begin{align}
& R_{{\rm out},-}\big|_{z\sim 0}=z^{-i\,\frac{\omega-m\,\Omega_{-}}{2\kappa_-}}\hat{R}_{{\rm out},-}(\omega,m;z)\,, \\
& R_{{\rm in},-}\big|_{z\sim 0}=z^{+i\,\frac{\omega-m\,\Omega_{-}}{2\kappa_-}}\hat{R}_{{\rm in},-}(\omega,m;z)\,.
\end{align}
\end{subequations}
where $\hat{R}_{{\rm out},-}(\omega,m;z)$ and  $\hat{R}_{{\rm in},-}(\omega,m;z)$ are analytic at $z=0$ with $\hat{R}_{{\rm out},-}(\omega,m;0)=\hat{R}_{{\rm in},-}(\omega,m;0)=1$.

We will denote the resulting mode solutions as $\Phi_{\rm out,-}$ and $\Phi_{\rm in,-}$. To investigate the behaviour of these solutions at ${\cal CH}_R^+$ we convert to the outgoing Eddington-Finkelstein coordinates $(u,r,\phi'')$ defined in section \ref{sec:BTZ}. This gives, near $z=0$,
\begin{subequations}\label{Cauchy:decayCHR}
\begin{align}
& \Phi_{\rm out,-}= e^{-i\omega (u-u_0)}e^{i m (\phi''-\phi''_0)}\left[1+\mathcal{O}\left(z\right)\right],\\
& \Phi_{\rm in,-}= e^{-i\omega (u-u_0)}e^{i m (\phi''-\phi''_0)}z^{+i\,\frac{\omega-m\,\Omega_{-}}{\kappa_-}}\left[1+\mathcal{O}\left(z\right)\right],
\end{align}
\end{subequations}
where $u_0$ and $\phi''_0$ are real constants depending only on the black hole parameters.
Thus, we see that $\Phi_{\rm out,-}$ is smooth ({\it i.e.} ``outgoing") at ${\cal CH}_R^+$ but $\Phi_{\rm in,-}$ is not. Similarly to investigate the behaviour at ${\cal CH}_L^+$ we convert to ingoing coordinates $(v,r,\phi')$ to obtain 
\begin{subequations}\label{Cauchy:decayCHL}
\begin{align}
& \Phi_{\rm out,-}= e^{-i\omega (v-v_0)}e^{i m (\phi'-\phi'_0)}z^{-i\,\frac{\omega-m\,\Omega_{-}}{\kappa_-}}\left[1+\mathcal{O}\left(z\right)\right], \\
& \Phi_{\rm in,-}= e^{-i\omega (v-v_0)}e^{i m (\phi'-\phi'_0)}\left[1+\mathcal{O}\left(z\right)\right].
\end{align}
\end{subequations}
Hence $\Phi_{\rm in,-}$ is smooth ({\it i.e.} ``ingoing") at ${\cal CH}_L^+$ but $\Phi_{\rm out,-}$ is not.

We now consider solutions in a neighbourhood of the event horizon ($z=1$). A basis for solutions $R(z)$ is  
\begin{subequations}\label{Event:basis2}
\begin{align}
& R_{{\rm out},+}=z^{-\frac{1}{2} (1-c)} |1-z|^{-\frac{1}{2} (a+b-c)} \, _2F_1(c-b,c-a;-a-b+c+1;1-z)\,, \\
& R_{{\rm in},+}=z^{-\frac{1}{2} (1-c)} |1-z|^{\frac{1}{2} (a+b-c)} \, _2F_1(a,b;a+b-c+1;1-z)\,.
\end{align}
\end{subequations}
At $r=r_+$ ($z=1$) this basis behaves as 
\begin{subequations}\label{event:decay}
\begin{align}
& R_{{\rm out},+}\big|_{z\sim 1}=|1-z|^{+i\,\frac{\omega-m\,\Omega_{+}}{2\kappa_+}}\hat{R}_{{\rm out},+}(\omega,m;z) \,, \\
& R_{{\rm in},+}\big|_{z\sim 1}=|1-z|^{-i\,\frac{\omega-m\,\Omega_{+}}{2\kappa_+}}\hat{R}_{{\rm in},+}(\omega,m;z)\,,
\end{align}
\end{subequations}
where $\hat{R}_{{\rm out},+}(\omega,m;z)$ and $\hat{R}_{{\rm in},+}(\omega,m;z)$ are analytic at $z=1$ with $\hat{R}_{{\rm out},+}(\omega,m;1)=\hat{R}_{{\rm in},+}(\omega,m;1)=1$. $R_{{\rm in},+}$ gives mode solutions $\Phi_{{\rm in},+}$ that are smooth at ${\cal H}_R^+$ while $R_{{\rm out},+}$ gives mode solutions $\Phi_{{\rm out},+}$ smooth at  ${\cal H}_L^+$ and ${\cal H}_R^-$ (see Fig. \ref{fig:penrose}).


Near the asymptotic timelike boundary $\mathcal{I}_R$ ($z\to \infty$) a basis for $R(z)$  is 
\begin{subequations}\label{bdry:basis}
\begin{align}
& R_{{\rm vev},\infty}=z^{-\frac{1}{2} (2 a-c+1)} (z-1)^{\frac{1}{2} (a+b-c)} \, _2F_1\left(a,a-c+1;a-b+1;\frac{1}{z}\right)\,, \\
& R_{{\rm source},\infty}=z^{-\frac{1}{2} (2 b-c+1)} (z-1)^{\frac{1}{2} (a+b-c)} \, _2F_1\left(b,b-c+1;-a+b+1;\frac{1}{z}\right)\,.
\end{align}
\end{subequations}
At $z\to \infty$ these decay as
\begin{subequations}\label{bdry:decay}
\begin{align}
& R_{{\rm vev},\infty}\big|_{z\sim \infty}=z^{-\Delta/2}\left[ 1+\mathcal{O}\left(1/z\right)\right] \,, \\
& R_{{\rm source},\infty}\big|_{z\sim \infty}=z^{-(2-\Delta)/2}\left[1+\mathcal{O}\left(1/z\right)\right]\,.
\end{align}
\end{subequations}
The corresponding mode solutions will be denoted $\Phi_{{\rm vev},\infty}$ and $\Phi_{{\rm source},\infty}$.

In later sections, we will be interested in studying how a solution written in the event horizon basis $(R_{{\rm in},+},R_{{\rm out},+})$ behaves near the Cauchy horizon. For that, it will be useful to rewrite the event horizon wavefunctions \eqref{Event:basis2} in the Cauchy horizon basis \eqref{Cauchy:basis},
\begin{subequations} \label{ABdefTot}
\begin{align} 
 & R_{{\rm out},+} = {\cal A}(\omega,m) R_{{\rm out},-} + {\cal B}(\omega,m) R_{{\rm in},-} 
\,,  \label{ABdef} \\
&R_{{\rm in},+} = \tilde{\cal A}(\omega,m) R_{{\rm in},-} + \tilde{\cal B}(\omega,m) R_{{\rm out},-}  \,,  \label{tildeABdef}
\end{align}
\end{subequations}
where ${\cal A}$ and ${\cal B}$ are transmission and reflection coefficients for fixed frequency scattering of waves propagating out from ${\cal H}_L^+$ and $\tilde{\cal A}$, $\tilde{\cal B}$ are transmission and reflection coefficients for scattering of waves propagating in from ${\cal H}_R^+$. Applying a hypergeometric transformation formulae \cite{Abramowitz:1964} to \eqref{Event:basis2} one finds that 
\begin{subequations} \label{ABexp}
\begin{align} 
 &  {\cal A}(\omega,m)=\frac{\Gamma (1-c) \Gamma (1-a-b+c)}{\Gamma (1-a) \Gamma (1-b)}\,, \qquad  {\cal B}(\omega,m)=\frac{\Gamma (c-1) \Gamma (-a-b+c+1)}{\Gamma (c-a) \Gamma (c-b)}\,;   \\
& \tilde{\cal A}(\omega,m)=\frac{\Gamma (c-1) \Gamma (a+b-c+1)}{\Gamma (a) \Gamma (b)}\,,  \qquad \tilde{\cal B}(\omega,m)=\frac{\Gamma (1-c) \Gamma (a+b-c+1)}{\Gamma (a-c+1) \Gamma (b-c+1)}  \,.
\end{align}
\end{subequations}


We will also need to know how to write $R_{{\rm in},+}$ in terms of the basis at infinity and how to write $R_{{\rm vev},\infty}$ in terms of the event horizon basis. We define the coefficients in this basis transformation as\footnote{Our notation is modelled on a similar calculation in \cite{Dias:2018etb}.}
\begin{subequations} \label{TRdefTot}
\begin{align} 
 &   R_{{\rm in},+} = \frac{1}{{\cal T}(\omega,m)} R_{{\rm source},\infty} + \frac{{\cal R}(\omega,m)}{{\cal T}(\omega,m)} R_{{\rm vev},\infty}\,,  \label{TRdef} \\
&  R_{{\rm vev},\infty} =\frac{1}{\tilde{\cal T}(\omega,m)} R_{{\rm out},+}+ \frac{\tilde{\cal R}(\omega,m)}{\tilde{\cal T}(\omega,m)} R_{{\rm in},+} \label{tildeTRdef} \,.
\end{align}
\end{subequations}
The coefficients defined in \eqref{TRdef}-\eqref{tildeTRdef}  can be obtained using hypergeometric transformation formulae, giving
\begin{subequations} \label{TRexpressions}
\begin{align} 
 &  {\cal T}(\omega,m)=\frac{\Gamma (a) \Gamma (a-c+1)}{\Gamma (a-b) \Gamma (a+b-c+1)}\,, \qquad  
 {\cal R}(\omega,m)=\frac{\Gamma (a) \Gamma (b-a) \Gamma (a-c+1)}{\Gamma (b) \Gamma (a-b) \Gamma (b-c+1)}\,;   \\
& \tilde{\cal T}(\omega,m)=\frac{\Gamma (a) \Gamma (a-c+1)}{\Gamma (a-b+1) \Gamma (a+b-c)}\,,  \qquad 
\tilde{\cal R}(\omega,m)=\frac{\Gamma (a) \Gamma (a-c+1) \Gamma (-a-b+c)}{\Gamma (1-b) \Gamma (c-b) \Gamma (a+b-c)}  \,.
\end{align}
\end{subequations}

Finally we will discuss the analyticity properties of our radial solutions in the complex $\omega$ plane. 
To do this we use the fact that the hypergeometric function $_2F_1(\alpha,\beta;\gamma;\zeta)$ is analytic in $\alpha,\beta,\gamma$ except for simple poles at $\gamma=-N$ where $N=0,1,2,\ldots$. From this it follows that $R_{{\rm in},-}(\omega,m;z)$ can be analytically continued to the complex $\omega$ plane, except for simple poles when $\omega-m\Omega_{-}$ is a positive integer multiple of $ i \kappa_-$. Similarly $R_{{\rm out},-}(\omega,m;z)$ has simple poles when $\omega-m\Omega_{-}$ is a negative integer multiple of $i \kappa_-$. $R_{{\rm in},+}(\omega,m;z)$ has simple poles when $\omega-m\Omega_{+}$ is a negative integer multiple of $ i \kappa_+$ and $R_{{\rm out},+}(\omega,m;z)$ has simple poles when  $\omega-m\Omega_{+}$ is a positive integer multiple of $i \kappa_+$. These results are similar to those  for 4d black holes \cite{chandra,Kehle:2018upl}. $ R_{{\rm vev},\infty}$ and  $R_{{\rm source},\infty}$ are entire functions of $\omega$. 

\subsection{Exterior quasinormal modes \label{sec:extQNM}}

The (standard, {\it {\it i.e.},  exterior}) black hole quasinormal mode frequencies of a massive scalar field in the rotating BTZ black hole background have been studied in detail in \cite{Birmingham:2001hc}. For completeness we will review them here.

Exterior quasinormal modes are linear mode solutions in region I that satisfy the no-source boundary condition at the asymptotic boundary ${\cal I}_R$ and are smooth at the future event horizon ${\cal H_R^+}$. The no-source condition implies that the radial function must be proportional to $R_{{\rm vev},\infty}$ and regularity at ${\cal H_R^+}$ implies that the radial function must be proportional to $R_{{\rm in},+}$. Hence quasinormal frequencies are defined by the condition $R_{{\rm in},+} \propto R_{{\rm vev},\infty}$.  From \eqref{TRdef} we see this is equivalent to ${\cal T}(\omega,m) = \infty$. Alternatively, from \eqref{tildeTRdef} it is equivalent to $\tilde{\cal T}(\omega,m)=\infty$.
\footnote{Note ${\cal R}/{\cal T}$ and $\tilde{\cal R}/\tilde{\cal T}$ are finite when ${\cal T}\to \infty$ or  $\tilde{\cal T}\to \infty$.}

It follows from \eqref{TRexpressions} that quasinormal frequencies are given by $a=-p$ or $a-c+1=-p$ for $p=0,1,2,\cdots$. Using the definitions \eqref{abc} for $(a,b,c)$, this yields two sectors of exterior quasinormal mode frequencies \cite{Birmingham:2001hc}:
\begin{subequations} \label{QNM}
\begin{align} 
 &  \omega_{\rm p} L=m-i\,\frac{r_+-r_-}{L}\big(\Delta +2 p\big), \qquad \hbox{for $p\in \mathbb{N}_0=\{0,1,2,\ldots\}$},   \\
& \omega_{\rm r} L=-m-i\,\frac{r_++r_-}{L}\big(\Delta +2 p\big), \qquad \hbox{for $p\in \mathbb{N}_0$}\,.
\end{align}
\end{subequations}
We refer to quasinormal modes with frequency $\omega_{\rm p}$ as {\it prograde} modes, {\it i.e.} modes that co-rotate with the black hole, while quasinormal modes with frequency $\omega_{\rm r}$ are  {\it retrograde} modes, {\it i.e.} modes that counter-rotate with the black hole.\footnote{This is an abuse of language when $m=0$.} 

For completeness, we note that we can define a second family of exterior quasinormal modes by demanding smoothness at ${\cal H}_R^-$ instead of at ${\cal H}_R^+$. This corresponds to imposing the boundary condition $\tilde{\cal R}(\omega,m)/\tilde{\cal T}(\omega,m)=0$ in \eqref{tildeTRdef}. This occurs if $1-b=-p$ or if $c-b=-p$ for $p=0,1,2,\cdots$. These are the ``white hole" quasinormal mode family. Its frequency spectrum is quantized as 
\begin{subequations} \label{QNMwhite}
\begin{align} 
 &  \omega_{\rm p,{\small WH}} = \bar{\omega}_{\rm p}\,, \\
& \omega_{\rm r,{\small WH}}=\bar{\omega}_{\rm r} \,,
\end{align}
\end{subequations}
where the bar denotes complex conjugation. That is, the exterior white hole quasinormal mode frequencies are just the complex conjugates of the exterior black  hole quasinormal mode frequencies. The white hole quasinormal frequencies have positive imaginary part so these modes grow exponentially with time. 


\subsection{Interior quasinormal modes  \label{sec:intQNM}}

In this subsection we will introduce another class of quasinormal modes defined in region II, the black hole interior. Therefore we call then {\it interior quasinormal modes}. Actually, there are four families of interior  quasinormal modes. One of these families will play a fundamental role in our discussion of strong cosmic censorship.

To introduce these interior modes we revisit \eqref{ABdefTot} where we write the event horizon wavefunctions \eqref{Event:basis2} in the Cauchy horizon basis \eqref{Cauchy:basis}. These two relations can be viewed as describing scattering of modes propagating out of ${\cal H}_{L,R}^+$ into modes that propagate across ${\cal CH}_{L,R}^+$ (see Fig. \ref{fig:penrose}). Equation 
\eqref{tildeABdef} describes {\it ``in-scattering"} in which an ingoing wave, proportional to $\Phi_{in,+}$, comes {\it in} through ${\cal H}_{R}^+$ and propagates to the Cauchy horizon ${\cal CH}_{L,R}^+$. Equation \eqref{ABdef} describes {\it``out-scattering"} in which a wave proportional to $\Phi_{{\rm out},+}$ enters region II coming {\it out} from ${\cal H}_{L}^+$ and then propagates to the Cauchy horizon ${\cal CH}_{L,R}^+$. 

We can define the four families of interior quasinormal modes (QNMs) as follows:
\begin{subequations} \label{def:interiorQNMfamilies}
\begin{align} 
& \hbox{1)  in$-$out interior QNMs:} \qquad  \tilde{\cal A}(\omega,m)=0\,;   \label{intQNMinout} \\
&\hbox{2) in$-$in interior QNMs:} \qquad  \tilde{\cal B}(\omega,m)=0\,; \label{intQNMinin} \\
& \hbox{3) out$-$in interior QNMs:}   \qquad {\cal A}(\omega,m)=0\,;  \label{intQNMoutin} \\
& \hbox{4)  out$-$out interior QNMs:}  \qquad {\cal B}(\omega,m)=0 \,.  \label{intQNMoutout} 
\end{align}
\end{subequations}
The first pair is associated with the ``in-scattering" while the second pair is associated with the  ``out-scattering".

Family 1), the in$-$out interior QNMs, describes modes that come {\it in} from ${\cal H}_{R}^+$ and are completely reflected {\it outwards} ${\cal CH}_R^+$, {\it i.e.}, these quasinormal modes occur in the special case where $R_{{\rm in},+}$ and $R_{{\rm out},-}$ become linearly dependent. Equivalently, the in$-$out interior QNMs are smooth at the ``right" event horizon ${\cal H}_R^+$ and at the ``right'' Cauchy horizon ${\cal CH}_R^+$. An inspection of \eqref{ABexp}, where we write the transmission/reflection coefficients in terms of $(a,b,c)$ which are themselves functions of the parameters $(\kappa_\pm,\Omega_\pm,\omega,m)$ via \eqref{abc}, indicates that $\tilde{\cal A}(\omega,m) = 0$ when $a=-p$ or if $b=-p$ for $p=0,1,2,\cdots$. This determines the spectrum of in$-$out interior QNMs as
\begin{subequations} \label{inoutQNMinterior}
\begin{align} 
 &  \omega_{\hbox{\small in-out,1}} L=m-i\,\frac{r_+-r_-}{L}\big(\Delta +2 p\big), \qquad \hbox{for $p\in \mathbb{N}_0 =\{0,1,2,\ldots\}$},  \\
& \omega_{\hbox{\small in-out,2}} L=m-i\,\frac{r_+-r_-}{L}\big(2-\Delta +2 p\big), \qquad \hbox{for $p\in \mathbb{N}_0$}\,.
\end{align}
\end{subequations} 
 
Family 2), the in$-$in interior QNMs, describes modes that come {\it in} from ${\cal H}_{R}^+$ and are completely transmitted {\it inwards} to ${\cal CH}_L^+$. So these modes are smooth both at the ``right" event horizon ${\cal H}_R^+$ and at the ``left'' Cauchy horizon ${\cal CH}_L^+$. From \eqref{ABexp}, $\tilde{\cal B}(\omega,m)=0$ occurs when $a-c+1=-p$ or if $b-c+1=-p$ for $p=0,1,2,\cdots$. This determines the spectrum of in$-$in interior QNMs as
\begin{subequations} \label{ininQNMinterior}
\begin{align} 
 &  \omega_{\hbox{\small in-in,1}} L=-m-i\,\frac{r_++r_-}{L}\big(\Delta +2 p\big), \qquad \hbox{for $p\in \mathbb{N}_0$} \,,   \\
& \omega_{\hbox{\small in-in,2}} L=-m-i\,\frac{r_++r_-}{L}\big(2-\Delta +2 p\big), \qquad \hbox{for $p\in \mathbb{N}_0$}\,.
\end{align}
\end{subequations} 
 
 Family 3), the out$-$in interior QNMs, describes modes that come {\it out} from ${\cal H}_{L}^+$ and are completely reflected {\it inwards} to ${\cal CH}_L^+$. Thus, these modes are smooth at the ``left" event horizon ${\cal H}_L^+$ and at the ``left'' Cauchy horizon ${\cal CH}_L^+$. From \eqref{ABexp}, ${\cal A}(\omega,m)=0$ occurs when $1-b=-p$ or when $1-b=-p$ for $p\in \mathbb{N}_0$. This determines the spectrum of out$-$in interior QNMs as
\begin{subequations} \label{outinQNMinterior}
\begin{align} 
 &  \omega_{\hbox{\small out-in,1}} = \bar{\omega}_{\hbox{\small in-out,1}} \,,   \\
& \omega_{\hbox{\small out-in,2}} = \bar{\omega}_{\hbox{\small in-out,2}} \,,
\end{align}
\end{subequations} 
where the bar stands for complex conjugate.

Family 4), the out$-$out interior QNMs, describes modes that come {\it out} from ${\cal H}_{L}^+$ and are completely transmitted {\it outwards} to ${\cal CH}_R^+$. Thus, these modes are smooth at the ``left" event horizon ${\cal H}_L^+$ and at the ``left'' Cauchy horizon ${\cal CH}_L^+$. From \eqref{ABexp}, ${\cal A}(\omega,m)=0$ occurs when $1-b=-p$ or when $1-b=-p$ for $p\in \mathbb{N}_0$. This determines the spectrum of out$-$in interior QNMs as
\begin{subequations} \label{outoutQNMinterior}
\begin{align} 
 &  \omega_{\hbox{\small out-out,1}} = \bar{\omega}_{\hbox{\small in-in,1}} \,,   \\
& \omega_{\hbox{\small out-out,2}} = \bar{\omega}_{\hbox{\small in-in,2}} \,.
\end{align}
\end{subequations} 
From \eqref{outinQNMinterior}-\eqref{outoutQNMinterior}, we see that the ``out-scattering" frequencies are related to ``in-scattering" frequencies by complex conjugation.

Note that some of the interior quasinormal frequencies have positive imaginary part. This does {\it not} mean that they can be interpreted as describing instabilities of region II because $t$ is not a time coordinate in this region, and because the boundary conditions defining the interior QNMs are not those appropriate to a study of stability. Only the quasinormal frequencies with negative imaginary part will be relevant for us. 

We can now present the coincidence that will have significant consequences for strong cosmic censorship: 
one of the in-out interior quasinormal mode frequencies in \eqref{inoutQNMinterior} is the same as the prograde {\it exterior} quasinormal frequency given in \eqref{QNM}, namely
\be
\label{coincidence}
\omega_{\hbox{\small in-out,1}} =  \omega_{\rm p}\,.
\ee
Similarly, one of the in-in interior QNM frequencies in \eqref{ininQNMinterior} coincides with the retrograde exterior QNM frequency given in \eqref{QNM}: $\omega_{\hbox{\small in-in,1}}= \omega_{\rm r}$. The out-in and out-out quasinormal frequencies are similarly related to the white hole quasinormal frequencies defined above.

This coincidence is a remarkable property of the BTZ black hole. We do not expect such a coincidence to occur for more general black hole {\it e.g.} in 4d. This coincidence appears to be related to the fact that the 
radial equation reduces to a hypergeometric equation, which has only three regular singular points. Perturbation equations about 4d black hole backgrounds are of the Heun type with more than three regular singular points. 


\subsection{Strong Cosmic Censorship violation in BTZ \label{sec:sccBTZviolation}}

We can now apply the results above to investigate strong cosmic censorship. Our strategy is as follows (see Fig. \ref{fig:scattering}). Consider characteristic initial data on ${\cal H}_L^+ \cup  {\cal H}_R^-$ consisting of a smooth outgoing wavepacket on ${\cal H}_L^+$ and a smooth outgoing wavepacket on ${\cal H}_R^-$. At timelike infinity ${\cal I}_R$ we turn out a source consisting of a smooth wavepacket, {\it i.e.},  the ``non-normalizable" part of the field is a wavepacket. (Of course this encompasses standard ``normalizable" boundary conditions by taking this wavepacket to vanish.) These initial-boundary conditions uniquely determine a solution $\Phi$ throughout regions I and II. We can then investigate the smoothness of $\Phi$ at the Cauchy horizon.

We start by determining the solution in region I. Consider the following superposition of mode solutions (where $x$ denotes all three coordinates)
\be
\label{Phisol}
 \Phi(x) = \sum_m \int \mathrm{d} \omega \left[ X(\omega,m)\tilde{\cal T}(\omega,m) \Phi_{{\rm vev},\infty} (\omega,m;x) + \tilde{X}(\omega,m) {\cal T}(\omega,m) \Phi_{{\rm in},+}(\omega,m;x) \right],
\ee
where $X(\omega,m)$ and $\tilde{X}(\omega,m)$ are chosen so that the above sum and integral converge. Recall that $\Phi_{{\rm vev},\infty}$ is the mode solution whose radial part is $R_{{\rm vev,}\infty}$ and similarly for $\Phi_{{\rm in},+}$ and that the coefficients ${\cal T}$ and $\tilde{\cal T}$ were defined in \eqref{TRdefTot}. Clearly $\Phi$ satisfies the equation of motion in region I. We will now determine the initial-boundary value data that gives rise to this solution. 

Using the first equation of \eqref{TRdefTot} we have (suppressing the $(\omega,m)$ arguments)
\be
  \Phi(x) = \sum_m \int \mathrm{d} \omega \left[ (X\tilde{\cal T} + \tilde{X} {\cal R} )\Phi_{{\rm vev},\infty} (x) + \tilde{X} \Phi_{{\rm source},\infty}(x) \right]
\ee
and hence as $r \rightarrow \infty$ we have
\be
 \Phi(x) \approx z^{-(2-\Delta)/2} \sum_m \int \mathrm{d} \omega e^{-i \omega t} e^{i m \phi} \tilde{X}(\omega,m)\,,
\ee
with $z$ the radial coordinate defined in \eqref{zcoord}. Hence we see that at infinity, the above solution obeys the boundary condition corresponding to switching on a source with Fourier transform $\tilde{X}(\omega,m)$. If $\tilde{X}(\omega,m)=0$ then the solution obeys normalizable boundary conditions. Note that if the source is compactly supported in $t$ then $\tilde{X}$ is an entire function of $\omega$. 

Using the second equation of \eqref{TRdefTot} gives
\be
\label{Phisol2}
 \Phi(x) =  \sum_m \int \mathrm{d} \omega \left[ X \Phi_{{\rm out},+} (x) + (\tilde{X} {\cal T} + X \tilde{\cal R} ) \Phi_{{\rm in},+} (x) \right].
\ee 
Recall that $\Phi_{{\rm out},+}$ describes waves propagating out of ${\cal H}_R^-$ and $\Phi_{{\rm in},+}$ describes waves propagating into ${\cal H}_R^+$. The second term above describes a superposition of these ingoing waves. We claim that this superposition vanishes on ${\cal H}_R^-$ . To see this we assume that $X$ and $\tilde{X}$ are analytic in $\omega$ in a neighbourhood of the real axis. We've already seen this is the case for $\tilde{X}$ if the source at ${\cal I}_R$ is compacty supported in $t$. It will remain true if the source is not compactly supported but decays sufficiently rapidly as $t \rightarrow \pm \infty$. We will justify this assumption about $X$ below, where we will also show that ${\cal T}$ and $\tilde{\cal R}$ are analytic in a neighbourhood of the real axis. The analyticity properties of $R_{{\rm in},+}$ discussed at the end of section \ref{sec:bases} imply that $\Phi_{{\rm in},+}$ is analytic in a neighbourhood of the real $\omega$ axis. Therefore to evaluate the second term above we can deform the contour of integration to a line ${\rm Im}(\omega) = \epsilon>0$. Now ${\cal H}_R^-$ has $r=r_+$ and $v \rightarrow -\infty$. As $r \rightarrow r_+$ we have $\Phi_{{\rm in},+} \propto e^{-i \omega v}$, which vanishes as $v \rightarrow -\infty$ because $\epsilon>0$. This justifies the claim. 

It now follows that
\be
  \Phi(x)|_{{\cal H}_R^-} =  \sum_m \int \mathrm{d} \omega X(\omega,m) \Phi_{{\rm out},+} (\omega,m;x) |_{{\cal H}_R^-} 
\ee
so we see that on ${\cal H}_R^-$, the above solution reduces to an outgoing wavepacket with profile specified by $X(\omega,m)$. Note that on ${\cal H}_R^-$ we have $\Phi_{{\rm out},+} = e^{-i \omega (u-\tilde{u}_0)} e^{i m (\phi''-\tilde{\phi}''_0)}$ for some constants $\tilde{u}_0$ and $\tilde{\phi}''_0$ (cf \eqref{Cauchy:decayCHR}) and so $X(\omega,m)$ is essentially the Fourier transform (w.r.t. $u$) of the wavepacket on ${\cal H}_R^-$. Hence if this wavepacket is compactly supported in $u$, or decays sufficiently rapidly as $u \rightarrow \pm \infty$ then $X$ will be analytic in a neighbourhood of the real $\omega$ axis, which justifies our assumption above.  

In summary, we have shown that \eqref{Phisol} is the solution of the Klein-Gordon equation that obeys initial-boundary conditions corresponding to an outgoing wavepacket on ${\cal H}_R^-$ with profile specified by $X(\omega,m)$ and a non-normalizable wavepacket on ${\cal I}_R$ with profile specified by $\tilde{X}(\omega,m)$. 

We will need to know the behaviour of this solution on ${\cal H}_R^+$. The first term in \eqref{Phisol2} can be shown to vanish at ${\cal H}_R^+$ (using a similar argument to the one below \eqref{Phisol2}). Hence we have
\be
\label{PhiHR}
  \Phi(x)|_{{\cal H}_R^+} =   \sum_m \int \mathrm{d} \omega    \tilde{Z}(\omega,m)   \Phi_{{\rm in},+} (x) |_{{\cal H}_R^+}
\ee
where
\be
\label{tildeZfinal}
  \tilde{Z}(\omega,m) =  \tilde{X}(\omega,m) {\cal T}(\omega,m) + X(\omega,m) \tilde{\cal R}(\omega,m)
\ee
{\it i.e.},  the solution on ${\cal H}_R^+$ consists of waves transmitted from ${\cal I}_R$ (the part proportional to $\tilde{X}$) and outgoing waves from ${\cal H}_R^-$ that are reflected back into ${\cal H}_R^+$ (the part proportional to $X$).

Now we turn to the solution in region II. As before, we start by writing down a solution in this region and then investigate how it behaves on ${\cal H}_{L,R}^+$. The solution is
\be
\label{Phisol3}
 \Phi(x) =  \sum_m \int \mathrm{d} \omega \left[ Z(\omega,m)\Phi_{{\rm out},+}(\omega,m;x) + \tilde{Z}(\omega,m)  \Phi_{{\rm in},+}(\omega,m;x) \right],
\ee
where $\tilde{Z}$ is defined in \eqref{tildeZfinal} and $Z$ is a new free function. In region II, $\Phi_{{\rm out},+}$ describes waves propagating out from ${\cal H}_L^+$ and  $\Phi_{{\rm in},+}$ describes waves propagating in from ${\cal H}_R^+$. Assuming that $Z$ is analytic in $\omega$ in a neighbourhood of the real axis then we can argue as we did below \eqref{Phisol2} to show that the first term in the above integral vanishes at ${\cal H}_R^+$ and hence evaluating the above solution at ${\cal H}_R^+$ reproduces \eqref{PhiHR} (recall that the mode functions $\Phi_{{\rm in},+}$ are smooth at ${\cal H}_R^+$). Thus the solution we have defined in region II matches continuously to the solution in region I. 

Arguing once again as we did below \eqref{Phisol2} one can show that the second term of \eqref{Phisol3} vanishes at ${\cal H}_L^+$. Hence we have
\be
  \Phi(x)|_{{\cal H}_L^+} =  \sum_m \int \mathrm{d} \omega Z(\omega,m)\Phi_{{\rm out},+}(\omega,m;x) |_{{\cal H}_L^+}
\ee
so on ${\cal H}_L^+$ the solution reduces to an outgoing wavepacket with profile specified by $Z(\omega,m)$. This is essentially the Fourier transform (w.r.t. $u$) of the wavepacket on ${\cal H}_L^+$. Hence if this wavepacket is compactly supported in $u$, or decays sufficiently rapidly as $u \rightarrow \pm \infty$ then $Z$ will be analytic in a neighbourhood of the real $\omega$ axis, as assumed above.   

In summary, we have defined a solution throughout regions I and II that arises from a wavepacket with profile $Z$ on ${\cal H}_L^+$, a wavepacket with profile $X$ on ${\cal H}_R^-$ and a source with profile $\tilde{X}$ at ${\cal I}_R$. We have shown that this solution is continuous at ${\cal H}_R^+$.  For suitable initial data, it is actually {\it smooth} there. To show this, we require that the data on ${\cal H}_L^+ \cup {\cal H}_R^-$ is smooth at the bifurcation surface $B$ (as will be the case for data compactly supported in $u$). Then the resulting solution will be smooth throughout regions I and II. We need to show that this smooth solution is the same as our solution. Well-posedness of the initial-boundary value problem implies that the smooth solution will agree with our solution in region I (as it has the same data) and therefore reduces to \eqref{PhiHR} on ${\cal H}_R^+$. Well-posedness of the characteristic value problem implies that the data on ${\cal H}_L^+ \cup {\cal H}_R^+$ uniquely determines the solution throughout region II. Since our solution has the same data as the smooth solution, they must be the same in region II, which concludes the argument.

We will now investigate the behaviour of our solution at the Cauchy horizon. We are mainly interested in the ``right" component of the Cauchy horizon, ${\cal CH}_R^+$ for the reason discussed in section \ref{sec:BTZ}. We use equations \eqref{ABdefTot} to write the solution in region II as
\be
\Phi(x)=  \Phi_{\rm out}(x) + \Phi_{\rm in}(x)\,,
\ee
where
\begin{subequations}\label{phioutin}
\begin{align} 
 &  \Phi_{\rm out} (x) \equiv \sum_m \int \mathrm{d}\omega\, \left[ Z(\omega,m) {\cal A}(\omega,m) + \tilde{Z}(\omega,m) \tilde{\cal B}(\omega,m) \right] \Phi_{{\rm out},-} (\omega,m;x) \,,\\
& \Phi_{\rm in} (x) \equiv  \sum_m \int \mathrm{d}\omega\, \left[ Z(\omega,m) {\cal B}(\omega,m) + \tilde{Z}(\omega,m) \tilde{\cal A}(\omega,m) \right] \Phi_{{\rm in},-} (\omega,m;x) \,.
\end{align}
\end{subequations}
$\Phi_{\rm out}$ is smooth ({\it i.e.} ``outgoing") at ${\cal CH}_R^+$ and $\Phi_{\rm in}$ is smooth ({\it i.e.} ``ingoing") at ${\cal CH}_L^+$. We need to be careful when we split $\Phi$ into these two parts. This is because examining the expressions \eqref{ABexp} for ${\cal A}(\omega,m)$ etc we find that all of these scattering coefficients have poles on the real axis at $\omega = m \Omega_-$, {\it i.e.},  at $c=1$. However, when the solution is written in the form \eqref{Phisol3}, no such poles appear because $\Phi_{{\rm out},+}$ and $\Phi_{{\rm in},+}$ do not have poles on the real axis, as discussed at the end of section \ref{sec:bases}. Hence these poles appear only when we write $\Phi_{{\rm out},+}$ and $\Phi_{{\rm in},+}$ in terms of $\Phi_{{\rm out},-}$ and $\Phi_{{\rm in},-}$.

The reason for the appearance of these poles is that the mode solutions $\Phi_{{\rm out},-}$ and $\Phi_{{\rm in},-}$ become linearly dependent when $\omega = m \Omega_-$. This is clear from \eqref{Cauchy:basis} which gives $R_{{\rm out},-} = R_{{\rm in},-}$ when $c=1$. To obtain a basis of solutions which is valid for $c=1$ we define (supressing $(\omega,m)$)
\be
 P_- = \frac{1}{2} \left( R_{{\rm out},-} + R_{{\rm in},-} \right)\,, \qquad Q_- = \frac{1}{2(c-1)} \left( R_{{\rm out},-} - R_{{\rm in},-} \right).
\ee
Note that $Q_-$ is non-trivial in the limit $c \rightarrow 1$: this gives the second linearly independent solution of the radial equation. We can now write the $+$ basis in terms of this new $-$ basis:
\be
 R_{{\rm out},+} = {\cal C} P_- + {\cal D} Q_- \,,\qquad R_{{\rm in},+} = \tilde{\cal C} P_- + \tilde{\cal D} Q_-\,,
\ee
where the coefficients ${\cal C}$, ${\cal D}$ etc are all analytic at $c=1$, {\it i.e.},  at $\omega = m \Omega_-$. From \eqref{ABdefTot} we then obtain
\be
 {\cal A} = \frac{\cal C}{2} + \frac{\cal D}{2(c-1)}\,, \qquad  {\cal B} = \frac{\cal C}{2} - \frac{\cal D}{2(c-1)}\,,
\ee
and similarly for the quantities with a tilde. We now see the appearance of the poles at $c=1$. They arise simply because the notions of ``ingoing" and ``outgoing" become degenerate for $\omega = m \Omega_-$. 

Since the integrand of \eqref{Phisol3} has no poles on the real axis, it follows that the {\it sum} of $\Phi_{\rm out}$ and $\Phi_{\rm in}$ can be written as an integral over real $\omega$ of an integrand with no poles on the real axis. We are free to indent the contour of integration so that it passes just below $\omega = m \Omega_-$ in the complex $\omega$ plane.\footnote{
Note that we do this before carrying out the sum over $m$.}
 Having done this, we can split the integral into $\Phi_{\rm out}$ and $\Phi_{\rm in}$ as above, with the same contour of integration. The poles now reappear in the integrands but they no longer lie on the contour of integration. 

Since $\Phi_{\rm out}$ is smooth at ${\cal CH}_R^+$, any non-smooth behaviour of $\Phi$ at ${\cal CH}_R^+$ must arise from $\Phi_{\rm in}$. We can use \eqref{Cauchy:decayCHR} to evaluate $\Phi_{\rm in}$ near ${\cal CH}_R^+$ using the outgoing Eddington-Finkelstein coordinates $(u,r,\phi'')$ regular at ${\cal CH}_R^+$. This gives
 \be
 \label{Phiin}
 \Phi_{\rm in} = \sum_m \int_{C_m} \mathrm{d}\omega \,  {\cal G}(\omega,m)
 e^{-i \omega (u-u_0) + i m (\phi''-\phi''_0)} \exp \left[  i \frac{(\omega - m \Omega_-)}{\kappa_-} \log z \right] \left( 1+ {\cal O}(z) \right) \ee
where $C_m$ is the contour of integration just discussed, $z$ is the radial coordinate defined in \eqref{zcoord} ($z=0$ at ${\cal CH}_R^+$), and 
\ba
\label{Gdef}
{\cal G}(\omega,m) &\equiv & 
Z(\omega,m) {\cal B}(\omega,m) + \tilde{Z}(\omega,m) \tilde{\cal A}(\omega,m)  \nonumber \\
  &=&  Z(\omega,m) {\cal B}(\omega,m) +
 \left[ \tilde{X}(\omega,m) {\cal T}(\omega,m) + X(\omega,m) \tilde{\cal R}(\omega,m) \right] \tilde{\cal A}(\omega,m) 
\ea
where we used \eqref{tildeZfinal} in the second line. 

The Cauchy horizon ${\cal CH}_R^+$ is at $z=0$ where $\log z \rightarrow - \infty$. To determine the behaviour of $\Phi_{\rm in}$ there we will deform the contour of integration to a line of constant ${\rm Im}(\omega)$ in the lower half complex $\omega$ plane. If we can do this then the integral will decay as $z \rightarrow 0$. The rate of decay, and hence the smoothness of $\Phi_{\rm in}$, depends on how far we can push the contour into the lower half-plane and the contribution from any poles that are crossed when the contour is deformed. This is determined by the analyticity properties of the quantity ${\cal G}(\omega,m)$, whose properties we will now investigate.

We start by investigating the scattering coefficients $\tilde{\cal A}(\omega,m)$ and ${\cal B}(\omega,m)$ appearing in ${\cal G}(\omega,m)$. Substituting $(a,b,c)$ as given in \eqref{abc} into \eqref{ABexp} we find
\be\label{finalAt}
 \tilde{\cal A}(\omega,m) = \frac{i \kappa_-}{\omega-m \Omega_{-}}
 \frac{\Gamma \left(1-i\frac{\omega-m \Omega_{-}}{\kappa_-}\right) \Gamma \left(1-i\frac{\omega-m \Omega_{+}}{\kappa_+}\right)}{\Gamma \left(\frac{\Delta }{2}+i\frac{L}{2}\frac{\omega L-m}{r_--r_+}\right) \Gamma \left(1-\frac{\Delta }{2}+i\frac{L}{2}\frac{\omega L-m}{r_--r_+}\right)}\,,
\ee
 \be\label{finalB}
 {\cal B}(\omega,m) = \frac{i \kappa_-}{\omega-m \Omega_{-}}
\frac{\Gamma \left(1-i\frac{\omega-m \Omega_{-}}{\kappa_-}\right) \Gamma \left(1+i\frac{\omega-m \Omega_{+}}{\kappa_+}\right)}{\Gamma \left(\frac{\Delta }{2}+i\frac{L}{2}\frac{\omega L+m}{r_-+r_+}\right) \Gamma \left(1-\frac{\Delta }{2}+i\frac{L}{2}\frac{\omega L+m}{r_-+r_+}\right)} \,.
\ee
Thus,  both $\tilde{\cal A}(\omega,m)$ and ${\cal B}(\omega,m)$  have a simple pole at $\omega-m \Omega_{-}=0$. This was discussed above. Both coefficients also have simple poles at $\omega-m \Omega_{-}=-i n \kappa_{-}$ for $n\in \mathbb{N}=\{1,2,3,\ldots \}$. These arise from the first gamma function in the numerators above. 
$\tilde{\cal A}(\omega,m)$  also has simple poles at $\omega-m \Omega_{+}=-i n \kappa_+$, and ${\cal B}(\omega,m)$ at  $\omega-m \Omega_{+}=i n \kappa_+$, for $n\in \mathbb{N}$. These arise from the second gamma function in each numerator.\footnote{The 
location of the complex poles in $\tilde{\cal A}(\omega,m)$ and ${\cal B}(\omega,m)$ is a consequence of $\tilde{\cal A}(\omega,m)\propto W[R_{{\rm in},+},R_{{\rm out},-}]$ and  ${\cal B}(\omega,m)\propto W[R_{{\rm out},+},R_{{\rm out},-}]$ where $W$ denotes the Wronskian of two solutions. Thus the poles of $R_{{\rm in},+}$ and $R_{{\rm out},-}$ discussed at the end of section \ref{sec:bases} give poles in $\tilde{\cal A}$. Similarly the poles of $R_{{\rm out},+}$ and $R_{{\rm out},-}$ give poles in ${\cal B}$.
}

Now we consider the coefficients ${\cal T}(\omega,m)$ and $\tilde{\cal R}(\omega,m)$ appearing in \eqref{Gdef}. Inserting \eqref{abc} into \eqref{TRexpressions} yields
\begin{eqnarray}\label{TRexpFinal} 
\hspace{-1cm} {\cal T}(\omega,m) &=& \frac{\Gamma \left( \frac{\Delta }{2}-i\frac{L}{2} \frac{\omega L-m}{r_+-r_-} \right) \Gamma \left( \frac{\Delta }{2}-i\frac{L}{2} \frac{\omega L+m}{r_++r_-}\right)}{\Gamma \left(1-i\frac{\omega-m \Omega_{+}}{\kappa_+}\right)\Gamma (\Delta -1)}\,, \nonumber \\
\hspace{-1cm} \tilde{\cal R}(\omega,m) &=& -\frac{\Gamma \left(1+i\frac{\omega-m \Omega_{+}}{\kappa_+}\right)}{\Gamma \left(1-i\frac{\omega-m \Omega_{+}}{\kappa_+}\right)}\frac{\Gamma\left(\frac{\Delta }{2}-i\frac{L}{2} \frac{\omega L-m}{r_+-r_-}\right) \Gamma\left(\frac{\Delta }{2}-i\frac{L}{2} \frac{\omega L+m}{r_++r_-}\right)}{\Gamma\left(\frac{\Delta }{2}+i\frac{L}{2} \frac{\omega L-m}{r_+-r_-}\right) \Gamma\left(\frac{\Delta }{2}+i\frac{L}{2} \frac{\omega L+m}{r_++r_-}\right)}\,.
\end{eqnarray}  
${\cal T}(\omega,m)$ has simple poles at the exterior quasinormal mode frequencies $\omega = \omega_{\rm p}$  and $\omega=\omega_{\rm r}$ given in \eqref{QNM}. This is not a surprise: as explained in section \ref{sec:extQNM}, ${\cal T}(\omega,m)=\infty$ is the condition that defines the quasinormal frequencies. $\tilde{\cal R}(\omega,m)$ also has simple poles at $\omega=\omega_{\rm p},\omega_{\rm r}$ and has additional simple poles in the upper half-plane at $\omega-m \Omega_{+}=i n \kappa_+$, $n\in \mathbb{N}$.\footnote{One can show that $ \tilde{\cal R}(\omega,m) = - \frac{W[R_{{\rm out},+},R_{{\rm vev},\infty}]}{W[R_{{\rm in},+},R_{{\rm vev},\infty}]}$. Thus, $\tilde{\cal R}(\omega,m)$ inherites the poles of $R_{{\rm out},+}$ and it also has poles when $W[R_{{\rm in},+},R_{{\rm vev},\infty}]=0$, {\it i.e.}, at the location of the exterior quasinormal frequencies.} We also note that both ${\cal T}(\omega,m)$ and $\tilde{\cal R}(\omega,m)$ have zeroes at $\omega-m \Omega_{+}=-i n \kappa_+$, $n\in \mathbb{N}$. 

Table \ref{Table:poles} summarizes the location of the poles of $\tilde{\cal A}(\omega,m)$, ${\cal B}(\omega,m)$, $ {\cal T}(\omega,m)$ and $\tilde{\cal R}(\omega,m)$.
\begin{table}[th]
\begin{eqnarray}
\nonumber
\begin{array}{||c|||c|c||c|c||}\hline\hline
 \hbox{\qquad\qquad\qquad {\bf Simple poles of } $\bm \searrow$}  &    & 
 &  &  \\
 \hbox{ {\it Located at}  $\bm \downarrow$ ? \qquad\qquad\qquad\qquad}  &  \tilde{\cal A}(\omega,m)   & {\cal B}(\omega,m) 
 & {\cal T}(\omega,m)  & \tilde{\cal R}(\omega,m)
  \\
\hline \hline
 \omega-m\Omega_- =0 & {\rm yes} & {\rm yes} & {\rm no} & {\rm no}  \\ \hline
 \omega-m\Omega_- = -i n \kappa_- \:(n=1,2,\cdots)  &  {\rm yes} &  {\rm yes} & {\rm no} & {\rm no}   \\ \hline
  \omega-m\Omega_- = +i n \kappa_- \:(n=1,2,\cdots)  & {\rm no} & {\rm no} & {\rm no} & {\rm no}  \\ \hline
\omega-m\Omega_+ = -i n \kappa_+ \:(n=1,2,\cdots)  & {\rm yes} & {\rm no} & {\rm no} & {\rm no}   \\ \hline
\omega-m\Omega_+ =+ i n \kappa_+ \:(n=1,2,\cdots)  &  {\rm no}  & {\rm yes} & {\rm no} & {\rm yes}  \\ \hline
\omega = \omega_{\rm p}  & {\rm no} & {\rm no} & {\rm yes} & {\rm yes}  \\ \hline
\omega = \omega_{\rm r}  & {\rm no} & {\rm no} & {\rm yes} & {\rm yes}  \\ 
\hline\hline
\end{array}
\end{eqnarray}
\caption{Location of simple poles of relevant scattering coefficients. The prograde  ($\omega_{\rm p}$) and retrograde ($\omega_{\rm r}$) exterior quasinormal frequencies  are given in \eqref{QNM}.}
\label{Table:poles}
\end{table}

We now have enough information to determine the analyticity of ${\cal G}(\omega,m)$ defined in \eqref{Gdef}. We are interested only in its behaviour in the lower half plane since this is where we want to deform our contour of integration. For simplicity, we will start by assuming that the wavepackets on ${\cal H}_L^+$ and ${\cal H}_R^-$ are compactly supported functions of $u$ and the wavepacket on ${\cal I}_R$ is compactly supported in $t$. This implies that their Fourier transforms are entire functions of $\omega$, which implies that $Z$, $X$ and $\tilde{X}$ are entire functions of $\omega$. Hence any singularities of ${\cal G}$ must arise from singularities of ${\cal B}$, ${\cal T} \tilde{\cal A}$ or $\tilde{\cal R} \tilde{\cal A}$. We will discuss these three objects in turn.

${\cal B}$ has a pole at $\omega = m \Omega_-$. This was discussed above, and we have explained why we can choose our contour of integration to pass below this pole. Hence it does not affect our freedom to deform the contour into the lower half-plane. The only poles of ${\cal B}$ in the lower half-plane lie at $\omega - m \Omega_- = -in \kappa_-$ for $n \in \mathbb{N}$.  

${\cal T} \tilde{\cal A}$ has poles arising from ${\cal T}$ and from $\tilde{\cal A}$. $\tilde{\cal A}$ has a pole at $\omega=m\Omega_-$ but this is irrelevant for the reason just discussed. It also has poles at $\omega -m \Omega_- = -i n \kappa_\pm$ for $n \in \mathbb{N}$. However, ${\cal T}$ has a zero at $\omega -m \Omega_- = -i n \kappa_+$ and so only the poles at $\omega -m \Omega_- = -i n \kappa_-$ lead to poles in ${\cal T} \tilde{\cal A}$. ${\cal T}$ has poles at $\omega = \omega_p,\omega_r$. We now exploit the remarkable coincidence \eqref{coincidence}. Since $\omega_p = \omega_{\hbox{\small in-out,1}}$, it follows from the definition of the ``in-out" interior quasinormal modes that $\tilde{\cal A}$ has a zero at $\omega = \omega_p$. So the pole in ${\cal T} $ at $\omega=\omega_p$ does not give rise to a pole in ${\cal T}  \tilde{\cal A}$. 
No such cancellation occurs for the pole at at $\omega = \omega_r$. So we conclude that the poles of ${\cal T} \tilde{\cal A}$ in the lower half-plane are located at $\omega = \omega_r$ and at $\omega -m \Omega_- = -i n \kappa_-$, $n \in \mathbb{N}$. 

$\tilde{\cal R} \tilde{\cal A}$ has poles arising from $\tilde{\cal R}$ and from $\tilde{\cal A}$. The latter have just been discussed. As above, the pole of $\tilde{\cal A}$ at $\omega -m \Omega_- = -i n \kappa_+$ is cancelled by a zero of $\tilde{\cal R}$ at this location. Hence the only poles of $\tilde{\cal A}$ that give rise to poles in $\tilde{\cal R} \tilde{\cal A}$ are those located at $\omega -m \Omega_- = -i n \kappa_-$. The only poles of $\tilde{\cal R}$ in the lower half-plane are those at $\omega = \omega_p,\omega_r$. As above, the coincidence \eqref{coincidence} implies that the poles at $\omega=\omega_p$ correspond to zeroes of $\tilde{\cal A}$ and hence do not give rise to a pole in $\tilde{\cal R} \tilde{\cal A}$. So we conclude that the poles of $\tilde{\cal R} \tilde{\cal A}$ in the lower half-plane are located at the same places as the poles of  ${\cal T} \tilde{\cal A}$, {\it i.e.},  at
$\omega = \omega_r$ and at $\omega -m \Omega_- = -i n \kappa_-$, $n \in \mathbb{N}$. 

In summary, for compactly supported initial data, ${\cal G}$ is analytic in the lower half-plane except for simple poles at $\omega = \omega_r$ and at $\omega -m \Omega_- = -i n \kappa_-$, $n \in \mathbb{N}$. We emphasize once again that the remarkable coincidence \eqref{coincidence} implies that there are no poles at $\omega = \omega_p$. 

\begin{figure}
	\centering
	\includegraphics[width=0.7\textwidth]{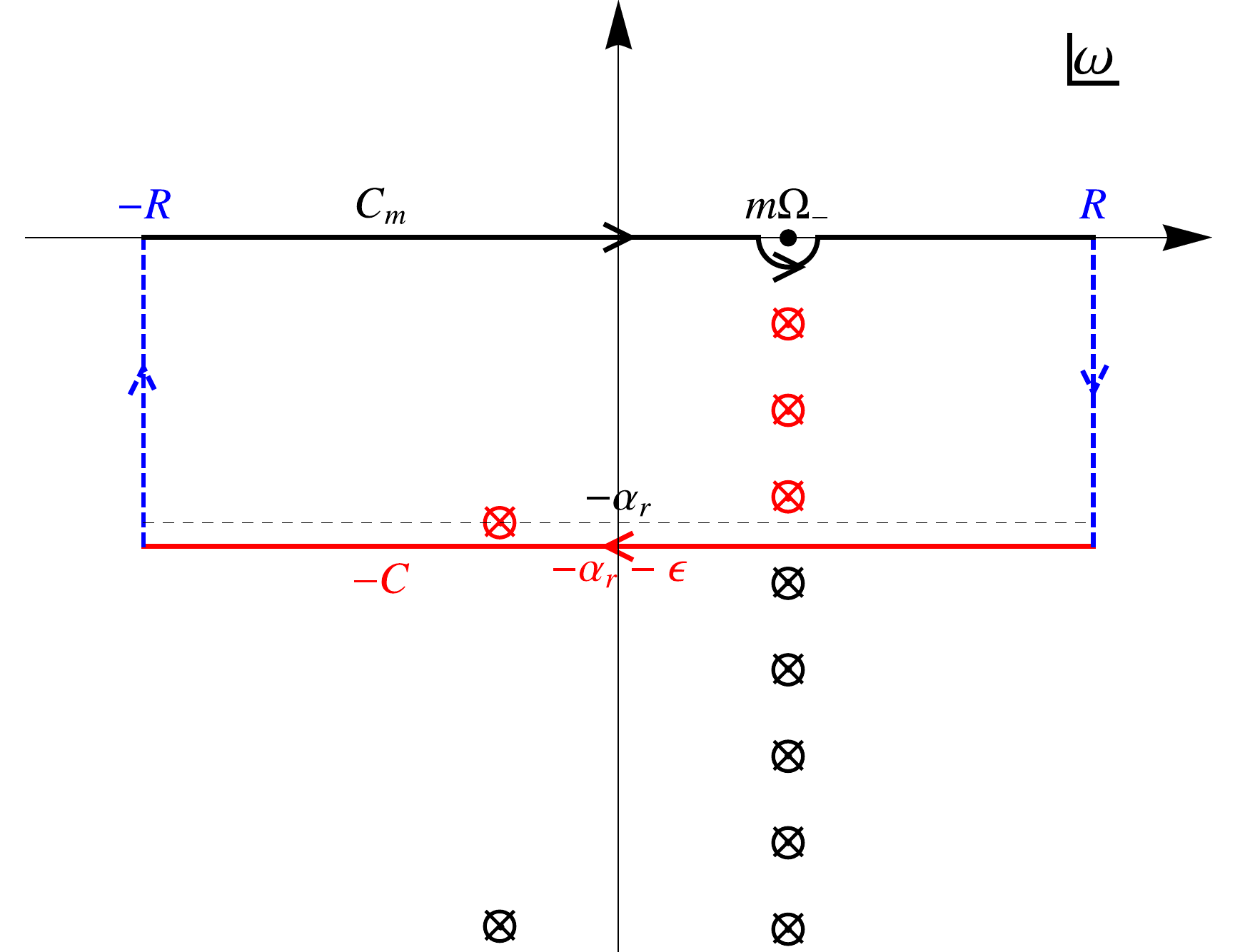}
	\caption{Contours of integration $C$ and $C_m$ in the complex $\omega$ plane and the relevant poles: poles marked in red contribute to the integral and poles marked in black don't. The pole with smallest $|\mathrm{Im}(\omega)|$ on the left is the fundamental retrograde quasinormal mode $\varpi_m$ and on the right we show the poles with $\omega -m \Omega_- = -i n \kappa_-$ which only give a smooth contribution at ${\cal CH}_R^+$.}
	\label{fig:contour}
\end{figure}

Let $\varpi_m$ be the frequency of the slowest decaying retrogade quasinormal mode with angular dependence $e^{i m \phi}$. From \eqref{QNM} we have
\be
 \varpi_m L=  - m   - i \frac{(r_+ + r_-)\Delta}{L }\,.
\ee
We define the ``retrograde spectral gap"
\be
 \alpha_r = -{\rm Im}(\varpi_m) = \frac{(r_+ + r_-)\Delta}{L^2}\,.
\ee 
We now deform the contour of integration $C_m$ in \eqref{Phiin} into a new contour $C$ defined as the straight line ${\rm Im}(\omega) =- \alpha_r- \epsilon$ (see Fig. \ref{fig:contour}), {\it i.e.},  we push the contour just beyond the pole at $\omega = \varpi_m$.\footnote{
More precisely, one considers the integral around a rectangle-shaped contour whose long edges are at ${\rm Im}(\omega)=0$ and ${\rm Im}(\omega) =- \alpha_r- \epsilon$, and whose short edges are at ${\rm Re}(\omega) = \pm R$. 
One can show that the contribution from the short edges vanishes as $R \rightarrow \infty$ using the asymptotic behaviour of $\Gamma(z)$ for large $|z|$ and the fact that smoothness of the initial-boundary data implies that $X$, $\tilde{X}$ and $Z$ decay faster than any power of $R$.} In doing this, we pick up a contribution from the pole at $\varpi_m$. We also pick up contributions from any poles with $\omega -m \Omega_- = -i n \kappa_-$ that lie between $C_m$ and $C$. Using the residue theorem, the contribution from a pole at $\omega -m \Omega_- = -i n \kappa_-$ to \eqref{Phiin} has $z$-dependence $z^n$, which is analytic in $z$ and hence analytic in $r$. Therefore these poles give a contribution to \eqref{Phiin} which vanishes {\it smoothly} at $z=0$, {\it i.e.},  vanishes smoothly at ${\cal CH}_R^+$, and is therefore irrelevant for strong cosmic censorship. 

The non-smooth part of \eqref{Phiin} arises from the poles at $\varpi_m$ and from the integral along $C$. Via the residue theorem, the contribution from the poles at $\varpi_m$ is
\be
\label{polecontrib}
 -2 \pi i z^\beta \sum_m {\cal G}_m e^{-i \varpi_m(u-u_0)} e^{im (\phi''-\phi''_0)} z^{-im\frac{(1+ L\Omega_-)}{L\kappa_-}}  ( 1 + {\cal O}(z)),
\ee
where ${\cal G}_m$ is the residue of ${\cal G}(\omega,m)$ at $\omega = \varpi_m$ and
\be
\label{betadef}
  \beta \equiv \frac{\alpha_r}{\kappa_-}\,.
\ee
Substituting in the values of $\alpha_r$ and $\kappa_-$ gives equation \eqref{betadef1} of the Introduction.

Now consider the integral along $C$. The integrand, and hence the integral, is ${\cal O}(z^{\beta + \epsilon/\kappa_-})$. Therefore, as $z\rightarrow 0$, the integral along $C$ is subleading compared to the ${\cal O}(z^\beta)$ contribution from the poles. We conclude that the dominant non-smooth contribution to $\Phi$ is given by \eqref{polecontrib}.\footnote{We could push the contour $C$ further into the lower half-plane and pick up subleading contributions from subdominant retrograde quasinormal frequencies $\omega_r$.}

Finally, we can address the question of strong cosmic censorship. We see from \eqref{polecontrib} that, generically, the gradient of the scalar field will diverge at ${\cal CH}_R^+$ if $\beta<1$. Thus $\beta<1$ guarantees that generically the energy-momentum tensor of the scalar field will diverge at ${\cal CH}_R^+$. This will cause a large backreaction, and potentially render ${\cal CH}_R^+$ singular, in agreement with strong cosmic censorship. From equation \eqref{betadef1} we see that $\beta<1$ if the black hole is sufficiently far from extremality. 

A stronger version of strong cosmic censorship, due to Christodoulou \cite{Christodoulou:2008nj}, requires that, generically, the scalar field should fail to be locally square integrable at ${\cal CH}_R^+$. This ensures that, when backreaction is included, the Einstein equation cannot be satisfied, even in the sense of weak solutions, at ${\cal CH}_R^+$ (see {\it e.g.} \cite{Dias:2018ynt}). From \eqref{polecontrib}, the condition for this is $\beta<1/2$, as in \cite{Cardoso:2017soq,Dias:2018ynt}. Again, equation \eqref{betadef1} shows that this condition is satisfied if the black hole is sufficiently far from extremality. 

Equation \eqref{betadef1} shows that $\beta$ diverges in the extremal limit $r_-\to r_+$. Hence near-extremal black holes have $\beta>1$. For such black holes, the gradient of scalar field perturbations is bounded at ${\cal CH}_R^+$ and so the energy-momentum tensor of such perturbations is bounded at ${\cal CH}_R^+$. Therefore such black holes violate strong cosmic censorship. In fact, for any given value of $k$, a sufficiently near-extremal black hole will have $\beta>k$, which implies that the $k$th derivative of the scalar field is bounded at ${\cal CH}_R^+$. In other words, scalar field perturbations can be made arbitrarily differentiable at the Cauchy horizon by taking the black hole sufficiently close to extremality. So strong cosmic censorship is badly violated.

Our conclusion disagrees with a previous study of classical scalar field perturbations of the BTZ black hole \cite{Balasubramanian:2004zu}, which concluded that strong cosmic censorship {\it does} hold. Therefore, we should explain where we think the error lies in Ref. \cite{Balasubramanian:2004zu}. We find that Ref. \cite{Balasubramanian:2004zu} obtains the same expressions for $\tilde{\cal A}(\omega,m)$, ${\cal B}(\omega,m)$, ${\cal T}(\omega,m)$ and  $\tilde{\cal R}(\omega,m)$ as us. The coincidence \eqref{coincidence} is also apparent in the results of  \cite{Balasubramanian:2004zu}. However, the final result, namely equation (5.14) of Ref. \cite{Balasubramanian:2004zu} for the behaviour near ${\cal CH}_R^+$, contains a contribution from a pole at $\omega = m \Omega_-$ as well as from the retrograde quasinormal modes. We explained above why the poles at $\omega = m \Omega_-$ arise simply from a bad choice of basis, and they do not lead to non-smooth behaviour at the Cauchy horizon. This subtle point is the reason for the disagreement with Ref. \cite{Balasubramanian:2004zu}.

The significance of the poles at $\omega = m \Omega_-$ is clarified by results of Ref. \cite{Kehle:2018upl}, where it was shown that there is an analogous pole at $\omega=0$ in the Reissner-Nordstr\"om-de Sitter spacetime (for an uncharged scalar field). However, this pole is absent for the Reissner-Nordstr\"om black hole (\emph{i.e.} vanishing cosmological constant). Ref. \cite{Kehle:2018upl} showed that, in the latter case, the absence of the pole implies that, for finite energy initial data on ${\cal H}_L^+ \cup {\cal H}_R^+$, the solution $\Phi$ always vanishes at the bifurcation sphere of the Cauchy horizon. But in the Reissner-Nordstr\"om-de Sitter case, the solution $\Phi$ is (generically) non-vanishing on this bifuration sphere. We can see how a similar result for BTZ arises from our discussion above. Our argument shows that $\Phi_{\rm in}$ vanishes on ${\cal CH}_R^+$. Hence the solution at ${\cal CH}_R^+$ is given by $\Phi_{\rm out}$. Early time on ${\cal CH}_R^+$ corresponds to $u \rightarrow \infty$. In this limit we can evaluate $\Phi_{\rm out}$ by displacing the contour of integration $C_m$ into the lower half-plane, as we did above for $\Phi_{\rm in}$, with the result that $\Phi_{\rm out} \rightarrow 0$. The bifurcation surface of the Cauchy horizon corresponds to late time on ${\cal CH}_R^+$, \emph{i.e.}, $u \rightarrow - \infty$. In this limit we can evaluate $\Phi_{\rm out}$ by deforming $C_m$ into the {\it upper} half-plane. But when we do this we pick up a contribution from the pole on the real axis at $\omega = m \Omega_-$. This contribution is proportional to $e^{- i m \Omega_- u} e^{i m \phi''}$, which is proportional to $e^{i m \phi_-}$ in the Kruskal coordinates $(U_-,V_-,\phi_-)$ regular at the Cauchy horizon. So the pole at $\omega = m\Omega_-$ determines the $m$th Fourier mode of the field at the bifurcation surface $U_-=V_-=0$.

We assumed above that the initial data on ${\cal H}_R^-$ and ${\cal H}_L^+$ as well as the source data on $\cal{I}_R$ (if present) have compact support. What happens if we relax this condition? Consider first the case in which we allow non-compact support of the initial data on ${\cal H}_R^-$ and/or ${\cal H}_L^+$. We still require that this data decays as $u \rightarrow \pm \infty$ in order that it has a well-defined Fourier transform. This implies that the initial data vanishes at the bifurcation surface $B$. Crucially, we demand that the initial data is {\it smooth} at $B$. This set-up was addressed in detail in a very similar context in Section 2.2 of \cite{Dias:2018etb} (where we analysed strong cosmic censorship in the 4d Reissner-Nordstr\"om de Sitter black hole). Our analysis there translates with only minor changes  to the BTZ background so we will not repeat the arguments here. Ultimately, this analysis shows that allowing smooth data with non-compact support on ${\cal H}_R^-$, and/or on ${\cal H}_L^+$ does not change the results.

Now consider allowing the source data on ${\cal I}_R$ to have non-compact support, but still demand that this data decay as $t \rightarrow \pm \infty$. For, example, let's assume that the data decays as $e^{-\gamma \kappa_- t}$ as $t \rightarrow \infty$, where $0<\gamma<1$. This implies that $\tilde{X}(\omega,m)$ has a pole at $\omega = -i \gamma \kappa_-$, which leads to a pole in ${\cal G}(\omega,m)$ at the same location. The effect of this additional pole is to generate a contribution that behaves as $z^\gamma$ near ${\cal CH}_R^+$. Since $\gamma<1$ this implies that the gradient of the scalar diverges at ${\cal CH}_R^+$. The same would be true if we considered a source with even slower decay {\it e.g.} power-law decay. So, unsurprisingly, if we perturb the black hole by sending in waves from infinity with sufficiently slow decay at late time then the solution is singular at the Cauchy horizon.\footnote{
This is relevant if the BTZ black hole is obtained as a decoupling limit of a higher-dimensional asymptotically flat black hole. In this case, power-law decay on ${\cal I}_R$ could arise from power-law decay in the asymptotically flat region.}

We emphasize that our conclusion that near-extremal BTZ black holes violate strong cosmic censorship depends on the coincidence \eqref{coincidence} between the interior and exterior quasinormal frequencies. Consider deforming the BTZ metric \eqref{BTZmetric} in the black hole interior by smoothly changing the function $f(r)$ by a small amount in the region $r_1 \le r \le r_2$ with $r_-<r_1<r_2<r_+$. This will not affect the exterior quasinormal frequencies but it will change the interior quasinormal frequencies so \eqref{coincidence} will no longer be true. For this deformed geometry, 
 the poles at $\omega=\omega_p$ will now give a contribution to \eqref{polecontrib} scaling as $z^{\beta_p}$ where 
\be
\label{betap}
 \beta_p \equiv \frac{\Delta}{r_+/r_- + 1}<\frac{\Delta}{2}\,.
\ee
Hence in this deformed geometry, strong cosmic censorship is respected by linear scalar field perturbations if the scalar field obeys $\Delta \le 2$ (or $\Delta \le 1$ for Christodoulou's version of strong cosmic censorship).\footnote{Of course backreaction of the scalar field will also deform the BTZ geometry, but not in the simple $t$-independent manner we have just discussed.}

A more interesting way of destroying the coincidence \eqref{coincidence} is to change the boundary conditions on the scalar field. For $\mu^2_{\rm BF}< \mu^2<0$ there are more general boundary conditions than the ones used above, corresponding to ``multi-trace" deformations in the dual CFT \cite{Witten:2001ua}. Such boundary conditions fix a relation between $A$ and $B$ in \eqref{KGdecays}. Consider a 1-parameter family of boundary conditions, parameterized by $\epsilon$, which reduces to one of the standard boundary conditions (corresponding to one of the two possible values of $\Delta$) when $\epsilon=0$. The exterior quasinormal frequencies will now depend on $\epsilon$ but the interior quasinormal frequencies are unaffected so \eqref{coincidence} is no longer satisfied. For small $\epsilon$, the exterior quasinormal frequencies will be close to those discussed above. It then follows, as in the previous paragraph, that the non-smooth part of the field near ${\cal CH}_R^+$ will scale as $z^{\beta_p+\ldots}$ where $\beta_p$ is given by \eqref{betap} and the ellipsis denotes a correction that vanishes as $\epsilon \rightarrow 0$. Since $\mu^2<0$ implies $\Delta<2$, it follows that $\beta_p <1$ and so, at least for small $\epsilon$, strong cosmic censorship is respected in a theory with such boundary conditions ($\Delta<1$ is needed for Christodoulou's version). 

\subsection{Non-smooth perturbations}

\label{sec:rough}

In the above discussion we have always assumed that our perturbation arises from {\it smooth} initial-boundary data. With this condition we have seen that strong cosmic censorship is badly violated by near-extremal BTZ black holes. This is similar to what happens for Reissner-Nordstr\"om-de Sitter black holes \cite{Cardoso:2017soq,Dias:2018etb}. In the latter case, it has been argued that strong cosmic censorship can be saved by widening the class of perturbations to allow perturbations arising from {\it rough} ({\it i.e.} non-smooth) initial data \cite{Dafermos:2018tha}. Specifically, if one has initial data with the minimum acceptable level of smoothness (in the sense of Sobolev spaces) then, generically, the solution at the Cauchy horizon will not have this minimum acceptable level of smoothness. Refs. \cite{Dafermos:2018tha} proved this using mode solutions with {\it complex} frequency. We will show that the same result holds for the BTZ black hole. The result relies crucially on equation \eqref{kappaineq}. 

What follows is a direct translation of our argument in Ref. \cite{Dias:2018etb} to the BTZ case. In region I we take the initial-boundary data to vanish, {\it i.e.},  $X = \tilde{X}=0$. It follows that the solution vanishes throughout region I. In region II we take initial data on ${\cal H}_L^+$ to coincide with the initial data for the mode solution $\Phi_{{\rm out},+}$ with frequency $\omega = \omega_1 - i \gamma \kappa_+$ where $\gamma>0$. Note that this vanishes as $u \rightarrow \infty$, {\it i.e.},  it vanishes at the bifurcation surface $B$.\footnote{
The data blows up as $u \rightarrow - \infty$, {\it i.e.},  at late time on ${\cal H}_L^+$. However, one is free to modify the data  by multiplying it by a smooth function that vanishes for $u \le u_1$ for some $u_1$ and is equal to $1$ for $u \ge u_2 > u_1$. This modification will not change the resulting solution near the early time portion of ${\cal CH}_R^+$ since this region does not lie in the domain of dependence of the region where we have modified the data.} The resulting solution in region II is simply $\Phi_{{\rm out},+}$. This vanishes as $u \rightarrow \infty$ so it vanishes at ${\cal H}_R^+$ and hence matches continuously to the solution in region I.

Now consider the behaviour of this initial data at $B$. To do this, we use the Kruskal coordinates $(U_+,V_+,\phi_+)$ which are smooth at $B$. The characteristic initial surface ${\cal H}_L^+\cup {\cal H}_R^-$ is the surface $V_+=0$. On this surface, for $U_+<0$ the initial data vanishes, and for $U_+>0$ it is proportional to $e^{-i\omega u}=(U_+)^{i \omega_1/\kappa_+ + \gamma}$. This data is continuous at $U_+=0$ but it is not necessarily smooth there {\it e.g.} it is not differentiable there if $\gamma<1$. Similarly, the resulting solution $\Phi$ is continuous but not smooth at ${\cal H}_R^+$. Nevertheless, since ${\cal H}_R^+$ is a characteristic ({\it i.e.} null) hypersurface, $\Phi$ still satisfies its equation of motion on ${\cal H}_R$, at least in the sense of weak solutions. 

The solution at the Cauchy horizon is now determined by \eqref{ABdef}:
\be
 \Phi = {\cal A} \Phi_{{\rm out},-} + {\cal B} \Phi_{{\rm in},-}\,.
\ee
By adjusting $\omega_1$ if necessary, we can ensure that $\omega$ is not a zero of ${\cal B}$. $\Phi_{{\rm out},-}$ is smooth at ${\cal CH}_R^+$ but $\Phi_{{\rm in},-}$ is not. Hence $\Phi$ is not smooth at ${\cal CH}_R^+$. From \eqref{Cauchy:decayCHR} we see that, in the outgoing Eddington-Finkelstein coordinates regular at ${\cal CH}_R^+$, it has radial dependence (for simplicity we now set $m=0$)
\be
 z^{i \omega/\kappa_-} = z^{\delta} z^{i \omega_1/\kappa_-}\,,
\ee
where $\delta = \gamma \kappa_+/\kappa_-$. Crucially \eqref{kappaineq} implies that $\delta<\gamma$ so the solution at ${\cal CH}_R^+$ is always less smooth than the initial data. For example, we can always choose $\gamma>1$ such that $\delta<1$. This gives a (weak) solution for which the initial data is $C^1$ but the solution is not $C^1$ at ${\cal CH}_R^+$. This behaviour is generic because, given any other solution, one can add to it a multiple of the solution just constructed. 

More rigorously, one should discuss the loss of smoothness using Sobolev spaces. The condition for the initial data to belong to the Sobolev space $H^k_{\rm loc}$ of functions whose first $k$ derivatives are locally square integrable is $\gamma>k-1/2$. We can always choose $\gamma>k-1/2$ so that $\delta<k-1/2$, which implies that the solution on a surface of constant $u$ intersecting ${\cal CH}_R^+$ does not belong to $H^k_{\rm loc}$. Thus, generically, initial data in $H^k_{\rm loc}$ gives a solution that does not belong to $H^k_{\rm loc}$ on a surface intersecting the Cauchy horizon. Thus if one decrees that ``acceptable" solutions should belong to $H^k_{\rm loc}$ for some value of $k$\footnote{
For example, in the non-linear context, local well-posedness of the initial value problem will only hold for $k \ge k_{\rm min}$ for some $k_{\rm min}$. So ``acceptable" might mean $k=k_{\rm min}$.} then, generically, acceptable initial data evolves to a non-acceptable solution at ${\cal CH}_R^+$. 

Note that this ``rough" version of strong cosmic censorship is weaker than the usual ``smooth" version in the sense that the smooth version implies the rough version. We will discuss this weaker version of strong cosmic censorship further in section \ref{sec:discussion}. 

\section{Other classical fields} 
\label{sec:fields} 

We have argued that, in the Einstein-scalar field system, the BTZ black hole violates the strong cosmic censorship conjecture (for smooth initial data). One might wonder whether the conjecture is restored by considering other classical fields. The question then arises: which fields should one consider?

Consider embedding the BTZ black hole in type IIB supergravity. There are many known such embeddings (see for instance \cite{Strominger:1996sh,Gauntlett:2006af}), with $\mathrm{BTZ}\times S^3\times X$, where $X$ can be $\mathbb{T}^4$ or $K3$, being the most well studied examples. We are now interested in studying fluctuations in the full ten-dimensional supergravity theory. The linearized field equations can be reduced to equations for fields in the 3d BTZ geometry. As well as scalar fields, one also obtains massive Chern-Simons gauge fields, Proca fields and Kaluza-Klein gravitons. (Massless gauge fields can be dualised to scalars, at least classically.) We are thus lead to investigate classical perturbations of such fields on fixed BTZ black backgrounds.
\subsection{Chern-Simons}
The equation of motion of Chern-Simons gauge theory is
\begin{equation}
\mathrm{d} \star F+\frac{\lambda}{L} F=0\,,
\label{eq:chern}
\end{equation}
where $F$ is the Maxwell 2-form. Note that the equation of motion implies $\mathrm{d}F=0$. Define a gauge-invariant 1-form
\begin{equation}
\label{tildeAdef}
\tilde{A}\equiv-\frac{L}{\lambda}\star F\,.
\end{equation}
The equation of motion becomes
\begin{equation}
\star \mathrm{d}\tilde{A}+\frac{\lambda}{L} \tilde{A}=0\,.
\label{eq:chern3}
\end{equation}
Since the BTZ black hole has an $\mathbb{R}_t\times S^1$ isometry, we can use Fourier decomposition along the $t$ and $\phi$ directions. We will use the non-coordinate basis
\begin{equation}
e^0\equiv \mathrm{d}t\,,\quad e^1\equiv \mathrm{d}r\,,\quad\text{and}\quad e^2\equiv \mathrm{d}\phi-\Omega\,\mathrm{d}t\,,
\label{eq:noncoordinate}
\end{equation}
in terms of which we can decompose $\tilde{A}$ as
\begin{equation}
\tilde{A} = [\tilde{A}_0(r)\,e^0+\tilde{A}_1(r)\,e^1+\tilde{A}_2(r)\,e^2]e^{-i\omega t+i m\phi}\,.
\end{equation}
Our aim is to solve for $\{\tilde{A}_0,\tilde{A}_1,\tilde{A}_2\}$.  From the first component of \eqref{eq:chern3} we find an algebraic relation amongst the components of $\tilde{A}$ of the form
\begin{equation}
i f r \lambda  \tilde{A}_1(r)+L m \tilde{A}_0(r)+L (\omega -m \Omega ) \tilde{A}_{2 }(r)=0\,.
\end{equation}
We can use this relation to express $\tilde{A}_1$ algebraically as a function of $\tilde{A}_{0 }(r)$ and $\tilde{A}_{2 }(r)$. The remaining two equations are first order ordinary differential equations for $\tilde{A}_{0 }(r)$ and $\tilde{A}_{2 }(r)$:
\begin{subequations}
\begin{align}
& r\,\frac{\mathrm{d}\tilde{A}_0}{\mathrm{d} r}-\frac{L m}{f\,\lambda}(\omega-m \Omega) \tilde{A}_0+\left[\frac{\lambda}{L}-\frac{L}{f\,\lambda}(\omega-m \Omega)^2-r\frac{\mathrm{d}\Omega}{\mathrm{d}r}\right]\tilde{A}_{2}=0\,,
\\
& f \frac{\mathrm{d}\tilde{A}_{2}}{\mathrm{d}r}+\frac{L\,m}{r\,\lambda}(\omega-m \Omega)\tilde{A}_{2}+\left(\frac{L m^2}{r \lambda}+\frac{r\,\lambda}{L}\right)\tilde{A}_{0}=0\,.
\end{align}
\label{eqs:ataphi}
\end{subequations}
We can reduce this system to a single second order equation by defining
\begin{equation}
R(r)=L\,\tilde{A}_0(r)+(\varepsilon_{\lambda}-\Omega L)\,\tilde{A}_{2}\,,
\label{eq:defphics}
\end{equation}
where $\varepsilon_{\lambda}\equiv \mathrm{sgn} \lambda$. Using \eqref{eqs:ataphi} it is a relatively simple exercise to show that $R(r)$ obeys the ODE
\begin{equation}
\frac{1}{r}\frac{\mathrm{d}}{\mathrm{d}r}\left(r f \frac{\mathrm{d}R}{\mathrm{d}r}\right)+\left[\frac{(\omega-m \Omega)^2}{f}-\frac{m^2}{r^2}-\mu^2\right]R=0\,,
\label{eq:scalarradial}
\end{equation}
where
\begin{equation}
L^2\mu^2\equiv |\lambda| (|\lambda|-2)\,.
\end{equation}
The equation for $R$ is \emph{precisely} the radial equation for a minimally coupled scalar field $\Phi$ with mass $\mu$ propagating on the BTZ black hole background. If we restore the $(t,\phi)$ dependence by defining
\be
\Phi = R(r) e^{-i \omega t + i m \phi}
\ee
then equation \eqref{eq:defphics} can be written covariantly as
\be
\label{PhiA}
 \Phi = L k^a \tilde{A}_a + \varepsilon_{\lambda}\,m^a \tilde{A}_a\,,
\ee
where $k^a$ is the Killing field that generates time translations and $m^a$ the Killing field that generates rotations ({\it i.e.} $k = \partial/\partial t$ and $m = \partial/\partial \phi$ in $(t,r,\phi)$ coordinates). Since $k,m$ are defined globally, this equation defines $\Phi$ globally.

This connection between $\tilde{A}$ and $\Phi$ will allow us to import results from section \ref{sec:sccBTZ} once we have determined the boundary conditions satisfied by $\Phi$. To do this we need to find a way to reconstruct $\tilde{A}$ from $\Phi$. This can be done as follows. From the definition of $R$ in \eqref{eq:defphics} we can extract $\tilde{A}_{0}$ as a function of $\tilde{A}_{2}$ and $R$. We then use this relation in equations \eqref{eqs:ataphi} to find an expression for $\tilde{A}_{2}$ and $\tilde{A}_{2}^\prime$ as a function of $R$ and $R^\prime$. Demanding that $\tilde{A}_{2}^\prime(R,R^\prime)$ follows from taking the derivative of $\tilde{A}_{2}(R,R^\prime)$ yields an integrability condition which is simply \eqref{eq:scalarradial}. This procedure yields a first order differential map from $R$ to $\tilde{A}_0$ and $\tilde{A}_{2}$, and thus also to $\tilde{A}_1$:
\begin{subequations}
\label{APhi}
\begin{align}
&\tilde{A}_0 = -\frac{L}{\varpi}\left\{\lambda(\varepsilon_{\lambda}-\Omega L) r f R^\prime+\left[f \lambda^2-L(L\,\omega-\varepsilon_{\lambda}m)(\omega-m\,\Omega)\right]R\right\}\,,
\\
& \tilde{A}_1=\frac{i L^2}{\varpi}\left\{(L\,\omega-\varepsilon_{\lambda}m)R^\prime-\frac{\lambda}{r}\left[m-\frac{r^2}{L\,f}(\varepsilon_{\lambda}-\Omega L)(\omega-m \Omega)\right]R\right\}\,,
\\
& \tilde{A}_2 = \frac{L^2}{\varpi}\left\{r \lambda f R^\prime-\left[m(L\,\omega-\varepsilon_{\lambda}m)-\frac{r^2\lambda^2}{L^2}(\varepsilon_{\lambda}-\Omega L)\right]R\right\}\,.
\end{align}
\end{subequations}
where we defined
\begin{equation}
\varpi \equiv L^2 (L\,\omega-\varepsilon_{\lambda}m)^2 + (r_+-\varepsilon_{\lambda} r_-)^2\lambda^2\,.
\end{equation}
Now we want to relate $R$ to the components of $F$, since this will render the discussion of boundary conditions more transparent. To do this we write
\begin{equation}
F \equiv [F_{01}(r)\,e^0\wedge e^1+F_{02}(r)\,e^0\wedge e^2+F_{12}(r)\,e^1\wedge e^2]e^{-i\omega t+im \phi}\,.
\end{equation}
From \eqref{tildeAdef}, we find $F=\lambda \star \tilde{A}/L$, and from \eqref{APhi} we obtain
\begin{subequations}
\begin{align}
& F_{01} = \frac{L\,\lambda}{\varpi}\left\{\lambda f R^\prime-\left[\frac{m(L\,\omega-\varepsilon_{\lambda}m)}{r}-\frac{r\,\lambda^2}{L^2}(\varepsilon_{\lambda}-\Omega L)\right]R\right\}\,,
\\
& F_{02} = -\frac{i \lambda}{\varpi}\left\{L r(L\,\omega-\varepsilon_{\lambda}m)f R^\prime-L \lambda f\left[m-\frac{r^2}{L f}(\varepsilon_{\lambda}-\Omega L)(\omega-m \Omega)\right]R\right\}\,,
\\
& F_{12} = \frac{r \lambda}{\varpi}\left\{r \lambda(\varepsilon_{\lambda}-\Omega L)R^\prime+\left[\lambda^2-\frac{L}{f}(L\,\omega-\varepsilon_{\lambda}m)(\omega-m \Omega)\right]R\right\}\,.
\end{align}
\label{eqs:componentsF}%
\end{subequations}
To determine the asymptotic behaviour of $F$, we need to find the asymptotic behaviour of $R$ first. This can be done via a Frobenius expansion close to the conformal boundary, where we find
\begin{equation}
R \approx \frac{H_+}{r^{|\lambda|}}\left[1+\mathcal{O}(r^{-2})\right]+\frac{H_-}{r^{2-|\lambda|}}\left[1+\mathcal{O}(r^{-2})\right]\,,
\end{equation}
where $H_{\pm}$ are constants. This translates into the following asymptotic decay for the components of $F$
\begin{subequations}
\begin{align}
& F_{01}\approx \frac{|\lambda|\,H_+}{2 L r^{1+|\lambda|}}\left[1+\mathcal{O}(r^{-2})\right]-\frac{2 H_- \lambda^2(\varepsilon_{\lambda}-\lambda)}{L\,\varpi}r^{|\lambda|-1}\left[1+\mathcal{O}(r^{-2})\right]\,,
\\
& F_{02}\approx -\frac{i\,H_+(\varepsilon_{\lambda}\,L\,\omega+m)}{2 L r^{|\lambda|}}\left[1+\mathcal{O}(r^{-2})\right]+\frac{2i\,H_- \lambda(\varepsilon_{\lambda}-\lambda)(L\,\omega-m)}{L\varpi}r^{|\lambda|}\left[1+\mathcal{O}(r^{-2})\right]\,,
\\
& F_{12}\approx -\frac{\lambda\,H_+}{2 r^{1+|\lambda|}}\left[1+\mathcal{O}(r^{-2})\right]-\frac{2 H_- \lambda^2(\varepsilon_{\lambda}-\lambda)}{\varpi}r^{|\lambda|-1}\left[1+\mathcal{O}(r^{-2})\right]\,.
\end{align}
\end{subequations}
Boundary conditions for Chern-Simons fields were studied in \cite{Andrade:2011sx}. It was argued that if  $0<|\lambda|<1$ then both fall-offs $H_{\pm}$ preserve the local AdS asymptotics and that the requirement that boundary conditions preserve the asymptotic conformal symmetry, implies either $H_+=0$ or $H_-=0$. It was also shown that the absence of ghosts selects the condition $H_-=0$. For $|\lambda|\geq1$, it is clear that we must also set $H_-=0$, since the decay associated with $H_-$ is non-normalisable. So for any $\lambda$ we must set $H_-=0$. Therefore, the boundary conditions on $F$ implies that $\Phi$ obeys the boundary conditions appropriate to a massive scalar field with 
\be
\label{Deltalambda}
\Delta = |\lambda|. 
\ee

The final thing we need to study is how regularity of $F$ (or equivalently $\tilde{A}$) is related to regularity of $\Phi$ at the event and Cauchy horizons. From equation \eqref{PhiA} we see immediately that smoothness of ${\tilde A}$ at a horizon implies smoothness of $\Phi$ at that horizon. What about the converse? Converting to the ingoing Eddington-Finkelstein coordinates $(v,r,\phi')$ defined in section \ref{sec:BTZ} gives\footnote{To simplify the exposition, in \eqref{tildeAcs} and below we do not rewrite the factor $e^{-i \omega t + i m \phi}$ in Eddington-Finkelstein coordinates.}
\be \label{tildeAcs}
 \tilde{A} = \left[ \tilde{A}_0 \mathrm{d}v + \left( \tilde{A}_1 - \frac{\tilde{A}_0}{f} \right) \mathrm{d}r + \tilde{A}_2 \left( \mathrm{d}\phi' - \Omega \mathrm{d}v \right) \right] e^{-i \omega t + i m \phi}.
\ee
Now take $\Phi$ to be a mode solution smooth at ${\cal H}_R^+$, {\it i.e.},  the radial function $R$ is $R_{{\rm in},+}$ defined in section \ref{sec:bases}. Using \eqref{APhi} and \eqref{event:decay} reveals that the above expression for $\tilde{A}$ is also smooth at ${\cal H}_R^+$.

Similarly, if we work in the coordinates $(u,r,\phi'')$ regular at ${\cal CH}_R^+$ we obtain
\be
 \tilde{A} = \left[ \tilde{A}_0 \mathrm{d}u + \left( \tilde{A}_1 + \frac{\tilde{A}_0}{f} \right) \mathrm{d}r + \tilde{A}_2 \left( \mathrm{d}\phi'' - \Omega \mathrm{d}u \right) \right] e^{-i \omega t + i m \phi}.
\ee
The generic non-smooth behaviour of the above components at ${\cal CH}_R^+$ can be evaluated by substituting the generic behaviour of the non-smooth part of $\Phi$ given by \eqref{polecontrib} into \eqref{APhi} to obtain the non-smooth part of $\tilde{A}$ as
\be
\label{tACH}
\tilde{A}_{\rm non-smooth} =  {\cal O}((r-r_-)^\beta) \mathrm{d}u + {\cal O}((r-r_-)^{\beta-1}) \mathrm{d}r +  {\cal O}((r-r_-)^\beta) \left( \mathrm{d}\phi'' - \Omega \mathrm{d}u \right) 
\ee
where we are using \eqref{Deltalambda} in the definition \eqref{betadef1} of $\beta$. From this it follows that $\tilde{A}$, and hence $F$, extends continuously to ${\cal CH}_R^+$ if $\beta>1$. Thus the energy-momentum tensor of the Chern-Simons field is finite at ${\cal CH}_R^+$ if $\beta>1$. Note that this is exactly the same as the condition for a scalar field to have finite energy momentum tensor at ${\cal CH}_R^+$. Similarly, the $k$th derivative of $F$ extends continuously to ${\cal CH}_R^+$ if $\beta >k+1$. The energy-momentum tensor of $F$ is {\it integrable} at ${\cal CH}_R^+$ if $\beta>1/2$ thus Christodoulou's version of strong cosmic censorship is violated by the Chern-Simons field if $\beta>1/2$. This is exactly the same condition as for the scalar field. 

In summary, we have shown that, as far as strong cosmic censorship is concerned, the Chern-Simons field behaves exactly as a scalar field with conformal dimension $\Delta = |\lambda|$. 

\subsection{Proca fields}
For Proca fields we will reach a similar conclusion, \emph{i.e.} everything will reduce down to studying properties of solutions of the Klein-Gordon equation. The difference is that for the Chern-Simons field, we only had to solve one equation, and for the Proca we will find two decoupled scalars are needed. This counting makes sense, since we expect to have two degrees of freedom associated to a three-dimensional massive vector field.

Massive vector bosons obey the equation
\begin{equation}
\nabla^a F_{ab}-\frac{\lambda_{\gamma}^2}{L^2}A_a=0\,,
\label{eq:proca}
\end{equation}
with $F=\mathrm{d}A$ and we take $\lambda_\gamma>0$. The mass of $A_a$ is  $\lambda_{\gamma}/L$. This equation manifestly breaks the gauge symmetry $A\to A+\mathrm{d}\chi$, because $A$ appears in the associated equations of motion. Taking a total divergence of the Proca equation, and using the fact that $F$ is an antisymmetric two-tensor, gives
\begin{equation}
\nabla^a A_a=0\,
\end{equation}
so that a Lorenz type identity is automatically enforced via \eqref{eq:proca}.

Before proceeding let us introduce the following operator that acts on arbitrary 1-forms $\omega$
\begin{equation}
\mathcal{D}^{(1)}_M \omega \equiv \star \mathrm{d}\omega+\frac{M}{L}\omega\,,
\end{equation}
where $M$ is a constant to be determined in what follows. It is then a simple exercise to show that
\begin{equation}
\left[\mathcal{D}^{(1)}_{-M}(\mathcal{D}^{(1)}_M \omega)\right]_a= \Box \omega_a-\nabla_a (\nabla^b \omega_b)-R_{ab}\omega^b-\frac{M^2}{L^2}\omega_a\,,
\end{equation}
which gives exactly \eqref{eq:proca} provided we identify $\omega=A$ and $M^2=\lambda_{\gamma}^2$. This means the Proca equation can be studied by inspecting at the following two systems of coupled first order equations \footnote{Note that $\mathcal{D}^{(1)}_{-M}$ and $\mathcal{D}^{(1)}_{M}$ commute.}
\begin{subequations}
\begin{equation}
\left\{
\begin{array}{l}
\mathcal{D}^{(1)}_{\lambda_\gamma} A=A_{-\lambda_\gamma}
\\
\mathcal{D}^{(1)}_{-\lambda_\gamma} A_{-\lambda_\gamma}=0
\end{array}
\right.\,,
\label{eq:firstsytem}
\end{equation}
or
\begin{equation}
\left\{
\begin{array}{l}
\mathcal{D}^{(1)}_{-\lambda_\gamma} A=A_{\lambda_\gamma}
\\
\mathcal{D}^{(1)}_{\lambda_\gamma} A_{\lambda_\gamma}=0
\end{array}
\right.\,.
\label{eq:secondsystem}
\end{equation}
\end{subequations}

Let us focus on the first system. Imagine that, after Fourier decomposition, one has a non-zero solution of
\begin{equation}
\mathcal{D}^{(1)}_{-\lambda_\gamma} A_{-\lambda_\gamma}=0\,,
\label{eq:positivelambda}
\end{equation}
and its associated quasinormal modes $\omega_{-\lambda_{\gamma}}$. This solution will then source the first equation in \eqref{eq:firstsytem}. We will later show that the eigenfrequencies $\omega_{-\lambda_{\gamma}}\neq \omega_{\lambda_{\gamma}}$. It then follows that the eigenfrequency solutions of \eqref{eq:firstsytem} will have the same eigenfrequencies as \eqref{eq:positivelambda}, since $\mathcal{D}^{(1)}_{\lambda_\gamma}$ is invertible for $\omega =\omega_{-\lambda_{\gamma}}$. This argument assumes that we start with a solution of \eqref{eq:positivelambda} that is non-zero. If it is zero then $A$ will be annihilated by $\mathcal{D}^{(1)}_{\lambda_\gamma}$, which gives the eigenfrequencies $\omega_{\lambda_{\gamma}}$. This shows that the eigenfrequencies of the system \eqref{eq:firstsytem} are $\omega_{\pm \lambda_\gamma}$. 

It thus suffices to study solutions of the following equation
\begin{equation}
\mathcal{D}^{(1)}_M A_M=0
\label{eq:M}
\end{equation}
where $M=\pm\lambda_{\gamma}$. This latter equation has precisely the same form as \eqref{eq:chern3} if we make the substitution $\lambda\to\pm\lambda_{\gamma}$. This means that our analysis of the Chern-Simons gauge field can be employed in this section \emph{mutatis mutandis}. In particular, this implies that the analysis of the Proca field should reduce to the study of {\it two} scalar fields. 

In particular, after decomposing w.r.t. the basis \eqref{eq:noncoordinate}
\begin{equation}
A_{M} = [A^M_0(r) e^0+A^M_1(r) e^1+A^M_2(r) e^2]e^{-i\omega t+m \phi}
\end{equation}
we find that the whole system can be studied by inspecting the following scalar
\begin{equation}
\Phi^M_{\gamma} = L k^a A_a^M-m^aA^M_{a}\,.
\label{eq:phiM}
\end{equation}
As expected, $\Phi^M_{\gamma}$ obeys a Klein-Gordon equation for a massive scalar field, if we identify the mass term as
\begin{equation}
\mu^2_{\gamma} = \frac{M(M-2)}{L^2}\,.
\end{equation}
That is to say, the radial part of $\Phi^M_{\gamma}$ will satisfy the same equation as $R$ in Eq.~(\ref{eq:scalarradial}), if we replace $\mu^2$ with $\mu^2_{\gamma}$.

We now return to the thorny issue of boundary conditions, which should be imposed on $A_{M}$ directly and not on $\Phi^M_{\gamma}$. This means we need to find a way to express $A_M$ as a function of $\Phi^M_{\gamma}$ and its first radial derivative $(\Phi^M_{\gamma})^\prime$. This is something we have done in our previous section, so we just quote the final results written in terms of $M$ instead of $\lambda$ (and set $\varepsilon_\lambda=1$)
\begin{subequations}
\begin{align}
&A_0^M = -\frac{L}{\varpi}\Bigg\{M r f(1+L \Omega)(\Phi^M_{\gamma})^\prime+\left[M^2 f-L(\omega L+m)(\omega-m \Omega)\right]\Phi_M\Bigg\}\,,
\\
&A_1^M = \frac{i L^2}{\varpi}\Bigg\{(\omega L+m)(\Phi^M_{\gamma})^\prime+\frac{M}{r}\left[m+\frac{r^2}{L}\frac{1+L \Omega}{f}(\omega -m \Omega)\right]\Phi_M\Bigg\}\,,
\\
& A_2^M = -\frac{L^2}{\varpi}\Bigg\{M r f(\Phi^M_{\gamma})^\prime+\left[m(\omega L+m)+\frac{r^2}{L^2}(1+L \Omega)M^2\right]\Phi_M\Bigg\}\,.
\end{align}
\end{subequations}
This means that for a given $\Phi^M_{\gamma}$ we can uniquely reconstruct $A_M$. To proceed further, we need to analyse the asymptotic behaviour of $\Phi^M_{\gamma}$ and see how it translates to the asymptotic behaviour of $A_M$. Since $\Phi^M_{\gamma}$ obeys a Klein-Gordon equation, we can readily write down its boundary expansion based on our previous sections. We will first investigate the case with $M=\lambda_{\gamma}$, in which case we find
\begin{equation}
\Phi^{\lambda_{\gamma}}_{\gamma} \underset{r\to+\infty}{=} \frac{\tilde{H}_+^{\lambda_{\gamma}}}{r^{\lambda_{\gamma}}}\left[1+\mathcal{O}(r^{-2})\right]+\frac{\tilde{H}_-^{\lambda_{\gamma}}}{r^{2-\lambda_{\gamma}}}\left[1+\mathcal{O}(r^{-2})\right],
\end{equation}
where $\tilde{H}_{\pm}^{\lambda_{\gamma}}$ are constants of integration that parametrise each of the decays.

For $M=-\lambda_{\gamma}$ we find
\begin{equation}
\Phi^{-\lambda_{\gamma}}_{\gamma} \underset{r\to+\infty}{=} \tilde{H}_+^{-\lambda_{\gamma}} r^{\lambda_{\gamma}}\left[1+\mathcal{O}(r^{-2})\right]+\frac{\tilde{H}_-^{-\lambda_{\gamma}}}{r^{2+\lambda_{\gamma}}}\left[1+\mathcal{O}(r^{-2})\right],
\end{equation}
where $\tilde{H}_{\pm}^{-\lambda_{\gamma}}$ are constants of integration that parametrise each of the decays.

These decays for $\Phi^M_{\gamma}$ induce the corresponding decay on the physical fields. For instance, for the time component of $A_M$ we find
\begin{subequations}
\begin{equation}
A_0^{\lambda_{\gamma}} \underset{r\to+\infty}{=}\frac{\tilde{H}_+^{\lambda_{\gamma}}}{2Lr^{\lambda_{\gamma}}}\left[1+\mathcal{O}(r^{-2})\right]+\frac{2 \lambda_{\gamma}(1-\lambda_{\gamma})\tilde{H}_-^{\lambda_{\gamma}}}{L[L^2(\omega L+m)^2+M^2 (r_++r_-)^2]}r^{\lambda_{\gamma}}\left[1+\mathcal{O}(r^{-2})\right]
\end{equation}
and 
\begin{equation}
A_0^{-\lambda_{\gamma}} \underset{r\to+\infty}{=}-\frac{2 \lambda_{\gamma}(1+\lambda_{\gamma})\tilde{H}_-^{-\lambda_{\gamma}}}{L[L^2(\omega L+m)^2+M^2 (r_++r_-)^2]r^{\lambda_{\gamma}}}\left[1+\mathcal{O}(r^{-2})\right]+\frac{\tilde{H}_+^{-\lambda_{\gamma}}}{2L}r^{\lambda_{\gamma}}\left[1+\mathcal{O}(r^{-2})\right],
\end{equation}
\end{subequations}
which agree with the scaling dimensions reported in \cite{Aharony:1999ti,Andrade:2011sx}. We shall be interested in boundary conditions that preserve conformal invariance near the conformal boundary. This requires that either $\tilde{H}^{\pm\lambda_{\gamma}}_-$ or $\tilde{H}^{\pm\lambda_{\gamma}}_+$ vanish \cite{Aharony:1999ti,Andrade:2011sx} for $0<\lambda_{\gamma}<1$. Furthermore, the absence of ghosts dictates that $\tilde{H}^{\mp\lambda_{\gamma}}_\pm=0$ \cite{Andrade:2011sx}. For $\lambda_{\gamma}\geq1$ we are forced to choose $\tilde{H}_\pm^{\mp\lambda_{\gamma}}=0$. Either of these conditions lead to standard boundary conditions for $\Phi^M_{\gamma}$ of the form we discussed in section \ref{sec:BTZ}, provided we identify the dimensions of our two scalar fields as $\Delta = \{\lambda_{\gamma},2+\lambda_{\gamma}\}$.

The analysis of regularity of the Proca field at the various horizons proceeds analogously to the Chern-Simons case. In particular, smoothness of the Proca field implies smoothness of the two scalar fields $\Phi^M_{\gamma}$. Conversely, one can use the generic behaviour \eqref{polecontrib} of the non-smooth part of a solution $\Phi^M_{\gamma}$ near ${\cal CH}_R^+$ to determine the behaviour of the Proca field there, exactly as in the Chern-Simons case. The result is that, in coordinates regular at ${\cal CH}_R^+$, the components of the non-smooth part of the Proca field $A$ behave just as those of the Chern-Simons field $\tilde{A}$ given in equation \eqref{tACH}. Here we define $\beta$ using the smaller value of $\Delta$, \emph{i.e.}, we set $\Delta = \lambda_\gamma$ in \eqref{betadef1}, since this gives the least smooth behaviour. 

We see that the condition for the Proca field to extend continuously to ${\cal CH}_R^+$ is $\beta>1$. Note that the energy-momentum tensor of $A$ involves derivatives of $A$ only in the form $F=\mathrm{d}A$. Since the $r$ derivative of $A_r$ does not appear in $F$, it follows from \eqref{tACH} that all components of the non-smooth part of $F$ are ${\cal O}((r-r_-)^{\beta-1})$ at ${\cal CH}_R^+$. Hence $\beta>1$ guarantees that both $A$ and $F$ extend continuously to ${\cal CH}_R^+$ and hence that the energy-momentum tensor of the Proca field is finite at ${\cal CH}_R^+$. Christodoulou's version of strong cosmic censorship is violated by the Proca field if $\beta>1/2$. 

\subsection{Kaluza-Klein gravitons}
The case of Kaluza-Klein gravitons is slightly more complicated, but fortunately we can use some existing technology (see for instance \cite{Li:2008dq}). The equation we want to solve takes the following form
\begin{equation}
\Box h_{ab}+2 R_{acbd}h^{cd}-\frac{\lambda_{g}^2}{L^2} h_{ab}=0\,,
\label{eq:lichen}
\end{equation}
together with the consistency conditions
\begin{equation}
\nabla^a h_{ab}=0\qquad\text{and}\qquad g^{ab}h_{ab} =0\,.
\end{equation}

Just as before we introduce an operator $\mathcal{D}^{(2)}_M$ which acts on $h_{ab}$ as follows
\begin{equation}
(\mathcal{D}^{(2)}_M h)_{ab}=\varepsilon_{acd}\nabla^c h^d_{\phantom{d}b}+\frac{M}{L}h_{ab}\,.
\end{equation}
Let us imagine for a moment that we impose
\begin{equation}
(\mathcal{D}^{(2)}_M h)_{ab}=0
\label{eq:simplefirst}
\end{equation}
on our perturbations. Then, by taking the trace we automatically get $h=0$ and by taking its divergence we also obtain $\nabla^ah_{ab}=0$. This means the consistency conditions are automatically satisfied. All we need to do is to show that this operator can be used to write the second order equations of motion Eq.~(\ref{eq:lichen}). To do that, we look at the following simple identity
\begin{equation}
(\mathcal{D}^{(2)}_{-M}\mathcal{D}^{(2)}_M h)_{ab}=\Box h_{ab}-\nabla_a\nabla^c h_{cb}+\frac{3-M^2}{L^2}h_{ab}-\frac{h}{L^2}g_{ab}\,.
\label{eq:DG}
\end{equation}
If we recall that BTZ is a three dimensional constant curvature spacetime, we can identify Eq.~(\ref{eq:DG}) with Eq.~(\ref{eq:lichen}) provided $M=\pm \sqrt{1+\lambda_{g}^2}$. Since $[\mathcal{D}^{(2)}_M ,\mathcal{D}^{(2)}_{-M} ]=0$, we can find all solutions of Eq.~(\ref{eq:lichen}) by considering solutions of Eq.~(\ref{eq:simplefirst}) with $M=\pm\sqrt{1+\lambda_{g}^2}$. The advantage of solving Eq.~(\ref{eq:simplefirst}) is immense, since it is already written in a first order form.

Just as in our previous sections, we take advantage of the fact that BTZ possesses a stationary Killing vector field $\partial/\partial t$ and an axisymmetric Killing field $\partial/\partial \phi$ to Fourier decompose our perturbations with respect to $t$ and $\phi$ using the basis \eqref{eq:noncoordinate}
\begin{multline}
h^M_{ab}\mathrm{d}x^a \mathrm{d}x^b=e^{-i\omega t+i m \phi}\Big[h^M_{00}(r)(e^0)^2+2 h^M_{01}(r)e^0e^1+2 h^M_{02}(r)e^0e^2+
\\
h^M_{11}(r)(e^1)^2+2h^M_{12}(r)e^1e^2+h^M_{22}(r)(e^2)^2\Big]\,,
\end{multline}
where the index $M=\pm\sqrt{1+\lambda_{g}^2}$.

From \eqref{eq:simplefirst} we can express $\{h^M_{01},h^M_{02},h^M_{12},h^M_{11}\}$ as algebraic functions of $h^M_{00}$ and $h^M_{22}$. The remaining two equations are first order in $h^M_{00}$ and $h^M_{22}$. These equations are too lengthy to be presented here, and are not very illuminating. Nevertheless, if we consider the combination
\begin{equation}
\Phi^M_g = -L^2 k^ah^M_{ab}k^b+m^a h^M_{ab}m^b
\end{equation}
it obeys a simple radial Klein-Gordon equation with the same form as \eqref{eq:scalarradial} if we substitute $\mu^2$ by
\begin{equation}
\mu^2_{g} = \frac{M^2-1}{L^2} = \frac{\lambda_{g}^2}{L^2}\,.
\end{equation}

Just as in the previous cases, we can determine all the metric perturbations as a function of $\Phi^M_{\gamma}$. Again, the expressions are too lengthy to be presented here, but the procedure is very similar to the one detailed in the previous sections. Next, one studies the behaviour of $\Phi^M_{\gamma}$ near the conformal boundary. It is a simple exercise to show that
\begin{equation}
\Phi^M_{g}\underset{r\to+\infty}{=}\frac{\tilde{J}_{M}}{r^{1+M}}\left[1+\mathcal{O}(r^{-2})\right]+\tilde{J}_{-M} r^{M-1}\left[1+\mathcal{O}(r^{-2})\right]\,.
\end{equation}
The asymptotic expansion for $\Phi^M_{g}$ maps onto the following asymptotic behaviour for $h_{ab}$
\begin{subequations}
\begin{align}
&h_{00}= -\frac{\tilde{J}_M (1-M)M }{L^2[(1-M)^2(r_+-r_-)^2+L^2(\omega L-m)^2]r^{M-1}}\left[1+\mathcal{O}(r^{-2})\right]+(M\rightarrow -M)\,,
\\
&h_{01}= \frac{iL\tilde{J}_M (\omega L-m)M}{[(1-M)^2(r_+-r_-)^2+L^2(\omega L-m)^2]r^{M+2}}\left[1+\mathcal{O}(r^{-2})\right]+(M\rightarrow -M)\,,
\\
&h_{02}= \frac{\tilde{J}_M (1-M)M}{L[(1-M)^2(r_+-r_-)^2+L^2(\omega L-m)^2]r^{M-1}}\left[1+\mathcal{O}(r^{-2})\right]+(M\rightarrow -M)\,,
\\
&h_{11}= -\frac{\tilde{J}_M L^2[(1-M)(r_+-r_-)^2+L^2(\omega L-m)^2]}{[(1-M)^2(r_+-r_-)^2+L^2(\omega L-m)^2]r^{M+5}}\left[1+\mathcal{O}(r^{-2})\right]+(M\rightarrow -M)\,,
\\
&h_{12}= -\frac{i\tilde{J}_M L^2(\omega L-m)M}{[(1-M)^2(r_+-r_-)^2+L^2(\omega L-m)^2] r^{M+2}}\left[1+\mathcal{O}(r^{-2})\right]+(M\rightarrow -M)\,,
\\
&h_{22}= -\frac{\tilde{J}_M (1-M)M}{[(1-M)^2(r_+-r_-)^2+L^2(\omega L-m)^2]r^{M-1}}\left[1+\mathcal{O}(r^{-2})\right]+(M\rightarrow -M)\,.
\end{align}
\end{subequations}

We are interested in boundary conditions that preserve the action of the conformal group near the conformal boundary \cite{Liu:2009bk}. These correspond to either setting $\tilde{J}_{M}=0$ or $\tilde{J}_{-M}=0$. Demanding normalizability further sets $\tilde{J}_{M}=0$ for $M=-\sqrt{1+\lambda_{g}^2}$ or $\tilde{J}_{-M}=0$ for $M=\sqrt{1+\lambda_{g}^2}$  \cite{Liu:2009bk}. These boundary conditions, in turn, translate into scalar type boundary conditions for $\Phi^M_{g}$ of the exact form we studied in section \ref{sec:BTZ}, provided we identify $\Delta=1+M$ for positive $M$ and $\Delta = 2-M$ for negative values of $M$. Thus the equation of motion for Kaluza-Klein gravitons can be reduced to the equations of motion of two scalar fields, with conformal dimensions $\Delta = \{ 1+ \sqrt{1+\lambda_{g}^2},2+ \sqrt{1+\lambda_{g}^2}\}$. 

As before, smoothness of $h_{ab}$ at ${\cal H}_R^+$ implies smoothness of the scalar fields there, and vice versa. Substituting the generic non-smooth behaviour \eqref{polecontrib} of a scalar field at ${\cal CH}_R^+$ into the equations giving the components of $h_{ab}$ in terms of $\Phi_g^M$, one can determine the behaviour at ${\cal CH}_R^+$ of the  non-smooth part of $h_{ab}$. The result is that, in the coordinates $(u,r,\phi'')$ regular at ${\cal CH}_R^+$, the least smooth component is $h_{rr} = {\cal O}((r-r_-)^{\beta-2})$ where $\beta$ is defined by \eqref{betadef1} with  $\Delta = 1+ \sqrt{1+\lambda_{g}^2}$. Hence $\beta >2$ ensures that the field extends continuously to ${\cal CH}_R^+$. Once again, the field can be made arbitrarily differentiable at ${\cal CH}_R^+$ by taking the black hole sufficiently close to extremality.

\section{Quantum field theory calculation} 
\label{sec:quantum} 

\subsection{Introduction}

We have shown that classical linear perturbations are arbitrarily smooth near the Cauchy horizon of a near-extremal BTZ black hole. Since classical perturbations do not enforce strong cosmic censorship we will now investigate whether quantum effects do so. Does the backreaction of quantum fields render the Cauchy horizon singular?

In this section we will calculate $\langle 0|T_{ab} | 0 \rangle$, the renormalized expectation value of the energy-momentum tensor of a free scalar field (of arbitrary mass) in the Hartle-Hawking state in the BTZ geometry. We will show that, for a near-extremal black hole, $\langle 0|T_{ab} | 0 \rangle$ is {\it finite} at the Cauchy horizon. Hence vacuum polarization does not rescue strong cosmic censorship. 

This result contradicts statements in the literature that $\langle 0|T_{ab} | 0 \rangle$ always diverges at the Cauchy horizon of a BTZ black hole. These statements are based on a calculation of $\langle 0|T_{ab} | 0 \rangle$ by Steif \cite{Steif:1993zv}. So we will start by explaining why Steif's calculation does {\it not} demonstrate that $\langle 0|T_{ab} | 0 \rangle$ diverges at the Cauchy horizon. 

Steif considered a conformally coupled scalar field obeying ``transparent" boundary conditions. He obtained an expression for $\langle 0|T_{ab} | 0 \rangle$ expressed as an infinite sum. When analytically continued to $r<r_-$, the $n$th term in the sum diverges at $r=r_n<r_-$ with $r_n \rightarrow r_-$ as $n \rightarrow \infty$. Steif concluded that $\langle 0|T_{ab} | 0 \rangle$ diverges on surfaces $r=r_n$ lying behind the Cauchy horizon, which accumulate at the Cauchy horizon as $n \rightarrow \infty$. This is the basis of claims in the literature that Steif showed that $\langle 0|T_{ab} | 0 \rangle$ diverges on the Cauchy horizon. Such claims are incorrect for at least two reasons. 

First, the calculation just described assumes that the quantity $\langle 0|T_{ab} | 0 \rangle$ is well-defined for $r < r_-$. But this is not the case. We know how to define quantum field theory in globally hyperbolic spacetimes. So quantum field theory is well-defined in the region $r>r_-$ of the BTZ spacetime.\footnote{Strictly speaking, this region is not globally hyperbolic because it is asymptotically AdS but of course this is dealt with in the usual way by imposing boundary conditions at infinity.} But there is no unambiguous way of extending quantum field theory into the region $r<r_-$. Any attempt to do so will be plagued by ambiguities that are at least as bad as those of the classical theory. So $\langle 0|T_{ab} | 0 \rangle$ is simply not defined for $r<r_-$. Hence the above calculation is a purely formal manipulation devoid of physical content. 

Second, even if we did know how to define quantum field theory in the region $r<r_-$, the results would depend on the spacetime geometry in $r<r_-$. But, as we emphasized in section \ref{sec:BTZ}, the classical geometry in this region is not unique! There are infinitely many ways of extending the BTZ spacetime smoothly, as a solution of Einstein's equation, into the region $r<r_-$. A calculation along the lines described above would have to show that similar divergences occur for any such extension. 

Since $\langle 0|T_{ab} | 0 \rangle$ is well-defined only for $r>r_-$, we need to consider the {\it limiting} behaviour of $\langle 0|T_{ab} (x) | 0 \rangle$ as $x$ approaches the Cauchy horizon from {\it outside}. We will consider the quantum theory of a massive scalar field, obeying standard boundary conditions, in the BTZ geometry. We will show that, for a near-extremal black hole, $\langle 0|T_{ab} (x) | 0 \rangle$ remains bounded as $r \rightarrow r_-$. In fact we will show that this quantity {\it extends continuously} to the Cauchy horizon of a near-extremal black hole.\footnote{Steif also stated this in a parenthetical remark but without mentioning the near-extremal condition.} Here ``near-extremal" means
\be
 \beta>1\,,
\ee
where $\beta$ is given by \eqref{betadef1}. Surprisingly, this is the same as the condition for {\it classical} perturbations to be $C^1$, and hence to have finite energy momentum tensor, at the Cauchy horizon. We conclude that, for near-extremal black holes, $\langle 0|T_{ab} | 0 \rangle$ is finite at the Cauchy horizon. Thus backreaction of vacuum polarization does not rescue strong cosmic censorship. 

Our result disagrees with a recent claim that such backreaction always renders the Cauchy horizon singular \cite{Casals:2016odj,Casals:2019jfo}. This claim is based on a study of a conformally coupled field obeying transparent boundary conditions, \emph{i.e.}, the case studied by Steif. The result of Refs.  \cite{Casals:2016odj,Casals:2019jfo} appears to be a consequence of using Steif's results for $\langle 0|T_{ab} | 0 \rangle$ in the region $r<r_-$. However, as we explained above, one cannot trust Steif's results in this region. A calculation of backreaction at the Cauchy horizon should use only the behaviour of $\langle 0|T_{ab} | 0 \rangle$ in the region $r>r_-$ and, for a near-extremal black hole, this is finite as $r \rightarrow r_-$ so backreaction will not render the Cauchy horizon singular.\footnote{The 
authors of Refs. \cite{Casals:2016odj,Casals:2019jfo} argue that it might be possible to define quantum field theory in the region $r<r_-$ by imposing boundary conditions at the timelike singularity of the (analytically extended) BTZ solution. However, it is not clear whether this is possible, and whether there is a choice of boundary conditions that would lead to Steif's result. Furthermore, it begs the question of {\it which} boundary conditions to impose at the singularity: much of the physical motivation for the cosmic censorship conjectures stems from the fact that we don't know what happens at a singularity.}

\subsection{Hadamard regularization}

Consider the quantum theory of a real scalar field with action
\be
 \int \mathrm{d}^4 x \sqrt{-g} \left( -\frac{1}{2} \nabla_a \Phi \nabla^a \Phi - \frac{1}{2} \mu^2 \Phi^2 \right).
\ee
We will employ Hadamard regularization (see {\it e.g.} \cite{Wald:1995yp} for a review) to define $\langle 0|T_{ab} (x) | 0 \rangle$ and $\langle 0| \Phi(x)^2 |0\rangle$ where $|0\rangle$ is the Hartle-Hawking state. 

We start from the symmetrized 2-point function in a general state $|\psi\rangle$
\be
 G^{(1)}(x,x') = \langle \psi | \Phi(x) \Phi(x') + \Phi(x') \Phi(x)  | \psi \rangle .
\ee
This quantity diverges as $x' \rightarrow x$. We regulate this divergence by defining
\be
\label{G1reg}
  G^{(1)}_{\rm reg}(x,x') = G^{(1)}(x,x') - 2H(x,x')\,,
\ee
where the {\it Hadamard parametrix} $H(x,x')$ captures the state-independent short-distance behaviour of $G^{(1)}$. In a three dimensional spacetime it takes the form \cite{Decanini:2005eg}
\be
 H(x,x') = \frac{1}{4 \sqrt{2} \pi [ \sigma(x,x')+i\epsilon]^{1/2}} \sum_{n=0}^\infty U_n(x,x') \sigma(x,x')^n\,,
\ee
where $2\sigma(x,x')$ is the square of the geodesic distance between $x$ and $x'$ (which is well-defined when $x,x'$ are nearby) and the coefficients $U_n$ are determined recursively by demanding that the above expression solves the equation of motion. These coefficients depend only on the spacetime geometry. In an Einstein spacetime ($R_{ab} = (1/3) Rg_{ab}$), the first few are \cite{Decanini:2005eg}\footnote{Here we have used the relation $g^{ab} \sigma_{,a} \sigma_{,b} = 2 \sigma$  \cite{Decanini:2005eg}.}
\be
 U_0 = 1 + \frac{1}{18} R \,\sigma + {\cal O}(\sigma^2)\,, \qquad U_1 = \mu^2  -\frac{1}{6} R + {\cal O}(\sigma)\,.
\ee
We now define the renormalized expectation value of $\Phi^2$ by taking the coincident limit of $G^{(1)}_{\rm reg}(x,x')$:
\be
\langle \psi| \Phi(x)^2 |\psi\rangle \equiv \lim_{x' \rightarrow x} \frac{1}{2} G^{(1)}_{\rm reg}(x,x') \,.
\ee 
Similarly we define the renormalized expectation value of $T_{ab}$ via point-splitting \cite{Wald:1995yp} as
\be
\label{Tdef}
 \langle \psi|T_{ab} (x) | \psi \rangle \equiv \lim_{x' \rightarrow x} {\cal T}_{ab}(x,x') \frac{1}{2} G^{(1)}_{\rm reg}(x,x')\,, 
\ee
where \cite{Decanini:2005eg}
\be
 {\cal T}_{ab} = g_{b}{}^{c'} \nabla_a \nabla_{c'} -\frac{1}{2} g_{ab} g^{cd'} \nabla_c \nabla_{d'} - \frac{1}{2} \mu^2 g_{ab}
\ee
and $g_a{}^{b'}$ is the operator that parallelly transports covectors from $x'$ to $x$. 

The above renormalization scheme gives results consistent with Wald's axioms \cite{Wald:1977up} for defining $\langle \psi|T_{ab} |\psi \rangle$ in a general state $|\psi\rangle$. Any other scheme for defining $\langle \psi|T_{ab} |\psi \rangle$ consistently with these axioms gives a result that differs from the above only by a tensor that is identically conserved and a function only of the local geometry \cite{Wald:1977up}. Since the BTZ spacetime is locally isometric to AdS$_3$, the only such tensor in our case is a multiple of the metric. Hence, any prescription for defining $\langle \psi|T_{ab} |\psi \rangle$ that is consistent with Wald's axioms will, in the BTZ spacetime, differ from the prescription described above by a constant, state-independent, multiple of the metric. Clearly this small ambiguity does not affect the question of whether or not $\langle \psi|T_{ab} |\psi \rangle$ diverges at the Cauchy horizon. 

A similar scheme-dependence occurs in the definition of $\langle \psi| \Phi(x)^2 |\psi \rangle$. In this case, the ambiguity amounts to the freedom to add a term of the form $c_1 R + c_2 \mu^2$ where $c_1$ and $c_2$ are state-independent constants \cite{Hollands:2001nf}. In our case, this amounts to the freedom to shift $\langle \psi| \Phi(x)^2|\psi  \rangle$ by a constant.

\subsection{Calculation for BTZ}

In our case the difficult step in the above calculation reduces to calculating $G^{(1)}(x,x')$ for the Hartle-Hawking state $|0\rangle$ in the BTZ geometry. Fortunately this has already been done in Ref. \cite{Ichinose:1994rg}. This reference calculated the Feynman propagator
\be
 -i G_F(x,x') \equiv \langle 0| T(\Phi(x) \Phi(x')) | 0 \rangle\,,
\ee
where $T$ denotes time-ordering. Recall the result \cite{Birrell:1982ix}
\be
 -i G_F(x,x') = -\frac{i}{2} \left[ G_A(x,x') + G_R(x,x') \right] + \frac{1}{2} G^{(1)}(x,x')\,,
\ee
where $G_A$ and $G_R$ are the advanced and retarded Green functions. This implies that if $x'$ is spacelike separated from $x$ then 
\be
\label{G1GF}
\frac{1}{2} G^{(1)}(x,x')= -i G_F(x,x')\,, \qquad {\rm spacelike \;\; separation.}
\ee
Ref. \cite{Ichinose:1994rg} calculated $G_F$ using the method of images. The result is 
\be
\label{GFsum}
-i G_F(x,x') =  \sum_{n=-\infty}^\infty f(z_n)\,,
\ee
where
\be
f(z_n)= \frac{1}{4\pi L}(z_n^2-1)^{-1/2} \left[ z_n + (z_n^2-1)^{1/2}\right]^{(1-\Delta)} \, ,
\ee 
$\Delta$ is the conformal dimension defined in section \ref{sec:Delta}\footnote{For the range of $\mu$ for which two choices of $\Delta$ are possible, this result applies for both choices.} and, when $x,x'$ are both in region I, $z_n$ is defined by
\ba
 z_n - i \epsilon&=& \frac{1}{r_+^2 - r_-^2} \left[ \sqrt{r^2 - r_-^2} \sqrt{{r'}^2 - r_-^2} \cosh\left( \frac{r_-}{L^2}\delta t - \frac{r_+}{L} \delta \phi_n \right)\right. \nonumber \\ &-& \left.  \sqrt{r^2 - r_+^2} \sqrt{{r'}^2 - r_+^2} \cosh\left( \frac{r_+}{L^2}\delta t - \frac{r_-}{L} \delta \phi_n \right) \right]
 \label{eq:zns}
\ea
where
\be
 \delta t = t-t' \,, \qquad \delta \phi_n = \phi - \phi' + 2\pi n \,.
\ee
To extend the above result into region II we convert to the coordinates $(U_+,V_+,\phi_+)$ defined in section \ref{sec:BTZ} with the result
\ba
\label{znKruskal}
 z_n - i\epsilon &=& \frac{1}{r_+^2 - r_-^2} \left\{ \sqrt{r^2 - r_-^2} \sqrt{{r'}^2 - r_-^2}\cosh \left( \frac{r_+}{L} \delta \phi_{+n} \right) \right. \nonumber \\ &+& \left. \frac{1}{2} G_+(r) G_+(r') \left[ U_+'V_+ \exp\left(-\frac{r_-}{L} \delta \phi_{+n} \right) + U_+ V_+'  \exp\left(\frac{r_-}{L} \delta \phi_{+n} \right)\right]\right\}\,.
\ea
Here, 
\be
\delta \phi_{+n} = \delta \phi_+ + 2\pi n = \phi_+ - \phi_+' + 2\pi n\,,
\ee
$r(U_+,V_+)$ (and $r(U_+',V_+')$) is defined by \eqref{rUVplus} and
\be
 G_+(r) \equiv \sqrt{\frac{r^2 - r_+^2}{F_+(r)}}\,,
\ee
where $F_+$ is defined in \eqref{Fpdef}. Note that $G_+(r)$ is real and analytic for $r>r_-$. 

We will perform point-splitting in a spacelike direction so that $G^{(1)}$ is determined by \eqref{G1GF}.
To renormalize $G_F$ we need to subtract the Hadamard parametrix. The effect of this subtraction is to eliminate the non-analytic terms in $-iG_F$ that have the form of $\sigma^{-1/2}$ times a smooth function of the coordinates. So we just need to identify this non-analytic part of $-iG_F$ and discard it. 

Note that for coincident points (in either region) we have, from \eqref{znKruskal}
\be
\label{znlim}
 z_n (x,x) = \bar{z}_n(r) \equiv \frac{r^2-r_-^2}{r_+^2 - r_-^2} \left[ \cosh \left( \frac{2\pi n r_+}{L} \right) - \cosh \left( \frac{2\pi n r_-}{L} \right)\right] +\cosh \left( \frac{2\pi n r_-}{L} \right).
\ee
Since $r>r_-$ we have $z_n(x,x)>1$ for $n \ne 0$. However $z_0(x,x)=1$. 
It follows that the divergence (for coincident points) in \eqref{GFsum} must arise from the $n=0$ term, as in \cite{Steif:1993zv}. This term is just the Feynman propagator for global AdS$_3$ (with the other terms in the sum arising from this one via the method of images). It follows that, for $x'$ near to $x$, $z_0$ must be a function of $\sigma$ (since this is the case in AdS$_3$). Of course the AdS$_3$ propagator will exhibit the Hadamard non-analytic (in $\sigma$) behaviour when $x'$ is close to $x$. Since this behaviour depends only on the local geometry, it is the same in BTZ and in AdS$_3$. Hence all of the non-analytic behaviour of $G_F$ must arise from the $n=0$ term. Expanding this term around $z_0=1$ gives
\begin{multline}
f(z_0) = \frac{1}{4\pi L\sqrt{2(z_0-1)}}+ \frac{1-\Delta}{4\pi L}  + \frac{3-8\Delta + 4\Delta^2}{16 \pi L} \sqrt{\frac{z_0-1}{2}} -
\\
 \frac{\Delta(\Delta-1)(\Delta-2)}{12 \pi L}(z_0-1) + {\cal O}\big((z_0-1)^{3/2}\big).
\end{multline}
Matching this to the Hadamard parametrix gives $z_0-1 = (\sigma/L^2)(1+a_1\sigma+a_2 \sigma^2 + \ldots)$ for certain coefficients $a_i$.\footnote{Of course we could also obtain this result from the expression for $z_0$. But it is quicker to match to the known Hadamard parametrix.} Hence the non-analytic (in $\sigma$) part of the above expression is simply the part that is non-analytic in $z_0-1$. So to perform Hadamard renormalization we simply discard the terms that are non-analytic in $z_0-1$. Since we ultimately want to take the limit of coincident points after taking at most two derivatives, we only need to retain the analytic terms up to order $(z_0-1)$. In other words, to perform Hadamard renormalization we can take (for spacelike separation)
\be
\label{G1sum}
 \frac{1}{2}G^{(1)}_{{\rm reg}}(x,x') =  \frac{1-\Delta}{4\pi L}-\frac{\Delta(\Delta-1)(\Delta-2)}{12 \pi L}(z_0-1) + \sum_{n=1}^\infty \left[f(z_n) + f(z_{-n}) \right]\,.
\ee
Note that $z_n$ is not smooth as $x' \rightarrow x$: it approaches the limit \eqref{znlim} but there are subleading terms behaving as powers of $\sigma^{1/2}$. However, these terms cancel out in the combination $f(z_n)+f(z_{-n})$. (This is easy to see if one performs the point-splitting in the $\phi_+$ direction which gives $\sigma^{1/2} \approx r \delta \phi_+/\sqrt{2}$ and $n \rightarrow -n$ has the same effect as $\delta\phi_+ \rightarrow - \delta \phi_+$.) This had to be the case because, as discussed above, the non-analytic behaviour arises only from the $n=0$ term.  

\subsection{Results for  $\langle 0|\Phi(x)^2|0 \rangle$}

\label{sec:Phi2}

From \eqref{G1sum} we have
\be
\label{Phi2}
 \langle 0|\Phi(x)^2|0  \rangle = \frac{1-\Delta}{4\pi L}+\sum_{n=1}^\infty F_n(r)\,,
\ee
where
\be
F_n(r) =  f(\bar{z}_n(r)) + f(\bar{z}_{-n}(r) )
\ee
and $\bar{z}_n(r)$ is defined in \eqref{znlim}. We need to discuss the convergence of the sum over $n$. We have $\bar{z}_n(r) \rightarrow \infty$ as $n \rightarrow \pm \infty$ for any $r \ge r_-$. More precisely, for $r \ge r_-$ we can bound $\bar{z}_n(r)$ as
\be
\label{znbound}
 \bar{z}_n(r) \ge \cosh \left( \frac{2\pi n r_-}{L} \right) > \frac{1}{2} \exp \left( \frac{2\pi |n| r_-}{L} \right). 
\ee
For large positive $z$ we have $f(z) < Cz^{-\Delta}$ for some positive constant $C$. Combining these bounds, we learn that if $r \ge r_-$ then 
\be
0<  F_n(r)< A \exp \left( \frac{-2\pi n \Delta  r_-}{L} \right) 
\ee
for some constant $A>0$. If $r_->0$ and $\Delta>0$ then it follows that the sum in \eqref{Phi2} is {\it uniformly convergent} for $r \ge r_-$. Hence it defines a function of $r$ that is continuous on $r \ge r_-$. This proves that $ \langle 0|\Phi(x)^2|0\rangle$ (which is defined only for $r>r_-$) is bounded and extends continuously to $r=r_-$. 

The limiting value of $ \langle 0|\Phi(x)^2|0\rangle$ at the Cauchy horizon, and also the value at the event horizon, can be expressed in terms of special functions as follows. We first recall the series definition of the $q-$digamma function $\psi_q(w)$:
\begin{equation}
\psi_q(w)=\frac{\mathrm{d}}{\mathrm{d}w}\log \Gamma_q(w)=-\log(1-q)-\log q \sum_{k=0}^{+\infty}\frac{1}{1-q^{-k-w}}\,,
\end{equation}
where $\Gamma_q(w)$ is the $q-$gamma function.

Uniform convergence for $r \ge r_-$ implies that the limiting value of $\langle 0|\Phi(x)^2|0\rangle$ as $r \rightarrow r_-$ is given by setting $r=r_-$ in each term on the RHS of \eqref{Phi2}, which gives
\begin{equation}
\lim_{r\rightarrow r_-} \langle 0|\Phi(x)^2|0  \rangle  = \frac{1-\Delta }{4 \pi  L}+\frac{1}{2 \pi  L}\sum_{n=1}^{+\infty} \frac{e^{\alpha  (1-\Delta ) n \chi } }{ \sinh(\alpha  n \chi )}\,,
\end{equation}
where we defined
\be
\label{alphadef}
\alpha \equiv 2\pi r_+/L\,,\qquad \text{and}\qquad \chi\equiv r_-/r_+\,.
\ee
However, we know that
\begin{equation}
\frac{1}{\sinh (\alpha n \chi)}=\frac{2 e^{-\alpha n \chi}}{1-e^{-2\alpha n \chi}}=2 e^{-\alpha n \chi} \sum_{k=0}^{+\infty} e^{-2k\alpha n \chi}\,,
\end{equation}
so that
\begin{align}
\lim_{r\rightarrow r_-} \langle 0|\Phi(x)^2|0  \rangle & = \frac{1-\Delta }{4 \pi  L}+\frac{1}{\pi L}\sum_{n=1}^{+\infty}  \sum_{k=0}^{+\infty} e^{-(2k+\Delta)\alpha n \chi}
 = \frac{1-\Delta }{4 \pi  L}+\frac{1}{\pi L}\sum_{k=0}^{+\infty}\sum_{n=1}^{+\infty}  e^{-(2k+\Delta)\alpha n \chi}\nonumber
\\
& = \frac{1-\Delta }{4 \pi  L}+\frac{1}{\pi L}\sum_{k=0}^{+\infty}\frac{1}{e^{\alpha  \chi  (\Delta +2 k)}-1}\,.\label{eq:last}
\end{align}
The infinite sum appearing in the last line of Eq.~(\ref{eq:last}) is precisely the one appearing in the definition of the $q-$digamma provided we identify $q\equiv e^{-2 \alpha  \chi }$ and $w\equiv \Delta/2$. This gives the compact expression
\begin{equation}
\lim_{r\rightarrow r_-} \langle 0|\Phi(x)^2|0  \rangle=\frac{1-\Delta }{4 \pi  L}-\frac{1}{2 \pi  L \alpha \chi }\left[\log \left(1-e^{-2 \alpha  \chi }\right)+\psi _{e^{-2 \alpha  \chi }}\left(\frac{\Delta }{2}\right)\right]\,.
\end{equation}
A very similar calculation gives the value of $\langle 0|\Phi(x)^2|0\rangle$ at the event horizon. The result is the same as above with the replacement $\chi\to1$, yielding
\begin{equation}
\left. \langle 0|\Phi(x)^2|0  \rangle \right|_{r=r_+}=\frac{1-\Delta }{4 \pi  L}-\frac{1}{2 \pi  L \alpha }\left[\log \left(1-e^{-2 \alpha}\right)+\psi _{e^{-2 \alpha}}\left(\frac{\Delta }{2}\right)\right]\,.
\end{equation}

\subsection{Analytic results for $\langle 0|T_{ab} | 0 \rangle$}

Next we consider $\langle 0|T_{ab} | 0 \rangle$. Since we are interested in the behaviour of this quantity as we approach the Cauchy horizon, we need to calculate in coordinates that are regular at the Cauchy horizon. We will use the Kruskal coordinates $(U_-,V_-,\phi_-)$. In region II these are defined by \eqref{KruskalCauchy}. Converting \eqref{znKruskal} to these coordinates gives
\ba
z_n - i\epsilon &=& \frac{1}{r_+^2 - r_-^2} \left\{ \sqrt{r_+^2-r^2} \sqrt{r_+^2 - {r'}^2 }\cosh \left( \frac{r_-}{L} \delta \phi_{-n} \right) \right. \nonumber \\ &+& \left. \frac{1}{2} G_-(r) G_-(r') \left[ U_-V_-' \exp\left(\frac{r_+}{L} \delta \phi_{-n} \right) + U_-' V_-  \exp\left(-\frac{r_+}{L} \delta \phi_{-n} \right)\right]\right\}
\ea
where
\be
 \delta \phi_{-n} = \delta \phi_- + 2\pi n = \phi_- - \phi_-' + 2\pi n,
\ee
$r(U_-,V_-)$ is defined by \eqref{rUVminus} and
\be
 G_-(r) \equiv \sqrt{\frac{r^2 - r_-^2}{F_-(r)}}
\ee
where $F_-$ is defined in \eqref{Fmdef}. Note that $G_-(r)$ is real for $r<r_+$ and smooth at $r=r_-$. 

The expression for $\langle 0|T_{ab} | 0 \rangle$ obtained from \eqref{Tdef} contains a term without derivatives. This term will give a result proportional to $\langle 0|\Phi(x)^2|0\rangle$ and so it extends continuously to the Cauchy horizon. The only terms in $\langle 0|T_{ab} | 0 \rangle$ that might diverge at the Cauchy horizon are those containing derivatives. These terms have the form
\be
 \lim_{x' \rightarrow x} \partial_\mu \partial_{\nu'}  \frac{1}{2}G^{(1)}_{{\rm reg}}(x,x') =  \lim_{x' \rightarrow x} \partial_\mu \partial_{\nu'} \left\{-\frac{\Delta(\Delta-1)(\Delta-2)}{12 \pi L}( z_0-1) + \sum_{n=1}^\infty \left[f(z_n) + f(z_{-n}) \right]
 \right\}.
\ee
We have seen above that $z_0 - 1 \propto \sigma$, which is a smooth function of $x,x'$ so the first term above is finite at $r=r_-$. The potentially dangerous behaviour as $r \rightarrow r_-$ comes from the sum over $n$. 

The result of term by term differentiation is
\be
\label{limG}
 \lim_{x' \rightarrow x} \partial_\mu \partial_{\nu'}  \frac{1}{2}G^{(1)}_{{\rm reg}}(x,x') =  -\frac{\Delta(\Delta-1)(\Delta-2)}{12 \pi L} \lim_{x' \rightarrow x} \partial_\mu \partial_{\nu'} z_0 + \sum_{n=1}^\infty H_{\mu\nu}^{(n)}(x) \,,
\ee
where
\be
H^{(n)}_{\mu\nu}(x) =  f'(\bar{z}_n(r))  \lim_{x' \rightarrow x}\partial_\mu \partial_{\nu'} z_n + f''(\bar{z}_n(r)) \lim_{x' \rightarrow x} \partial_\mu z_n  \partial_{\nu'} z_n + (n \rightarrow -n)\,.
\ee
We need to understand the large $n$ behaviour of $H^{(n)}_{\mu\nu}(x)$. Recall the bound \eqref{znbound}. It is easy to bound the large $z$ behaviour of derivative of $f(z)$:
\be
 |f(z)| < Cz^{-\Delta}\,, \qquad |f'(z)| < C'z^{-\Delta-1}\,, \qquad |f''(z)| < C'' z^{-\Delta-2}\,,
\ee
for constants $C,C',C''$. It follows that, for $r \ge r_-$ we have (for positive or negative $n$)
\be
 | f'(\bar{z}_n(r))|< B'\exp \left(-\frac{2\pi |n| (\Delta+1) r_-}{L} \right) \,, \qquad  | f''(\bar{z}_n(r))|< B''\exp \left(-\frac{2\pi |n| (\Delta+2) r_-}{L} \right)
\ee
for constants $B,B',B''$. Next we calculate the first and second derivatives of $z_n$. We claim that the following bounds are satisfied for $r_- \le r \le r_0$ where $r_-<r_0<r_+$:
\be
\label{dznbound}
 |\lim_{x' \rightarrow x} \partial_{U_-} z_n | \le c_1 |V_-| e^{\alpha |n|}\,, \qquad
 |\lim_{x' \rightarrow x} \partial_{V_-} z_n | \le c_1 |U_-| e^{\alpha |n|}\,, \qquad
|\lim_{x' \rightarrow x} \partial_{\phi_-} z_n | \le c_2 e^{\alpha |n|}\,,
\ee
\ba
\label{ddznbound}
 |\lim_{x' \rightarrow x} \partial_{U_-} \partial_{U_-'} z_n | &\le& d_1 V_-^2 e^{\alpha |n|} \,,\qquad
 |\lim_{x' \rightarrow x} \partial_{V_-} \partial_{V_-'} z_n | \le d_1 U_-^2 e^{\alpha |n|}\,,\nonumber \\
 |\lim_{x' \rightarrow x} \partial_{U_-} \partial_{V_-'} z_n | &\le& d_2 e^{\alpha |n|} \,,\nonumber \\
 |\lim_{x' \rightarrow x} \partial_{U_-} \partial_{\phi_-'} z_n | &\le& d_3 |V_-| e^{\alpha |n|} \,,\qquad
 |\lim_{x' \rightarrow x} \partial_{V_-} \partial_{\phi_-'} z_n | \le d_3 |U_-| e^{\alpha |n|}\,, \nonumber \\
|\lim_{x' \rightarrow x} \partial_{\phi_-} \partial_{\phi_-'} z_n | &\le& d_4 e^{\alpha |n|}\,,
\ea
where $c_i$ and $d_i$ are positive constants depending only on $r_\pm$ and $r_0$. The proof of these results is given in the Appendix.\footnote{Note that the various powers of $U_-$ and $V_-$ on the RHS of these equations cancel out if one converts to Eddington-Finkelstein coordinates, leaving expressions that are functions of $r$ alone. This is a consequence of the fact that the Hartle-Hawking state shares the symmetries of the background.} Using these results we obtain
\be
\label{Hnbound}
 |H^{(n)}_{\mu\nu}(x)| \le |U_-|^\sigma |V_-|^\tau \left[ D e^{(\alpha - 2\pi (\Delta+1)r_-/L)n } + D'e^{(2\alpha - 2\pi (\Delta+2)r_-/L)n} \right]
\ee
where $D,D'$ are positive constants depending only on $r_\pm$, $r_0$ and $\Delta$; the constants $\sigma,\tau$ are defined by $\sigma=1$ for $\{\mu\nu\} = \{V_-,\phi_-\}$, $\sigma=2$ for $\{\mu\nu\} = \{V_-,V_-\}$ and $\sigma=0$ otherwise; $\tau=1$ for $\{\mu\nu\} = \{U_-,\phi_-\}$, $\tau=2$ for $\{\mu\nu\} = \{U_-,U_-\}$ and $\tau=0$ otherwise.

The bound \eqref{Hnbound} implies that the sum over $n$ in \eqref{limG} is uniformly convergent for $r_- \le r \le r_0$ if we have
\be
\label{alpha_cond}
 \alpha<2\pi (\Delta+1)r_-/L \qquad {\rm and}  \qquad 2\alpha<2\pi (\Delta+2)r_-/L\,.
\ee
The latter condition is more restrictive, and this condition can be rearranged to
\be
\beta>2\,,
\ee
where $\beta$ is defined in \eqref{betadef1}. For any $\Delta>0$, this condition can be satisfied by taking $r_-$ close enough to $r_+$, {\it i.e.}, by taking the black hole close enough to extremality.  

We have shown that, for any given $\Delta>0$, if the black hole is close enough to extremality then the sums defining $\langle 0|T_{ab}|0\rangle$ are uniformly convergent for $r_- \le r \le r_0$. It follows that the resulting expression for $\langle 0|T_{ab}|0\rangle$ is {\it continuous} on this interval. In particular this proves that  $\langle 0|T_{ab}|0\rangle$ (which is defined only for $r>r_-$) is bounded for $r_-<r<r_0$ and {\it extends continuously} to $r=r_-$, {\it i.e.},  $\lim_{r\rightarrow r_-}\langle 0|T_{ab}|0\rangle$ exists. {\it Hence, for a massive scalar field in a near-extremal BTZ black hole spacetime, the renormalized expectation value of the energy momentum tensor in the Hartle-Hawking state is finite at the Cauchy horizon.}


Let us now discuss the case $\beta \le 2$ for which the above proof of uniform convergence on $r_- \le r \le r_0$ no longer holds. In this case we can still prove uniform convergence on any compact subset of $r_-<r<r_+$ as follows. Choose $r_1$ and $r_0$ so that $r_-<r_1<r_0<r_+$ and assume $r_1 \le r \le r_0$. Previously we used the lower bound \eqref{znbound} for $\bar{z}_n(r)$. Since we are assuming that $r_1$ is strictly greater than $r_-$ we now have a stronger bound:
\be
 \bar{z}_n(r) \ge C \exp(\alpha |n|)\,,
\ee
for some positive constant $C$ depending on $r_\pm$, $r_0$ and $r_1$, with $\alpha$ defined in \eqref{alphadef}. We can then repeat the steps described above using this stronger bound. The result is to replace $r_-$ with $r_+$ in the exponents of \eqref{Hnbound}. So the condition for uniform convergence is given by replacing $r_-$ with $r_+$ on the RHS of \eqref{alpha_cond}. These inequalities are trivially satisfied because $\Delta>0$. Hence we have uniform convergence for $r_0 \le r \le r_1$, {\it i.e.}, uniform convergence on any compact subset of $r_-<r<r_+$. Hence $\langle 0|T_{ab}|0\rangle$ is finite for $r_-<r<r_+$.\footnote{
Of course $\langle 0|T_{ab}|0\rangle$ is always finite at $r=r_+$. This is not apparent from the above analysis because we used coordinates adapted to the Cauchy horizon that break down at $r=r_+$.} However, since we no longer have uniform convergence on a compact set containing $r=r_-$ we can no longer conclude that $\langle 0|T_{ab}|0\rangle$ will be bounded as $r \rightarrow r_-$.

\subsection{Improving the bound}

We have proved that $\langle 0|T_{ab}|0\rangle$ extends continuously to the Cauchy horizon if $\beta>2$. In this section we are going to show that this result can be improved to $\beta>1$. To do this we need to make use of the detailed structure of $T_{ab}$. This makes the calculations more difficult so we will not attempt to proceed with the same level of rigour used above, and the results below were obtained with the use of {\it Mathematica}.

We work in {\it co-rotating} outgoing coordinates $(u,r,\phi''_-)$ defined in terms of the $(t,r,\phi)$ coordinates of section \ref{sec:BTZ} as
\begin{equation}
\mathrm{d}t=\mathrm{d}u+\frac{\mathrm{d}r}{f}\,,\qquad\text{and}\qquad \mathrm{d}\phi=\mathrm{d}\phi^{\prime\prime}_-+\Omega(r)\frac{\mathrm{d}r}{f}-\Omega_-\,\mathrm{d}u\,,
\end{equation}
in terms of which the BTZ metric reads
\begin{equation}
\mathrm{d}s^2=-f\,\mathrm{d}u^2-2\,\mathrm{d}u\,\mathrm{d}r+r^2\left(\mathrm{d}\phi^{\prime\prime}_-+\frac{r^2-r_-^2}{r^2}\Omega_- \mathrm{d}u\right)^2\,.
\end{equation}
In these coordinates, ${\cal CH}_R^+$ is at $r=r_-$  and ${\cal H}_L^+$ is at $r=r_+$. 

The expression we want to study takes the form
\be
\label{Tsum}
\langle 0|T_{\mu\nu} | 0 \rangle = T^0_{\mu\nu}(r)+\sum_{n=1}^{+\infty} T^n_{\mu\nu}(r)\,,
\ee
where
\be
T^0_{\mu\nu}(r)=-\frac{5 (2-\Delta ) (1-\Delta ) \Delta}{24 \pi }  \left(
\begin{array}{ccc}
\frac{\alpha ^2 \left(1-\chi ^2\right) \left(\zeta^2-\chi ^2\right)}{4\pi^2\chi ^2} & -1 & \frac{\alpha ^2 \left(\zeta^2-\chi^2\right)}{4\pi^2\chi }
\\
 -1 & 0 & 0 \\
\frac{\alpha ^2 \left(\zeta^2-\chi^2\right)}{4\pi^2\chi }& 0 &\frac{\alpha ^2 \zeta^2}{4\pi^2} \\
\end{array}
\right)
\ee
comes from the the term proportional to $(z_0-1)$ in \eqref{G1sum}. Here we have defined
\begin{equation}
\label{zetadef}
\zeta\equiv \frac{r}{r_+}\,,
\end{equation}
and $(\alpha,\chi)$ are given in \eqref{alphadef}. 
The $T^n_{\mu\nu}(r)$ are too cumbersome to be presented here, but we are only interested in their large $n$ behaviour. For large $n$, things simplify considerably. The leading order behaviour is
\begin{multline}
T^n_{\mu\nu}(r)\approx -\frac{4 \pi  \Delta ^2 \left(1-\chi ^2\right)^{\Delta }}{\alpha ^2}\frac{X_n^5 \left(\zeta^2-\chi ^2\right)^4 }{\left[X_n^{\chi } \left(1-\zeta^2\right)+X_n \left(\zeta^2-\chi ^2\right)\right]^{5+\Delta }}
\\
\left(
\begin{array}{ccc}
 \frac{\alpha ^4 (1-\Delta ) \left(1-\chi ^2\right) \left(\zeta^2-\chi ^2\right)^2}{16 \pi ^4 \Delta  \chi ^2} & -\frac{\alpha ^2 (1-\Delta ) \left(\zeta^2-\chi ^2\right)}{4 \pi ^2 \Delta } & \frac{\alpha ^4
   (1-\Delta ) \left(\zeta^2-\chi ^2\right)^2}{16 \pi ^4 \Delta  \chi } \\
 -\frac{\alpha ^2 (1-\Delta ) \left(\zeta^2-\chi ^2\right)}{4 \pi ^2 \Delta } & 1 & \frac{\alpha ^2 \chi }{4 \pi ^2} \\
 \frac{\alpha ^4 (1-\Delta ) \left(\zeta^2-\chi ^2\right)^2}{16 \pi ^4 \Delta  \chi } & \frac{\alpha ^2 \chi }{4 \pi ^2} & \frac{\alpha ^4}{16 \pi ^4 \Delta } \left[\zeta^2 \left(\zeta^2-\chi ^2\right)-\left(\zeta^4-\chi ^2\right)
   \Delta \right]
   \\
\end{array}
\right)
\label{eq:crazy}
\end{multline}
where we defined $X_n \equiv e^{2\pi n r_+/L}$. Recall that ${\cal CH}_R^+$ is located at $\zeta=\chi$ and ${\cal H}_L^+$ at $\zeta=1$. The region of interest is thus $\chi\leq \zeta\leq 1$. We now note that the only dependence in $n$ in the $(\mu,\nu)$ component is given by a factor of the form
\be
\frac{ X_n^5\left(\zeta^2-\chi ^2\right)^{p_{\mu\nu}} }{\left[X_n^{\chi } \left(1-\zeta^2\right)+X_n \left(\zeta^2-\chi ^2\right)\right]^{5+\Delta }}\,,
\ee
where $p_{\mu \nu}$ is symmetric in $\mu$ and $\nu$ and
\be
p_{rr}=p_{r\phi^{\prime\prime}_-}=p_{\phi^{\prime\prime}_-\phi^{\prime\prime}_-}=4\,,\qquad p_{u r}=5\,,\qquad\text{and}\qquad p_{uu}=6.
\ee
The following series of inequalities let us bound the behaviour of the above terms:
\begin{align}
\label{Xnbound}
0 &\leq \frac{X_n^5 \left(\zeta^2-\chi ^2\right)^{p_{\mu\nu}}}{\left[X_n^{\chi } \left(1-\zeta^2\right)+X_n \left(\zeta^2-\chi ^2\right)\right]^{5+\Delta }}\leq \frac{{X_n}^{5-{p_{\mu\nu}}} \left[X_n^{\chi } \left(1-\zeta^2\right)+X_n \left(\zeta^2-\chi ^2\right)\right]^{p_{\mu\nu}}}{ \left[X_n^{\chi } \left(1-\zeta^2\right)+X_n \left(\zeta^2-\chi ^2\right)\right]^{5+\Delta }}\nonumber
\\
& =\frac{{X_n}^{5-p_{\mu\nu}}}{\left[X_n^{\chi } \left(1-\zeta^2\right)+X_n \left(\zeta^2-\chi ^2\right)\right]^{5+\Delta-{p_{\mu\nu}}}}\leq \frac{1}{X_n^{\chi(5+\Delta-p_{\mu\nu})+p_{\mu\nu}-5} \left(1-\zeta^2\right)^{5+\Delta-p_{\mu\nu}}}\,.
\end{align}
We will restrict to $\zeta \in [\chi,\zeta_0]$ where $\chi<\zeta_0<1$ so that the factor of $(1-\zeta^2)$ in the denominator is bounded.\footnote{Of course nothing funny happens at $\zeta=1$ but we would have to bound the terms differently in order to demonstrate convergence at $\zeta=1$.} Equation \eqref{Xnbound} implies that when we sum \eqref{eq:crazy} over $n$,
the terms in the sum are uniformly bounded by $C e^{-\alpha q_{\mu\nu} n}$ for some constant $C>0$ where $q_{\mu\nu} = \chi(5+\Delta-p_{\mu\nu})+p_{\mu\nu}-5$. Hence, provided there are no surprises coming from subleading (in $n$) corrections to \eqref{eq:crazy}, the sum in \eqref{Tsum} converges uniformly for $\zeta \in  [\chi,\zeta_0]$ if $q_{\mu\nu}>0$. This condition can be rearranged to
\begin{equation}
 \beta>5-p_{\mu\nu}\,.
\end{equation}
The strongest bound thus follows from the smallest value of $p_{\mu\nu}$ which occurs for $p_{rr}=p_{r\phi^{\prime\prime}_-}=p_{\phi^{\prime\prime}_-\phi^{\prime\prime}_-}=4$, which gives $\beta>1$. 

To complete the argument we need to inspect the subleading (in $n$) correction to \eqref{eq:crazy} to see if more stringent conditions appear. Rather remarkably, that is not the case. All other terms in the large $n$ expansion not shown in \eqref{eq:crazy} are proportional to terms of the form
\begin{subequations}
\begin{align}
& \frac{X_n^a(\zeta^2-\chi^2)^b}{\left[X_n^{\chi } \left(1-\zeta^2\right)+X_n \left(\zeta^2-\chi ^2\right)\right]^{5+\Delta }}\,,\;\text{with}\; a-b-(5+\Delta-b)\chi<0\,,\;\text{provided}\;\beta>1
\\
\text{or}\nonumber
\\
& \frac{X_n^a}{\left[X_n^{\chi } \left(1-\zeta^2\right)+X_n \left(\zeta^2-\chi ^2\right)\right]^{5+\Delta }}\,,\;\text{with}\; a -(5+\Delta ) \chi<0\,,\;\text{provided}\;\beta>1\,,
\end{align}
\end{subequations}
where the proportionality coefficients depend on $\chi$, $\zeta$, $\Delta$ and $y_+$ but not on $n$. Furthermore, these proportionality coefficients do not vanish when $\zeta=\chi$. The first type of term is precisely of the same form as the leading term appearing in \eqref{eq:crazy}, and it is easy to show that, at most, it provides the same bound. The second type of term can also be easily bounded via
\begin{equation}
\frac{X_n^a}{\left[X_n^{\chi } \left(1-\zeta^2\right)+X_n \left(\zeta^2-\chi ^2\right)\right]^{5+\Delta }}<\frac{X_n^a}{\left[X_n^{\chi } \left(1-\zeta^2\right)\right]^{5+\Delta }}=\frac{X_n^{a-(5+\Delta)\chi}}{\left(1-\zeta^2\right)^{5+\Delta }}< C' e^{ \alpha [a-(5+\Delta)\chi] n }\,.
\end{equation}
In summary, we have found that the sum over $n$ appearing in our expression for $\langle 0|T_{\mu\nu} | 0 \rangle$ is uniformly convergent for $r_- \le r \le r_0$ (where $r_-<r_0<r_+$) when $\beta>1$. Hence, for $\beta>1$, $\langle 0|T_{ab} | 0 \rangle$ extends continuously to ${\cal CH}_R^+$. 

\subsection{Numerical results for $\langle 0|T_{ab} | 0 \rangle$}

\label{sec:Tnum}

We have argued that $\langle 0|T_{ab} | 0 \rangle$ extends continuously to ${\cal CH}_R^+$ for $\beta>1$. In this section we will explore what happens for $\beta \le 1$. We will evaluate the sum in \eqref{Tsum} numerically. Recent results for $\langle 0|T_{ab} | 0 \rangle$ for a 4d Reissner-Nordstr\"om black hole found that non-trivial behaviour can arise {\it very} close to the Cauchy horizon: $(r-r_-)/M < 10^{-175}$ \cite{Lanir:2018vgb}. Therefore we need to ensure that our numerics is accurate enough to see features at this scale. To do this we have written a small {\it Mathematica} code that uses extended precision\footnote{We also used a Fortran code with octuple precision, and the results were unchanged.}. Numerically, we can only sum a certain number of terms in the series, say $N$. In order to test the convergence of our numerical results we increase $N$ and see if the outcome of our calculations changes. All the plots generated in this manuscript were done using $N=10^6$, and convergence was checked by repeating the same calculations with $N=10^5$. The global error is well under $10^{-10}\%$.  In this section we continue to work in the coordinates $(u,r,\phi''_-)$ of the previous section. 

\begin{figure}
	\centering
	\includegraphics[width=0.5\textwidth]{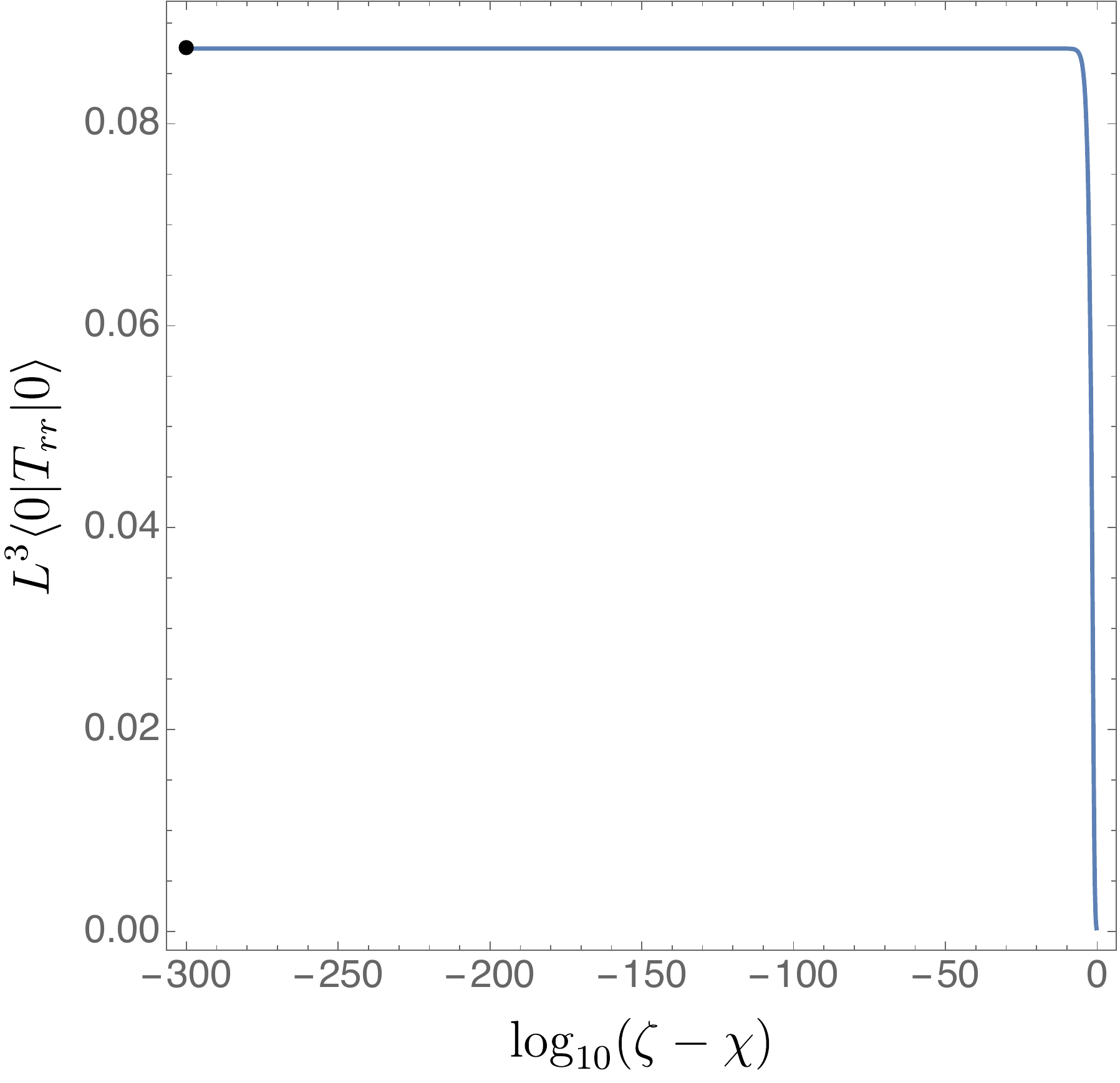}
	\caption{Plot of $L^3\langle 0|T_{rr} | 0 \rangle$ as a function of $\log_{10}(\zeta-\chi)$ computed for $\Delta = 1$, $\alpha=2\pi$ and $\beta=3/2$. The black dot denotes $L^3 t_{rr}$.
	}
	\label{fig:check1}
\end{figure}

First we use our numerics to confirm the analytical prediction of the previous section. Let $t_{\mu\nu}$ be the result obtained by substituting $r=r_-$ into the RHS of \eqref{Tsum}. Our prediction is that, for $\beta>1$, $t_{\mu\nu}$ is finite and $\lim_{r\rightarrow r_-} \langle 0|T_{\mu\nu} | 0 \rangle = t_{\mu\nu}$. This is confirmed numerically in Fig.~\ref{fig:check1} where we plot $L^3\langle 0|T_{rr} | 0 \rangle$ as a function of $\log_{10}(\zeta-\chi)$, for $\Delta = 1$, $\beta=3/2$ and $\alpha = 2\pi$. (Recall that $(\alpha,\chi)$ are defined in \eqref{alphadef}.) The plot shows no interesting features as we approach the Cauchy horizon $(\zeta=\chi)$. The black dot denotes $L^3 t_{rr}$. We see that indeed $\lim_{r\rightarrow r_-} \langle 0|T_{rr} | 0 \rangle = t_{rr}$. We have repeated this exercise for many values of $\Delta$ and $\beta>1$ with similar results.

\begin{figure}
	\centering
	\includegraphics[width=\textwidth]{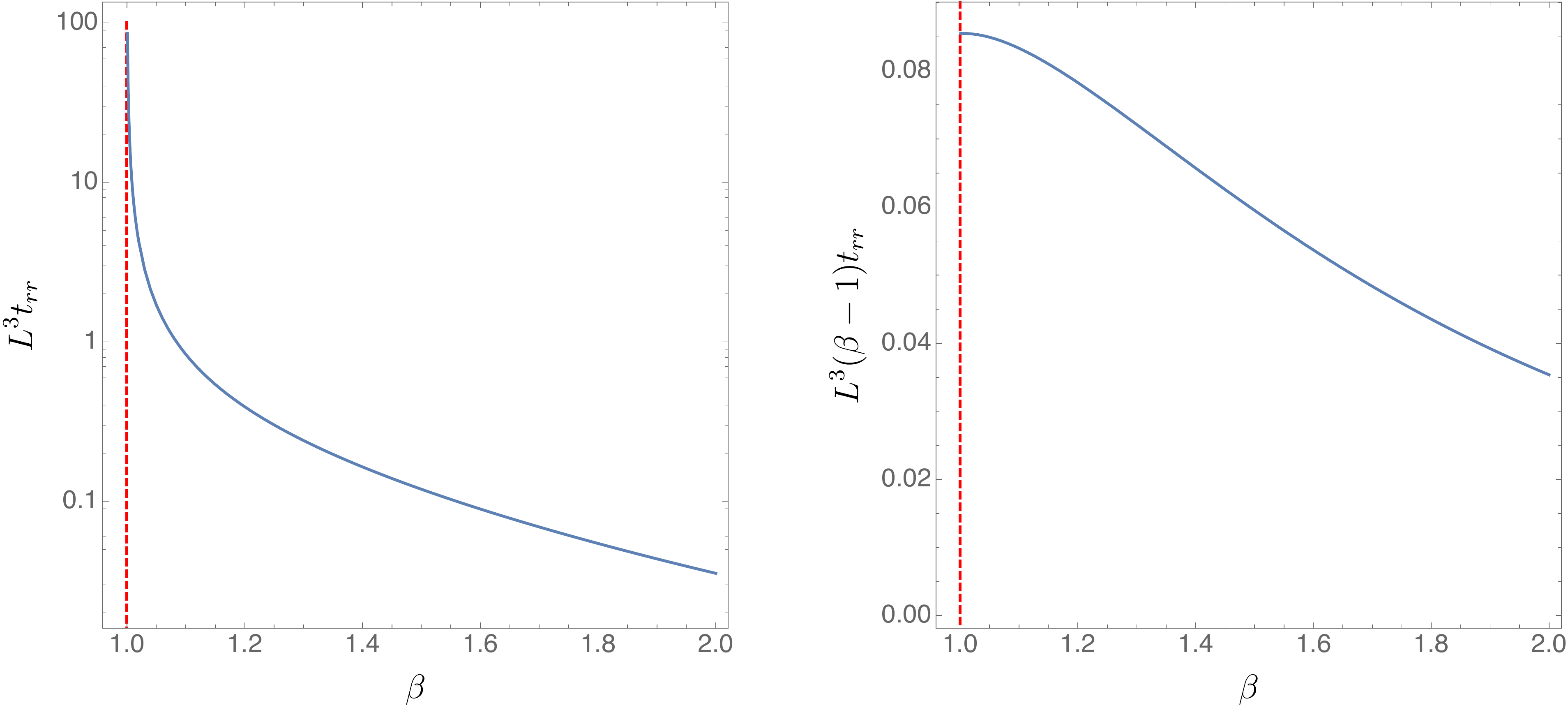}
	\caption{{\bf Left panel:} Logarithmic plot of $L^3 t_{rr}$ as a function of $\beta$. {\bf Right panel:} $L^3(\beta-1) t_{rr}$ as a function of $\beta$. Both plots are for $\Delta=2$ and $\alpha=2\pi$. The vertical red dashed line marks $\beta=1$. The plots show that $\lim_{r\rightarrow r_-} \langle 0|T_{rr} | 0 \rangle$ diverges as $1/(\beta-1)$ as $\beta\to1^+$.}
	\label{fig:check2}
\end{figure}

Next we will investigate the sharpness of the bound $\beta>1$ for $\langle 0|T_{\mu\nu} | 0 \rangle$ to extend continuously to the Cauchy horizon. If $\beta>1$ then  $\lim_{r\rightarrow r_-} \langle 0|T_{\mu\nu} | 0 \rangle=t_{\mu\nu}$. We now investigate what happens to $t_{\mu\nu}$ if we take $\beta \rightarrow 1^+$ with $\Delta$ and $\alpha$ fixed. This is what we plot on the left panel of Fig.~\ref{fig:check2} in a logarithmic scale for $\Delta=2$ and $\alpha=2\pi$. This plot shows that there is a divergence as $\beta \rightarrow 1^+$. On the right panel of Fig.~\ref{fig:check2} we plot $L^3(\beta-1)\lim_{r\rightarrow r_-} \langle 0|T_{rr} | 0 \rangle$, which remains finite and non-zero in the limit $\beta\to1^{+}$. This demonstrates that $\lim_{r\rightarrow r_-} \langle 0|T_{rr} | 0 \rangle$ diverges as $1/(\beta-1)$ as $\beta\to1^+$. We have checked this is the case for many values of $\Delta$. So our numerical results indicate that the bound $\beta>1$ is indeed sharp. 

\begin{figure}
	\centering
	\includegraphics[width=\textwidth]{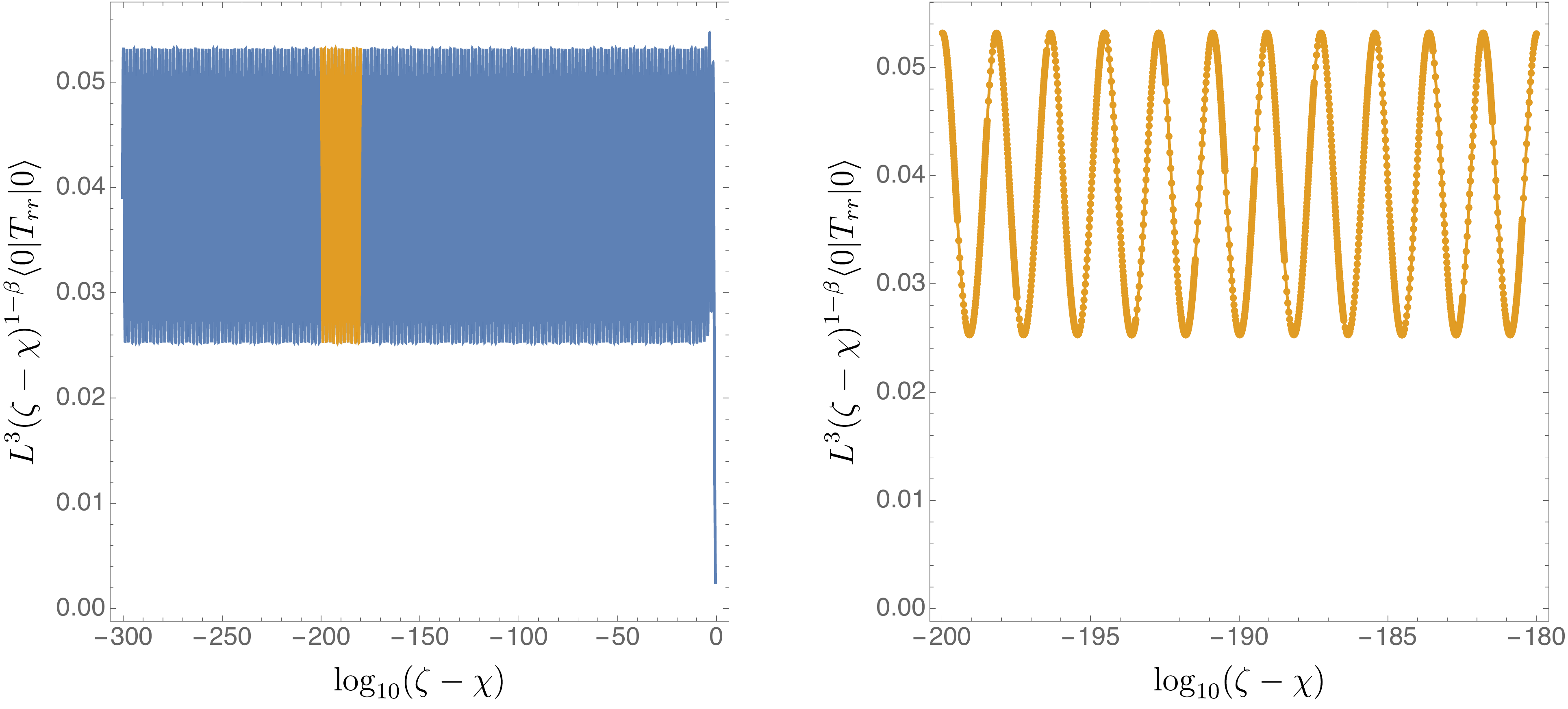}
	\caption{{\bf Left panel:} Plot of $(\zeta-\chi)^{1-\beta}L^3\langle 0|T_{rr} | 0 \rangle$ as a function of $\log_{10}(\zeta-\chi)$ showing the complete range of $\zeta$ that we probed. {\bf Right panel:} Plot of $(\zeta-\chi)^{1-\beta}L^3\langle 0|T_{rr} | 0 \rangle$ in the range $-200<\log_{10}(\zeta-\chi)<-180$ corresponding to the orange region on the left panel. Both panels are for $\Delta=1$, $\beta=1/2$ and $\alpha=2\pi$.}
	\label{fig:crazy}
\end{figure}

Now consider the case $\beta<1$. In all of the many cases we have investigated, we find that $L^3\langle 0|T_{rr} | 0 \rangle$ diverges as $r\rightarrow r_-$ as $1/(r-r_-)^{1-\beta}$, modulated by an oscillatory behaviour characteristic of discrete self-similarity with logarithm in $\zeta-\chi$ behaviour. In Fig.~\ref{fig:crazy} we plot $L^3(\zeta-\chi)^{1-\beta}\langle 0|T_{rr} | 0 \rangle$ as a function of $\log_{10}(\zeta-\chi)$ for $\Delta = 1$, $\beta=1/2$ and $\alpha=2\pi$. The plot on the left panel shows the range of $\zeta$ that we have probed, and on the right panel we show a zoom around the region $-200<\log_{10}(\zeta-\chi)<-180$ showing the discrete self-similar behaviour.
For $\log_{10}(\zeta-\chi)\lesssim-12$, the oscillations shown on the right panel can be fitted with the expression
\be
a_0 +b_0 \cos\left[\Omega_0 \log_{10}(\zeta-\chi)-c_0\right]\,,
\label{eq:self}
\ee
with $a_0$, $b_0$, $c_0$ and $\Omega_0$ all dependent on $\beta$, $\Delta$ and $\alpha$. For instance, for $\Delta=1$, $\beta=1/2$ and $\alpha=2\pi$ we obtain
\be
a_0\approx 0.0387\,,\qquad b_0\approx-0.0140\,,\quad c_0 \approx 3.5802\,,\qquad \text{and}\qquad \Omega_0 \approx 3.4534\,.
\ee
It would be interesting to understand analytically the origin of these oscillations. 

This divergence in $\langle 0|T_{ab} | 0 \rangle$ will backreact on the metric via the semi-classical Einstein equation $G_{ab} = 8 \pi \langle 0|T_{ab} | 0 \rangle$. It is interesting to note that the divergence is {\it integrable} and hence sufficiently mild that the Einstein equation can still be satisfied {\it weakly} at the Cauchy horizon. So, even for $\beta<1$, vacuum polarization does not enforce Christodoulou's version of strong cosmic censorship (although, as we have seen, classical perturbations enforce it for $\beta<1/2$).

In summary, we have shown analytically that $\langle 0|T_{ab} | 0 \rangle$ extends continuously to the Cauchy horizon $r=r_-$ if $\beta>1$ and our numerical results show that $\langle 0|T_{ab} | 0 \rangle$ diverges as $r \rightarrow r_-$ for $\beta<1$, where $\beta$ is defined by \eqref{betadef1}. 
\subsection{\label{subsec:Hadamardviolation}The Hartle-Hawking state is not smooth at the Cauchy horizon}

We have seen that, for a near-extremal BTZ black hole, $\langle 0|T_{ab} | 0 \rangle$ extends continuously to the Cauchy horizon. This is similar to what happens for classical perturbations. In the classical case, a generic perturbation is never smooth at the Cauchy horizon, although it can be made arbitrarily differentiable there by taking the black hole close enough to extremality. So one might wonder whether one can see a similar effect in our quantum field theory calculations for the Hartle-Hawking state. In this section we will show that there is indeed an analogous effect.

We will say that a state satisfies the {\it Hadamard condition to order $N$} if, in a convex normal neighbourhood ${\cal U}$ of any point $p$, the quantity $G^{(1)}_{\rm reg}(x,x')$ defined by \eqref{G1reg} is a $C^N$ function on ${\cal U} \times {\cal U}$. Consider some smooth extension of the BTZ geometry beyond the Cauchy horizon and, in this extended geometry, let ${\cal U}$ be a convex normal neighbourhood of a point on the Cauchy horizon. Now let ${\cal V} = {\cal U} \cap \{r \ge r_-\}$, which is independent of the choice of extension. If $r>r_-$ then the Hartle-Hawking state satisfies the Hadamard condition for any order $N$, so $G^{(1)}_{\rm reg}(x,x')$ is smooth for $x,x' \in {\rm int}{\cal V}$. We will say that the Hartle-Hawking state violates the order $N$ Hadamard condition at the Cauchy horizon if $G^{(1)}_{\rm reg}(x,x')$ cannot be extended to a $C^N$ function for $x,x' \in {\cal V}$. 

If the Hadamard condition is satisfied to order $2$ then $\langle 0|T_{ab} | 0 \rangle$ must be finite. Thus our results above demonstrate that if $\beta<1$ then the Hartle-Hawking state violates the order $2$ Hadamard condition at the Cauchy horizon. We now want to show that, for any given $\beta$, the Hartle-Hawking state violates the order $N$ Hadamard condition at the Cauchy horizon if $N$ is sufficiently large. 

We will restrict attention to a scalar field with $\Delta=1$. We will show that if $\beta<N$ then the Hartle-Hawking state violates the order $N$ Hadamard condition at the Cauchy horizon.\footnote{If $1<\beta<2$ then the Hartle-Hawking state violates the order $2$ Hadamard condition at the Cauchy horizon but $\langle 0|T_{ab} | 0 \rangle$ extends continuously to the Cauchy horizon. Violation of the order $2$ Hadamard condition is a necessary, but not sufficient, condition for $\langle 0|T_{ab} | 0 \rangle$ to diverge.} To do this, we will calculate 
\be
Z_N(r) \equiv \langle 0 |  \left\{ \partial_r^N \Phi(x), \Phi(x) \right\} |0  \rangle \equiv \lim_{x' \rightarrow x} \partial_r^N  G^{(1)}_{\rm reg}(x,x')
\ee 
where $\{,\}$ denotes the anticommutator, and we are using the coordinates $(u,r,\phi'')$ defined in \eqref{uphi''}, which are smooth at ${\cal CH}_R^+$. Note that symmetry implies that $Z_N$ is a function only of $r$. A divergence in $Z_N(r)$ at $r=r_-$ implies that the order $N$ Hadamard condition is violated at the Cauchy horizon. We will show (numerically) that $Z_N(r)$ indeed diverges as $r\rightarrow r_-$ if $\beta<N$. 
 
We will also prove that $Z_N(r)$ extends continuously to $r=r_-$ if $\beta>N$. Hence, for any given $N$, $Z_N(r)$ is finite at the Cauchy horizon if the black hole is close enough to extremality. Thus there is a close similarity with our results for classical perturbations. This result suggests that, for $\beta>N$, the Hartle-Hawking state satisfies the order $N$ Hadamard condition at the Cauchy horizon although we will not attempt to prove that here.

As mentioned above, we will focus on the case with $\Delta=1$, for which $G^{(1)}_{\rm reg}(x,x^\prime)$ simplifies considerably. Assuming $x,x'$ are spacelike separated, \eqref{G1sum} gives
\begin{equation}
\frac{1}{2}G^{(1)}_{\rm reg}(x,x')=\sum_{n=1}^\infty \left[f(z_n) + f(z_{-n}) \right]\,,
\label{eq:Gsum}
\end{equation}
where
\begin{equation}
\label{fDelta1}
f(z_n)=\frac{1}{4\pi L}\frac{1}{\sqrt{z_n^2-1}} \, ,
\end{equation}
with $z_n$ given in Eq.~(\ref{eq:zns}). In the $(u,r,\phi'')$ coordinates we have
\begin{subequations}
\begin{multline}
z_n-i \epsilon = \frac{1}{r_+^2-r_-^2}\Big[(r r^\prime-r_-^2)\cosh \xi_n^+-(r r^\prime-r_+^2)\cosh \xi_n^-
\\
-(r-r^\prime)(r_+\sinh \xi_n^--r_-\sinh \xi_n^+)\Big]\,,
\end{multline}
where we defined
\be
\xi_n^\pm = \frac{r_{\pm}}{L^2}(u-u^\prime)-\frac{r_\pm}{L}(\delta \phi''+2\,n\,\pi)\,.
\ee
\end{subequations}
We now differentiate the series term by term and let $x' \rightarrow x$ to obtain
\begin{equation}
Z_N(r) = \lim_{x' \rightarrow x} \frac{\partial^N}{\partial r^N}G^{(1)}_{\rm reg}(x,x')=2 \sum_{n=1}^\infty \left[f^{(N)}(z_n)  \left(\frac{\partial z_n}{\partial r}\right)^N + f^{(N)}(z_{-n})  \left(\frac{\partial z_{-n}}{\partial r}\right)^N \right]_{x'=x},
\label{eq:bigseries}
\end{equation}
where we used the fact that $z_n$ is a {\it linear} function of $r$. Now we need to study the convergence of the series. We have
\be
 \left( \frac{\partial z_n}{\partial r}\right)_{x'=x} = \frac{1}{r_+^2 - r_-^2} \left[  r \cosh (\alpha n) - r_- \sinh (\alpha n) - r\cosh (\alpha n \chi) + r_+ \sinh (\alpha n \chi) \right] 
\ee 
with $\alpha$ and $\chi$ defined in \eqref{alphadef}. In the region $r_- \le r \le r_0$ for any fixed $r_0>r_-$ we have
\be
 \left|  \left( \frac{\partial z_n}{\partial r}\right)_{x'=x}\right| < C e^{\alpha |n|}
\ee
for some constant $C$ depending only on $r_\pm$ and $r_0$. With $x'=x$ we know that $z_n=\bar{z}_n$ given by \eqref{znlim}. Since $\bar{z}_n \rightarrow \infty$ as $n \rightarrow \pm \infty$ we have
\be
 | f^{(N)} (\bar{z}_n) | < C' \bar{z}_n^{-(N+1)} < C'' e^{-(N+1) \alpha |n| \chi } 
\ee
where the first inequality follows from \eqref{fDelta1} and the second inequality from \eqref{znbound}. The constant $C''$ depends only on $N$, $r_\pm$ and $r_0$. Combining our estimates, we see that (for $x'=x$) the magnitude of the $n$th term in the sum on the RHS of \eqref{eq:bigseries} is bounded by $C''' e^{-(N+1) \alpha n \chi} e^{N \alpha n}$. Thus the series is uniformly convergent on $r_- \le r \le r_0$ if $\chi (N+1) >N$. This is equivalent to $\beta >N$ (as $\Delta =1$). Uniform convergence implies that the series defines a continous function on $r_- \le r \le r_0$. Hence if $\beta>N$ then $Z_N(r)$ extends continuously to $r=r_-$ as claimed above.

\begin{figure}
	\centering
	\includegraphics[width=\textwidth]{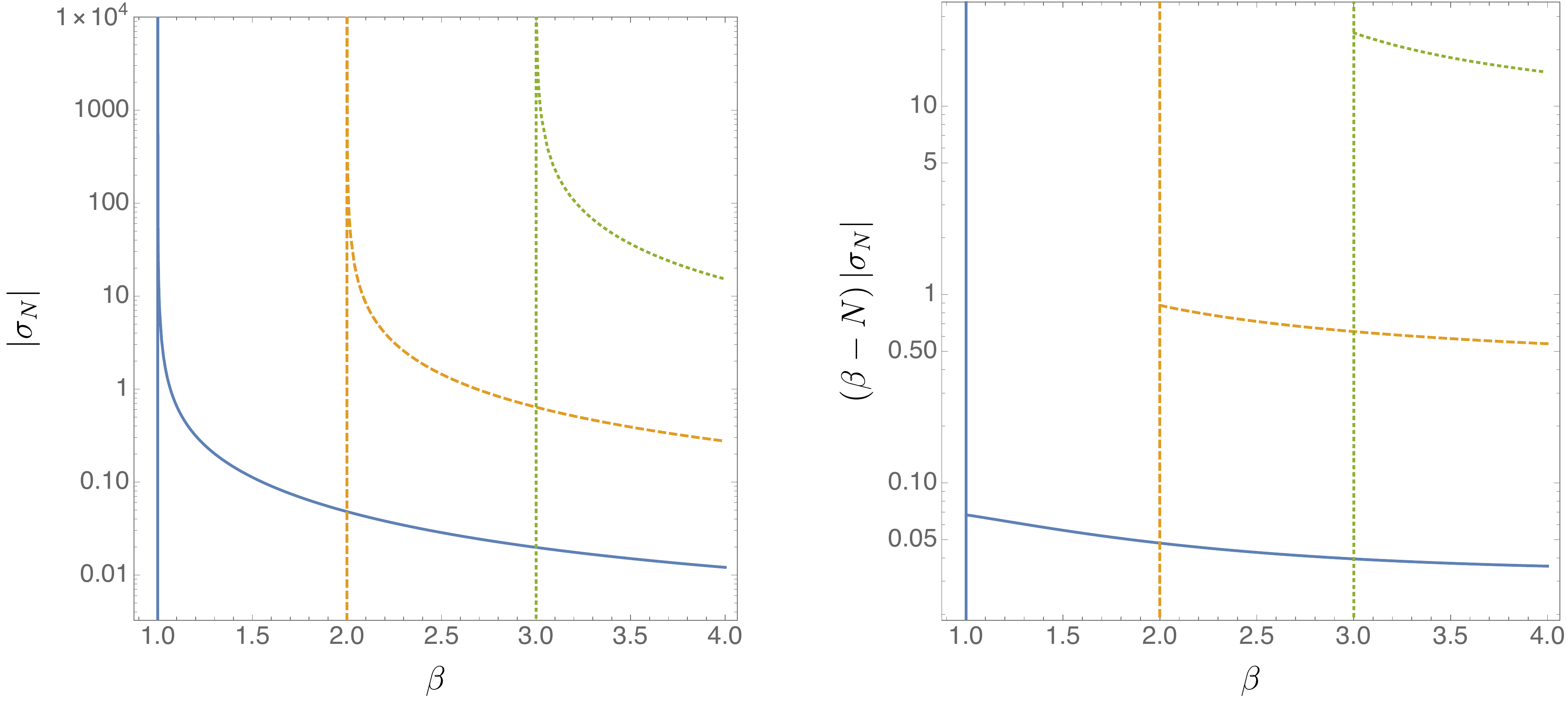}
	\caption{{\bf Left panel:} $\left|\sigma_N\right|$ computed for several values of $N$. {\bf Right panel:} $\left|(\beta-N)\sigma_N\right|$ as a function of $\beta$ computed for several values of $N$. In both panels, the solid blue curve corresponds to $N=1$, the orange dashed line to $N=2$ and the dotted green line to $N=3$. Finally, the vertical lines represent $\beta = 1,2,3$, from left to right. In all cases we fixed $\alpha=2\pi$.
	}
	\label{fig:last1}
\end{figure}

We now investigate numerically what happens as $\beta \rightarrow N^+$. We define $\sigma_N/L^N$ to be the result obtained by substituting $r=r_-$ into the RHS of \eqref{eq:bigseries}. Uniform convergence implies that, for $\beta>N$, we have 
\be
  \lim_{r \rightarrow r_-} L^N Z_N(r) =\sigma_N
\ee
The left panel of Fig. \ref{fig:last1} shows the behaviour, for $N=1,2,3$, of $\sigma_N$ as $\beta \rightarrow N^+$ with $\alpha$ fixed. We see that $\sigma_N$ diverges in this limit. From the right panel we see that $\left|(\beta-N)\sigma_N\right|$ remains finite and non-zero in this limit. Hence $\sigma_N$ diverges as $(\beta-N)^{-1}$ as $\beta \rightarrow N^+$ at fixed $\alpha$. This is very similar to the behaviour exhibited by $\langle 0|T_{ab} | 0 \rangle$ as $\beta \rightarrow 1^+$ (see Fig.  \ref{fig:check2}).

\begin{figure}
	\centering
	\includegraphics[width=\textwidth]{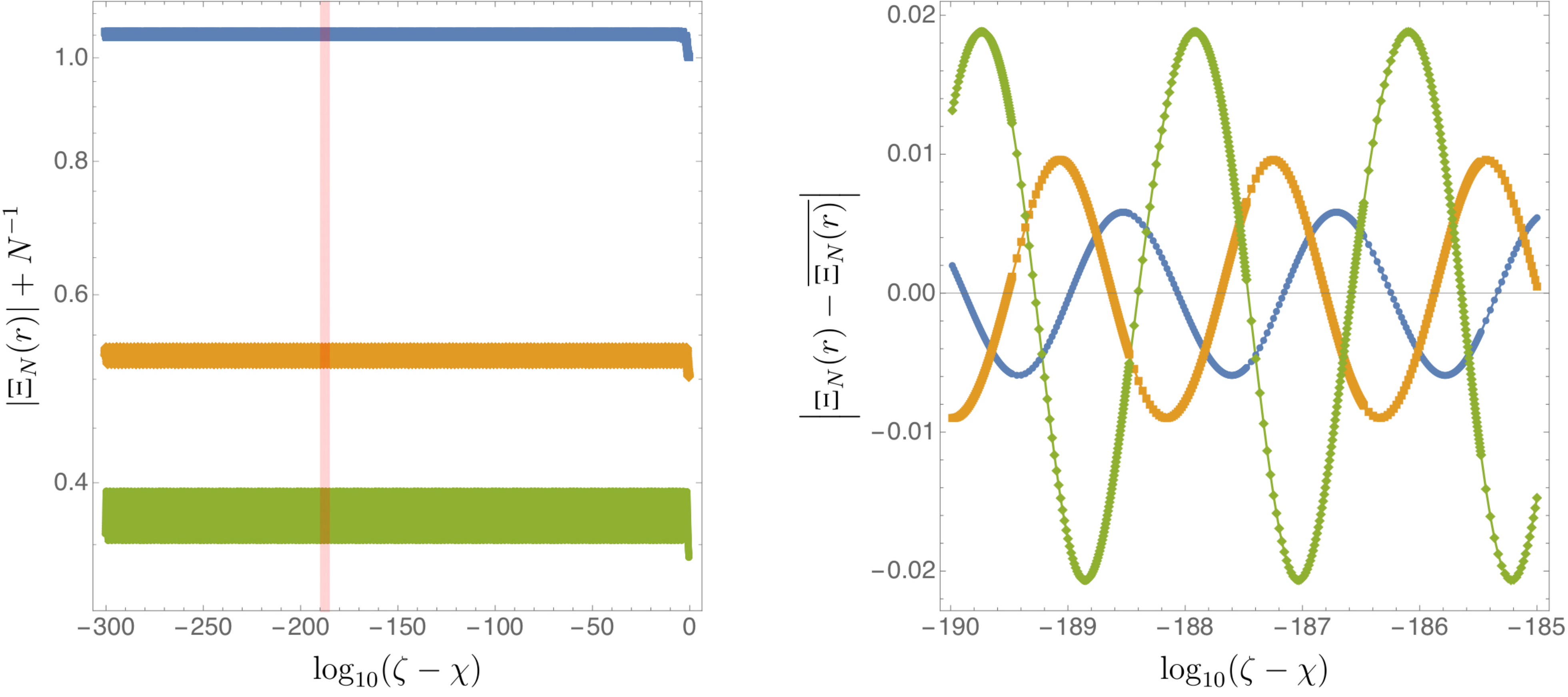}
	\caption{{\bf Left panel:} $\left|\Xi_N(r)\right|+N^{-1}$, computed for several values of $N$, as a function of $\log_{10}(\zeta-\chi)$. The $N^{-1}$ offset has been added to prevent the curves lying on top of each other. {\bf Right panel:} $\left|\Xi_N(r)-\overline{\Xi_N(r)}\right|$ in the region $-190<\log_{10}(\zeta-\chi)<-185$, corresponding to the red band on the left panel. In both panels, the blue disks corresponds to $N=1$, the orange squares to $N=2$ and the green diamonds to $N=3$. In all cases we we have $\alpha=2\pi$ and $\beta=1/2$.
	}
	\label{fig:last2}
\end{figure}

Finally we consider the behaviour of $Z_N(r)$ for $\beta <N$. Define (recall $\zeta=r/r_+$)
\begin{equation}
\Xi_{N}(r)\equiv L^N (\zeta - \chi)^{N - \beta} Z_N(r)\,,\quad \overline{\Xi_{N}(r)} \equiv \underset{\log_{10}(\zeta-\chi)<-100}{\mathrm{mean}}\Xi_{N}(r)\,.
\end{equation}
We find numerically that, at least near $r\gtrsim r_-$, $\Xi_{N}(x)$ is positive (negative) for even (odd) $N$. Fig.~\ref{fig:last2} shows the behaviour of $\Xi_N(r)$ as $r\rightarrow r_-$ (\emph{i.e.} $\zeta \rightarrow \chi$) with $\beta = 1/2$, $\alpha=2\pi$ and $N=1,2,3$. From the figure we see that $Z_N(r)$ exhibits divergent oscillations as $r\rightarrow r_-$. The amplitude of these oscillations diverges as $(r-r_-)^{\beta-N}$. We have repeated this calculation for several different values of $\beta$, and find similar results. Note that the period of the oscillations appears to be independent of $N$ and it matches the frequency of the oscillations seen in $\langle 0|T_{ab} | 0 \rangle$ for the same values of $\beta,\alpha$ (Fig. \ref{fig:crazy}).

\section{Discussion} 
\label{sec:discussion} 

We have shown that classical linear perturbations of a BTZ black hole can be made arbitrarily differentiable at the Cauchy horizon by taking the black hole sufficiently close to extremality. The high differentiability near extremality suggests that the linear approximation should be valid, {\it i.e.},  that nonlinearity will not modify our conclusion that strong cosmic censorship is violated badly by near-extremal BTZ black holes. 

The viewpoint of effective field theory is that classical equations should be supplemented by higher derivative terms that arise from ``integrating out" high energy degrees of freedom. So generically one would expect terms in the action involving higher derivatives of the scalar field that we studied in section \ref{sec:sccBTZviolation}. Generically this scalar field is only finitely differentiable at the Cauchy horizon so we expect that higher-derivative terms with sufficiently many derivatives will diverge at the Cauchy horizon. One might take this as an indication that the classical approximation is breaking down at the Cauchy horizon. However, we are reluctant to reach this conclusion. There are other situations in physics where one could say the same thing. For example, the formation of a shock in a compressible perfect fluid. In this case, first derivatives of the fields diverge, which is a much worse loss of smoothness than the one we have been discussing. Nevertheless, one can develop a theory of shocks without leaving the setting of a perfect fluid. The effect of ``higher derivative corrections", in this case viscosity, is simply to smooth out the shock without changing the qualitative predictions of the perfect fluid equations. 

Furthermore, in our case, the number of derivatives that one has to take to see a divergence can be made arbitrarily large by taking the black hole sufficiently close to extremality. If 2nd derivatives diverge then one might question the use of the classical approximation. But it seems far-fetched to object to the classical approximation because the first 99 derivatives are bounded but the 100th derivative diverges! Having said this, if the classical approximation is valid when we have non-smooth behaviour at the Cauchy horizon then, presumably, it should also be valid (physically) to allow non-smooth initial data. If we do this then, as we have described in section \ref{sec:rough}, there does seem to be a way of rescuing strong cosmic censorship \cite{Dafermos:2018tha}. So should we allow rough initial data? In the case of a perfect fluid, starting from smooth data we can form shocks dynamically and so one is forced to enlarge the class of permissible initial data to include rough data to describe pre-existing shocks. However, in our case, the lack of smoothness can occur only on the boundary of the region of predictability and so we are not forced to enlarge the class of permissible initial data. Therefore it is less clear that rough data is physical. 

Our result that, for a near-extremal black hole, $\langle 0|T_{ab} | 0 \rangle$ remains finite at the Cauchy horizon is surprising in view of the result for a conformally coupled scalar in a 2d toy model \cite{Birrell:1978th}, for which $\langle 0|T_{ab} | 0 \rangle$ always diverges at the Cauchy horizon. In the 4d case, a recent study \cite{Lanir:2018vgb} found that (for a massless scalar) $\langle 0|T^{\phantom{a}a}_a | 0 \rangle$ diverges at the Cauchy horizon of a specific non-extremal Reissner-Nordstr\"om black hole and stated that this divergence is actually present for {\it any} non-extremal Reissner-Nordstr\"om solution. This differs from what we find for the BTZ black hole. However, classically, the BTZ black hole exhibits behaviour more similar to Reissner-Nordstr\"om-{\it de Sitter}, for which strong cosmic censorship fails near extremality \cite{Cardoso:2017soq,Dias:2018etb}. In would be interesting to see whether this similarity holds also for the behaviour of $\langle 0|T_{ab} | 0 \rangle$ by extending the analysis of \cite{Lanir:2018vgb} to include a positive cosmological constant.

Another interesting comparison is with the behaviour of $\langle 0|T_{ab} | 0 \rangle$ in the BTZ background for a strongly coupled, large $N$, CFT with a holographic dual. This has been calculated in Ref. \cite{Hubeny:2009rc}, from which we can deduce the behaviour at the Cauchy horizon. The result is that, to leading order in $1/N$, $\langle 0|T_{ab} | 0 \rangle$ is finite (indeed smooth) at the Cauchy horizon of {\it any} rotating BTZ black hole. But this is only the leading order behaviour. Our results suggest that a lack of smoothness (or a divergence) at the Cauchy horizon should appear at subleading order in the $1/N$ expansion. 

Turning to other possibilities for future work, it would be nice to understand analytically the behaviour shown in Figs. \ref{fig:crazy} and \ref{fig:last2}. It would be interesting to calculate $\langle 0|T_{ab} | 0 \rangle$ in the rotating BTZ geometry for other types of free field, especially fermionic fields. A much harder calculation would be to go beyond leading order in the semi-classical Einstein equation and determine the backreaction of quantum fields to second order in $\hbar$.  This would probably require computing $\langle 0|T_{ab}(x) T_{cd}(y) | 0 \rangle$ in the BTZ geometry.

\subsection*{Acknowledgments}

We are very grateful to Vijay Balasubramanian and Christoph Kehle for comments on a draft of this paper. We thank Don Marolf for helpful discussions which led to the work of section \ref{subsec:Hadamardviolation}. We also thank Dejan Gajic and David Tong for helpful conversations. OJCD is supported by the STFC Ernest Rutherford Grant No. ST/K005391/1, and by the STFC ``Particle Physics Grants Panel (PPGP) 2016" Grant No. ST/P000711/1. HSR and JES were supported in part by STFC Grants No. PHY-1504541 and ST/P000681/1.

\appendix
\section{Bounds on derivatives of $z_n$}

In this Appendix we will explain the proof of the bounds \eqref{dznbound} and \eqref{ddznbound}. A straightforward calculation gives
\ba
\label{dVminuszn}
 \partial_{V_-} z_n &=& \frac{1}{r_+^2 - r_-^2} \left\{ \frac{1}{2} \dot{G}_-(r) G_-(r')\left[ U_- V_-' \exp\left( \frac{r_+}{L} \Delta \phi_{-n} \right) + U_-' V_-  \exp\left( -\frac{r_+}{L} \Delta \phi_{-n} \right) \right]  \partial_{V_-}r \right. \nonumber \\
 &+& \left. \frac{1}{2} G_-(r) G_-(r') U_-'  \exp\left( -\frac{r_+}{L} \Delta \phi_{-n} \right) - \frac{r\sqrt{r_+^2 - {r'}^2} }{\sqrt{r_+^2-r^2}}  \cosh \left( \frac{r_-}{L} \Delta \phi_{-n} \right) \partial_{V_- }r\right\}
\ea
where $\dot{G}_-$ denotes the derivative of $G_-$. The definition of $r(U_-,V_-)$ \eqref{rUVminus} gives
\be
 \partial_{V_-} r = \frac{U_-}{\dot{F}_-(r)}\,,
\ee
where $\dot{F}_-$ is the derivative of $F_-$. From the definition of $G_-(r)$ we obtain
\be
\dot{G}_- = \frac{r}{F_-G_-} - \frac{(r^2-r_-^2)\dot{F}_-}{2F_-^2 G_-}\,.
\ee
Using these results, together with \eqref{rUVminus} we obtain
\be
 \lim_{x' \rightarrow x}  \partial_{V_-} z_n=\frac{U_-}{r_+^2 - r_-^2} \left \{ \frac{r}{\dot{F}_-(r)} \left[ \cosh \left(\frac{2\pi r_+ n}{L}\right) - \cosh \left(\frac{2\pi r_- n}{L}\right) \right] - \frac{G_-(r)^2 }{2} \exp \left(\frac{2\pi r_+ n}{L}\right) \right\}
\ee
hence
\ba
| \lim_{x' \rightarrow x}  \partial_{V_-} z_n| &\le& \frac{|U_-|}{r_+^2 - r_-^2} \left \{ \frac{r}{|\dot{F}_-(r)|}  \cosh \left(\frac{2\pi r_+ n}{L}\right) + \frac{1}{2} G_-(r)^2 \exp \left(\frac{2\pi r_+ |n|}{L}\right) \right\} \nonumber \\
& \le &\frac{|U_-|}{r_+^2 - r_-^2} \left( \frac{2r}{|\dot{F}_-(r)|}+ \frac{G_-(r)^2}{2} \right) \exp \left(\frac{2\pi r_+ |n|}{L}\right).
\ea
From its definition, we see that $G_-(r)$ is continuous on the interval $[r_-,r_+]$ so $G_-(r)^2$ is bounded on this interval. One can show that $\dot{F}_-(r) \ne 0$ for $r \in [r_-,r_+)$ and $\dot{F}_-(r)$ diverges as $r \rightarrow r_+$. Hence $r/\dot{F_-}(r)$ is also bounded on $[r_-,r_+]$. This establishes the second equation in \eqref{dznbound}. The other equations in \eqref{dznbound} follow similarly.

Now we consider second derivatives of $z_n$. Taking the $V_-'$ derivative of \eqref{dVminuszn} and then taking the limit $x' \rightarrow x$ gives
\ba
\lim_{x' \rightarrow x} \partial_{V_-} \partial_{V_-'} z_n &=& \frac{U_-^2}{r_+^2-r_-^2} \left\{\frac{r^2}{\dot{F}_-(r)^2 (r_+^2 - r^2)}\cosh \left(\frac{2\pi r_- n}{L}\right) \right.  \\
&+& \left.
 \frac{1}{F_-(r) G_-(r)^2}\left( \frac{r}{\dot{F_-}(r)} - \frac{G_-(r)^2}{2} \right) \left( \frac{r}{\dot{F_-}(r)} +  \frac{G_-(r)^2}{2} \right) \cosh \left(\frac{2\pi r_+ n}{L}\right)  \right\}. \nonumber 
\ea
To bound this we first use $\cosh (2 \pi r_\pm n/L) \le 2 \exp(2 \pi r_+ |n|/L)$. We then bound the $r$-dependent coefficients as follows. First, using the properties of $\dot{F}_-$ discussed above, we see that $r^2/(\dot{F}_-^2(r_+^2-r^2))$ is bounded on $[r_-,r_0]$ because $r_-<r_0<r_+$. Second, we have
\be
\frac{1}{G_-(r)^2} \left| \frac{r}{\dot{F}_-(r)} +  \frac{G_-(r)^2}{2} \right| <C\,,
\ee
for some constant $C$ and $r \in [r_-,r_0]$ using the properties of $\dot{F}_-(r)$ and the fact that $G_-(r)$ is continuous and non-vanishing on $[r_-,r_0]$. Thirdly we have
\be
\frac{1}{F_-(r)}\left| \frac{r}{\dot{F_-}(r)} - \frac{G_-(r)^2}{2} \right|= \frac{1}{F_-(r)} \left|\frac{r}{\dot{F_-}(r)}  - \frac{r^2-r_-^2}{2F_-(r)} \right| < C'\,,
\ee
for some constant $C'$ and $r\in [r_-,r_0]$. This is because the expression inside the modulus has a first order zero at $r=r_-$ which cancels the corresponding zero of $F_-(r)$. Putting all of this together we obtain the second equation in \eqref{ddznbound}. 

Similarly, taking the $U_-'$ derivative of \eqref{dVminuszn} and then taking the limit $x' \rightarrow x$ gives (after using \eqref{rUVminus})
\ba
\lim_{x' \rightarrow x} \partial_{V_-} \partial_{U_-'} z_n &=& \frac{1}{r_+^2-r_-^2} \left\{ \frac{1}{G_-(r)^2} \left( \frac{r}{\dot{F}_-(r)} - \frac{G_-(r)^2}{2} \right)^2 \cosh  \left(\frac{2\pi r_+ n}{L}\right)  \right. \nonumber \\ &+& \ \frac{r}{\dot{F}_-(r)}  \exp \left(-\frac{2\pi r_+ n}{L}\right) + \left. \frac{r^2 F_-(r)}{(r_+^2-r^2) \dot{F}_-(r)^2} \cosh  \left(\frac{2\pi r_- n}{L}\right) \right\}.
\ea
Using the arguments above it is easy to show that this satisfies the bound in \eqref{ddznbound} for $r \in [r_-,r_0]$. The other equations in \eqref{ddznbound} are obtained similarly, most with slightly less work.

\bibliographystyle{JHEP}
\bibliography{btz}{}

\end{document}